\documentclass[11pt,letterpaper]{article}
\pdfoutput=1

\usepackage{jcappub}

\usepackage{amsmath,graphicx}
\usepackage{simplewick}
\usepackage{epsf}
\usepackage{graphicx,epsfig}
\usepackage{amsfonts}
\usepackage{amssymb}
\usepackage{setspace}
\usepackage{caption}
\usepackage{epstopdf}
\usepackage{color}
\newcommand{\nc}{\newcommand}
\nc{\ba}{\begin{eqnarray}}
\nc{\ea}{\end{eqnarray}}
\newcommand\be{\begin{equation}}
\newcommand\ee{\end{equation}}

\usepackage{slashed}
\nc{\x}{{\bf{x}}}
\nc{\bfk}{{\bf{k}}}

\usepackage{bm}
\usepackage{natbib}
\usepackage{epstopdf}
\usepackage{hhline}
\usepackage{relsize}
 \usepackage[titletoc,toc,title]{appendix}
 \usepackage{array}
\usepackage[T1]{fontenc}

\newcommand{\bea}{\begin{eqnarray}}
\newcommand{\eea}{\end{eqnarray}}
\newcommand{\barr}{\begin{array}}
\newcommand{\earr}{\end{array}}

\def\bal#1\eal{\begin{align}#1\end{align}}

\usepackage{tikz}
\usetikzlibrary{trees}
\usetikzlibrary{decorations.pathmorphing}
\usetikzlibrary{decorations.markings}
\tikzset{
    photon/.style={decorate, line width=0.15mm, decoration={snake,amplitude=3pt,segment length=8pt}, draw=black},
    wino/.style={draw=redwine},    
    fermion/.style={draw=black, line width=0.2mm, postaction={decorate},
        decoration={markings,mark=at position .55 with {\arrow[draw=black,scale=2,#1]{>}}}},
    scalar/.style={draw=black, dashed,postaction={decorate},
        decoration={markings,mark=at position .55 with {\arrow[draw=black,scale=2,#1]{>}}}},
    scalarline/.style={draw=black, postaction={decorate},
        decoration={markings,mark=at position .55 with {\arrow[draw=black,scale=2,#1]{>}}}},
    gluon/.style={decorate, draw=black,
        decoration={coil,amplitude=3pt, segment length=4pt}},
    graviton/.style={decorate, draw=black,
        decoration={zigzag,amplitude=3pt, segment length=4pt}}
}
\tikzstyle{blob}=[circle,
                              minimum size=20,
                              draw=black!80,
                              fill=black!80]   
\tikzstyle{redblob}=[circle,
                              thick,
                              minimum size=0.4cm,
                              draw=red!80,
                              fill=red!60]

\begin{document}
\title{Enhanced $n$-body annihilation of dark matter and its indirect signatures}

\author[a]{Mohammad Hossein Namjoo,}

\affiliation[a]{Center for Theoretical Physics, Massachusetts Institute of Technology, Cambridge, MA 02139, USA}

\author[a,b]{Tracy R. Slatyer,}

\affiliation[b]{School of Natural Sciences, Institute for Advanced Study, Einstein Drive, Princeton, NJ 08540, USA}

\author[a]{Chih-Liang Wu}

\emailAdd{namjoo@mit.edu}
\emailAdd{tslatyer@mit.edu}
\emailAdd{cliang@mit.edu}

\abstract{We examine the possible indirect signatures of dark matter annihilation processes with a non-standard scaling with the dark matter density, and in particular the case where more than two dark matter particles participate in the annihilation process. We point out that such processes can be strongly enhanced at low velocities without violating unitarity, similar to Sommerfeld enhancement in the standard case of two-body annihilation, potentially leading to visible signals in indirect searches. We study in detail the impact of such multi-body annihilations on the ionization history of the universe and consequently the cosmic microwave background, and find that unlike in the two-body case, the dominant signal can naturally arise from the end of the cosmic dark ages, after the onset of structure formation. We examine the complementary constraints from the Galactic Center, Galactic halo, and galaxy clusters, and outline the circumstances under which each search would give rise to the strongest constraints. We also show that if there is a population of ultra-compact dense dark matter clumps present in the Milky Way with sufficiently steep density profile, then it might be possible to detect point sources illuminated by multi-body annihilation, even if there is no large low-velocity enhancement. Finally, we provide a case study of a model where 3-body annihilation dominates the freezeout process, and in particular the resonant regime where a large low-velocity enhancement is naturally generated.}

\preprint{MIT-CTP/5052}

\maketitle
\section{Introduction}

In recent years there has been great interest in dark matter (DM) models with modified thermal histories, where DM annihilation is still responsible for setting the late-time abundance of DM, but the annihilation involves three or more DM (or DM-like) particles (e.g. \cite{1992ApJ...398...43C, Hochberg:2014dra, Hochberg:2014kqa, Bernal:2015bla, Lee:2015gsa, Hansen:2015yaa, Hochberg:2015vrg, Choi:2015bya, Choi:2016hid, Dey:2016qgf, Bernal:2017mqb, Choi:2017mkk, Cline:2017tka, Ho:2017fte, Choi:2017zww, Hochberg:2017khi, Hochberg:2018rjs}). This pushes the preferred mass scale for the DM below 1 GeV \cite{Hochberg:2014dra}, a range of interest for novel direct detection searches (see e.g. \cite{Battaglieri:2017aum} for a recent review) and which has interesting implications for DM self-interaction rates, while simultaneously evading indirect-detection constraints on sub-GeV DM whose abundance is set by ordinary two-body annihilation \cite{Slatyer:2015jla}. Such light DM models are quite challenging to probe with conventional direct detection and at high-energy colliders. Indirect detection signals could be present if the final state of the annihilation process involves Standard Model (SM) particles, or unstable particles which subsequently decay back to the SM; this feature is not present in all models with higher-body annihilation processes, but even when it is, the conventional wisdom is that these signals will always be unobservably tiny long after freezeout, since they are suppressed by an extra power of the DM density compared to the usual case of 2-body annihilation.

There are several possible caveats to this argument:
\begin{itemize}
\item For parameter regions and search channels where  2-body thermal relic cross sections are already very strongly excluded, it might be possible to observe a 3-body thermal relic signal.
\item In ultracompact minihalos or DM cusps around black holes, where the DM density could potentially be very large, the suppression of $n$-body processes relative to 2-body would be reduced, and it might be possible to observe a signal.
\item The DM annihilation rate may depend on parameters other than the DM density; in particular, the cross section may be a function of the relative velocity of the interacting particles, and could be strongly enhanced at low velocities, with larger enhancements for $n$-body (for $n>2$) annihilation compared to 2-body annihilation.
\end{itemize}

In this work, we will explore these three caveats, to address the question of whether it could ever be possible to detect SM particles produced by $n$-body annihilation in such models, with $n>2$, in cases where (at least some of) the immediate products of annihilation decay back to the SM if they are not SM particles themselves.

In particular, we note that the partial-wave unitarity bound on the annihilation rate has a much stronger velocity dependence as $n$ increases, permitting annihilation rates scaling as rapidly as $v^{-(1 + 3 (n-2))}$. In the presence of effective long-range interactions, multiple partial waves can contribute and the velocity scaling can be even stronger. We present an example model of a secluded dark sector where due to the presence of a resonance, the annihilation rate has a very strong velocity scaling down to some regulating velocity scale, leading to greatly enhanced signatures in low-redshift indirect probes (resonant $3\rightarrow 2$ interactions were also considered e.g. in Ref.~\cite{Choi:2016hid}). The phenomenology of such models is similar in some ways to Sommerfeld-enhanced \cite{Hisano:2004ds} or Breit-Wigner-enhanced \cite{Ibe:2008ye} DM annihilation in the two-body case, but the different density scaling and the much sharper velocity dependence necessitates a reconsideration of indirect-detection limits.

We will explore indirect-detection limits on DM at the keV-GeV scale from the cosmic microwave background and X-ray/$\gamma$-ray observations of galaxies and clusters. We will attempt to proceed in a maximally model-independent manner, estimating limits on the parameter which describes the amount of power injected through such annihilation processes. This will allow us to compare the constraining power of different search channels in a largely model-independent way. We focus here on indirect detection as it constrains the $n$-body annihilation process directly. 

Models of this type could also potentially be detectable at colliders or in direct detection experiments, depending on the strength of the coupling between the dark sector and the SM;  indirect detection signals are relatively insensitive to this coupling if the dominant annihilation channel involves only dark sector particles, as the coupling only controls the decay lifetime of unstable dark-sector particles into the SM (which can be quite long without affecting the constraints). However, appreciable couplings between the dark sector and the Standard Model will tend to generate observable signals from 2-body annihilation (directly into SM particles) that overwhelm any $n$-body signal for $n > 2$, so we are driven to consider models with very small SM couplings that in turn have very suppressed signatures in terrestrial experiments.

The paper is organized as follows: in Section \ref{sec:parametric}, we provide general parametric estimates for freezeout through velocity-independent $n$-body processes, the strength of such $n$-body processes relative to 2-body annihilation in regimes relevant for indirect detection, and upper bounds on the $n$-body annihilation signal arising from upper bounds on the DM density. In Section \ref{sec:sommerfeld}, we discuss upper bounds from unitarity on the $n$-body annihilation cross section at low velocities, for general $n$, and outline some implications if a large low-velocity enhancement is present. In Section \ref{sec:boost}, we discuss our methodology for estimating contributions to annihilation signals from DM halos and subhalos; this is already a major source of uncertainty for 2-body annihilation and we will see that the effect is amplified for $n$-body annihilations with $n > 2$. In Section \ref{sec:constraints}, we derive and compare the sensitivity of various indirect searches to 3- and 4-body annihilation, attempting to work as model-independently as possible. In Section \ref{sec:ucmh}, we discuss the potential for detecting annihilation signals from a population of UCMHs. In Section \ref{sec:models}, we perform a case study of one specific class of models, where $3$-body annihilation dominates freezeout and there is potential for large low-velocity enhancements. We present our conclusions in Section \ref{sec:conclusion}. 

\section{Parametric estimates for velocity-independent $n$-body annihilation}
\label{sec:parametric} 

\subsection{Cross sections for $n$-body processes}

Let us first briefly discuss notation. We will use the notation $\langle \sigma v^{n-1}\rangle$ to denote the annihilation rate coefficient for a given velocity distribution of $n$ annihilating particles. If the $n$ annihilating particles are all distinguishable, and have mass densities $\rho_{1}, \cdots, \rho_{n}$, and masses $m_{\chi_1}, \cdots, m_{\chi_n}$ (here we assume all initial-state particles are non-relativistic), then the annihilation rate per unit volume per unit time is $\langle \sigma v^{n-1}\rangle (\rho_{1}/m_{\chi_1}) \cdots (\rho_{n}/m_{\chi_n})$. If some of the annihilating particles are identical, this expression must be corrected by a combinatoric factor; each set of $j$ identical particles contributes a factor $1/j!$ to the annihilation rate. Note as a consequence that the units of $\langle \sigma v^{n-1}\rangle$ depend on $n$; in natural units, $\langle \sigma v^{n-1}\rangle$ has mass dimension $-(3 n - 4)$.

Suppose the annihilation has $m$ final-state particles (here $m$ is an index, not to be confused with the particle masses), and $S_f$ is the symmetry factor associated with identical particles in the final state, receiving a factor of $j!$ for every set of $j$ identical particles. Let us label the incoming momenta as $p_1\cdots p_n$ and the outgoing momenta as $p_{n+1}\cdots p_{n+m}$. Then the relation between the $n\rightarrow m$ matrix element $\mathcal{M}$ (as computed from Feynman diagrams) and $\langle \sigma v^{n-1}\rangle$ is given by (e.g. \cite{Cline:2017tka}):
\begin{equation}\langle \sigma v^{n-1}\rangle = \frac{1}{S_f} \frac{1}{\prod_{i=1..n} (\rho_i/m_{\chi_i})} \int \prod_{i=1}^{n+m} \frac{g_i d^3 p_i}{(2\pi)^3 (2 E_i)} (2\pi)^4 \delta^{4}\left(\sum_{j=1}^n p_j - \sum_{j=n+1}^{n+m} p_j \right) \left(\prod_{k=1}^n f_k  \right) \overline{|\mathcal{M}|^2}, \label{eq:temp_average} \end{equation}
where $\overline{|\mathcal{M}|^2}$ is the squared matrix element averaged over the degrees of freedom in both the initial and final states, $f_k$ is the distribution function of the $k$th particle in the initial state, and each $g_i$ factor counts the degrees of freedom (e.g. spin) in the state labeled by $i$.

In the two-body case, if the initial particles are highly non-relativistic and final-state particles are highly relativistic, $\sigma v$ is approximately independent of the relative velocity of the annihilating particles (and hence equal to $\langle \sigma v \rangle$, which is approximately independent of the velocity distribution) for $s$-wave annihilation, where $\mathcal{M}$ is velocity-independent. Likewise, in the general case we can separate the integral into parts pertaining to the initial state and final state:
\begin{equation}\langle \sigma v^{n-1}\rangle = \frac{1}{S_f} \frac{1}{\prod_{i=1..n} (\rho_i/m_{\chi_i})} \left(\int \prod_{i=1}^{n} \frac{g_i f_i d^3 p_i}{(2\pi)^3 (2 E_i)} \right) \left(\prod_{i=m}^{n+m} g_i \right) \int d\Pi_m \overline{|\mathcal{M}|^2},\end{equation}
Here $d\Pi_m$ is the standard $m$-body Lorentz-invariant phase space, 
\begin{equation} d\Pi_m = (2\pi)^4 \delta^{4}\left(\sum_{j=1}^n p_j - \sum_{j=n+1}^{n+m} p_j \right) \prod_{i=n+1}^{n+m} \frac{d^3 p_i}{(2\pi)^3 (2 E_i)} ,\label{eq:pils}\end{equation}
set by the total incoming momentum $\sum_{i=1}^n p_i$; in particular, for a 2-body final state in the center-of-momentum (COM) frame, it takes the form $\int d\Pi_m \overline{|\mathcal{M}|^2} = \frac{1}{16\pi^2} |\vec{p}|/\sqrt{s} \int d\Omega \overline{|\mathcal{M}|^2}$, where $\vec{p}$ is the 3-momentum of either of the two final-state particles, and $s = (\sum_{i=1}^n p_i)^2$. (Note that we do not include the $g_i$ factors in this phase space, but track them separately.) If the final-state particles are much less massive than the initial center-of-mass energy $\sqrt{s}$, and hence are relativistic, and $|\mathcal{M}|^2$ is momentum-independent, then $d\Pi_m \overline{|\mathcal{M}|^2}$ is approximately independent of the initial momenta when the initial particles are all non-relativistic and $\sqrt{s}$ is dominated by their masses. In this case, the integrals over the initial particle phase-space distributions can be performed trivially, yielding factors of the initial-particle number densities, and as in the 2-body case, we find that $\langle \sigma v^{n-1}\rangle$ is independent of the characteristic velocity of the colliding particles.

If the final-state particles are not relativistic, the two-body final-state phase space will contain a factor of their 3-momentum, leading to a phase-space suppression in the annihilation rate as usual. In this case, or if $\overline{|\mathcal{M}|^2}$ has a non-trivial momentum dependence, then the full integral must be performed.

\subsection{Thermal freezeout estimates} 

We now review how the standard thermal freezeout scenario is modified when the dominant interaction with the radiation bath is through a process involving $n$ DM particles, following the estimates of Ref.~\cite{Hochberg:2014dra}. For the moment, let us assume that the annihilation products are in thermal equilibrium with the SM bath. We will find it illustrative to write our results in terms of the DM mass, the temperature of matter-radiation equality (which sets the DM abundance), and the Planck mass.

Freezeout occurs when the annihilation rate is comparable to the Hubble rate. During radiation domination, the Hubble rate is given by $H^2 \sim \frac{1}{m_\text{Pl}^2} \rho_r \sim \frac{1}{m_\text{Pl}^2} g_* T^4$, where $\rho_r = (\pi^2/30) g_* T^4$ is the radiation energy density ($g_*$ counts the number of relativistic degrees of freedom as usual), i.e. $H \sim g_*^{1/2} T^2/m_\text{Pl}$. Thus the criterion for freezeout becomes $(g_*)_f^{1/2} T_f^2/m_\text{Pl} \sim H \sim (\rho_{\chi, f}/m_\chi)^{n-1} \langle \sigma v^{n-1}\rangle$, where $\rho_\chi$ is the density of the annihilating species, and $f$ subscripts indicate evaluation at the time of freezeout. If freezeout occurs due to the particle becoming non-relativistic and experiencing a Boltzmann suppression to its number density, then $x_f = m_\chi/T_f$ is expected to be a $\mathcal{O}(1)$ number (broadly defined; it could easily be 1-2 orders of magnitude larger) without a strong dependence on the other parameters of the problem. Thus it is convenient to write:
\begin{equation}  \langle \sigma v^{n-1}\rangle \sim \frac{(g_*)_\text{f}^{1/2}}{x_f^2 m_\text{Pl}} \rho_{\chi, f}^{1-n} m_\chi^{n+1}. \label{eq:sigmav} \end{equation} 

The DM density at matter-radiation equality (denoted by ``eq'' subscripts) is then given by $\rho_{\chi, f} (a_f/a_\text{eq})^3 = \rho_{\chi, f} s_\text{eq} / s_f = \rho_{\chi, f} (g_{*S})_\text{eq} T_\text{eq}^3/(g_{*S})_f T_f^3$. Here $g_{*S}$ follows the standard definition, with the entropy density $s = (2\pi^2/45) g_{*S} T^3$. While all relativistic species have the same temperature, $g_{*S} = g_*$ to a good approximation; the two diverge slightly (at the $20\%$ level) after the neutrino and photon temperature decouple.

 The radiation energy density at matter-radiation equality is given by $\rho_{r,\text{eq}} \sim (g_*)_\text{eq} T_\text{eq}^4$. Equating these two quantities (and ignoring the non-DM contribution to matter, since we are dropping $\mathcal{O}(1)$ factors) gives:
 \begin{equation}  \rho_{\chi, f}  = \dfrac{(g_*)_\text{eq} (g_{*S})_f T_\text{eq}T_f^3}{(g_{*S})_\text{eq}} =  \frac{(g_*)_\text{eq} (g_{*S})_f T_\text{eq} m_\chi^3}{(g_{*S})_\text{eq} x_f^3}, \label{eq:rhof} \end{equation}
and substituting this into Eq.~\eqref{eq:sigmav} above, we obtain:
 \begin{align}  \langle \sigma v^{n-1}\rangle & \sim \frac{(g_*)_\text{f}^{1/2}}{x_f^2 m_\text{Pl}} \left(\frac{(g_*)_\text{eq} (g_{*S})_f T_\text{eq} m_\chi^3 }{(g_{*S})_\text{eq} x_f^3} \right)^{1-n} m_\chi^{n+1}, \nonumber \\
 & =  \left[ (g_*)_\text{f}^{1/2} \left(\frac{(g_*)_\text{eq} (g_{*S})_f }{(g_{*S})_\text{eq} } \right)^{1-n} \right] x_f^{-5 + 3n}  \frac{m_\chi^{4 - 2n}  T_\text{eq}^{1-n} }{ m_\text{Pl}}. \end{align} 
For the estimates contained in this section, we will assume $g_{*S} \approx g_*$ at all times (however, any numerical calculations track both separately); thus we obtain:
  \begin{align}  \langle \sigma v^{n-1}\rangle & \sim  (g_*)_\text{f}^{\frac{3}{2} - n} x_f^{-5 + 3n}  \frac{m_\chi^{4 - 2n}  T_\text{eq}^{1-n} }{ m_\text{Pl}}. \label{eq:thermal} \end{align} 
 
 We see that in the familiar $n=2$ case, the required cross-section is essentially independent of $m_\chi$, being given roughly by $1/(m_\text{Pl} T_\text{eq})$, and more accurately by $(g_*)_f^{-1/2} x_f$ multiplied by this value. In the $n=3$ case the required cross section scales as $m_\chi^{-2}$, i.e. for smaller masses a larger cross section is needed; for $n=4$ the scaling is $m_\chi^{-4}$. 

If we estimate the rate as $ \langle \sigma v^{n-1}\rangle \sim \alpha^n/m_\chi^{3 n - 4}$, on dimensional grounds (assuming a tree-level cross section and that the only relevant mass scale for the annihilation is comparable to the DM mass), we obtain an estimate for the required DM mass:
  \begin{align} m_\chi \sim \alpha \sqrt[n]{ m_\text{Pl} T_\text{eq}^{n-1}} \left[ x_f^{\frac{5}{n} - 3} (g_*)_\text{f}^{1 - \frac{3}{2n}} \right].    \end{align}
For example, for 3-body annihilation, we obtain $m_\chi/\alpha \sim 2$ GeV ($x_f = 1$), 40 MeV ($x_f = 20$), 4 MeV ($x_f = 100$), ignoring the $g_*$ factors. In general we expect perturbative 3-body annihilation to require sub-GeV DM masses if the DM is a thermal relic, and 4-body annihilation to require sub-MeV DM masses.

It is also possible that the thermal annihilating component comprises only a small fraction $\eta$ of the DM. In this case, the calculation proceeds as previously, except that $T_\text{eq}$ in Eq.~\eqref{eq:rhof} must be replaced with $\eta T_\text{eq}$. Thus the thermal cross section becomes:
  \begin{align}  \langle \sigma v^{n-1}\rangle & \sim  (g_*)_\text{f}^{\frac{3}{2} - n} x_f^{-5 + 3n}  \frac{m_\chi^{4 - 2n} \eta^{1-n}  T_\text{eq}^{1-n} }{ m_\text{Pl}}. \end{align} 
 Since the annihilation rate scales as $\langle \sigma v^{n-1}\rangle (\eta \rho_\chi)^n$, we see that overall the annihilation rate for a thermal relic is expected to be directly proportional to $\eta$, independent of $n$, in contrast to the $\eta^n$ scaling if the annihilation rate is held fixed and not required to be thermal.

\subsection{Maximum density at late times}

We now outline some generic upper limits on the signal from $n$-body annihilation processes. Note that the annihilation rate per particle per unit time $\Gamma_\chi$ must be larger at freezeout than at any later time, if the mechanism for freezeout is annihilation and the DM density is not rapidly changing. In this case, at freezeout the annihilation rate is similar to $H_f$, and since $H$ is monotonically decreasing after inflation, $H_f$ is larger than $H$ at any later time. Thus if the annihilation rate at any time and in any region is comparable to its value at freezeout, it will also be faster than the Hubble expansion at that time, and the DM density in that region will be rapidly depleted (in a second ``freezeout''), reducing $\Gamma_\chi$ until it is below $H$. If the annihilation cross section does not decrease after freezeout, it follows that the equilibrium density in any region after freezeout must always be lower than the cosmological density at freezeout. 

From this argument, we can set an upper limit on the (steady-state) density for the DM in any region after freezeout, $\rho_\chi < \rho_f$, where $\rho_f$ is the density at freezeout, provided the annihilation cross section does not decrease at late times. In fact, from the same argument, the steady-state density at redshift $z$ satisfies $\rho_\chi(z)^{n-1} < \rho_f^{n-1} H(z)/H(z_f)$.

Furthermore, if the annihilation rate scales only with density, it follows that the requirement that $n$-body annihilation dominates over other annihilation channels at late times is always more stringent than the requirement that it dominates at freezeout (except possibly for transient periods when the DM density is being rapidly depleted). However, there can be exceptions to this latter rule if the annihilation rate does not only scale with density (e.g. if it also scales with velocity).

For fermionic DM we can set another generic constraint on the maximum DM density from phase-space considerations; this is a form of the well-known Tremaine-Gunn bound \cite{Tremaine:1979we}. The phase-space distribution function satisfies $ f_c \le 2$ from the Pauli exclusion principle. Consider a region of density $\rho_\chi$, corresponding to a number density $n_\chi = \rho_\chi/m_\chi$; then the phase-space distribution function can be approximated as $f_c \sim \rho_\chi/(m_\chi^4 \langle v^2\rangle^{3/2})$ (in the non-relativistic regime). Thus we find $\rho_\chi \lesssim m_\chi^4 \langle v^2\rangle^{3/2}$.

Note that if the two-body and $n$-body cross sections have ``natural'' scaling, i.e. $\langle \sigma v^{n-1}\rangle \sim \alpha^n/m_\chi^{3n - 4}$, then it follows that the ratio of annihilation rates between the $n$-body and 2-body cases satisfies:
\begin{align} \Gamma_{\chi, n}/\Gamma_{\chi, 2} & \equiv (\rho_\chi/m_\chi)^{n-1} \langle \sigma v^{n-1}\rangle / (\rho_\chi/m_\chi) \langle\sigma v\rangle, \nonumber  \\ 
& \sim (\rho_\chi/m_\chi)^{n-2} \alpha^{n-2} m_\chi^{-3(n-2)},  \nonumber  \\ 
& \lesssim  \alpha^{n-2}\langle v^2\rangle^{3(n-2)/2}.\end{align}
Thus we see that the natural suppression of $n$-body annihilation relative to 2-body, at least for fermionic DM, is a strictly stronger effect than the suppression of $p$-wave annihilation relative to $s$-wave, which corresponds to a factor of $\langle v^2\rangle$. However, this argument can break down if the $n$-body annihilation rate increases at low velocities, or otherwise does not satisfy the ``natural'' scaling estimate.

\section{Low-velocity enhancements for $n$-body processes}
\label{sec:sommerfeld}

Above we have assumed that the annihilation is dominated by tree-level processes, and the mass scale for the effective operator is similar to the mass of the DM. However, there could be a large enhancement to the annihilation rate due to resonance effects or the presence of lighter particles in the spectrum; in the latter case, attractive interactions between the DM particles could produce a low-velocity enhancement to the annihilation rate (as discussed for Sommerfeld enhancement of two-body annihilation in e.g. \cite{Hisano:2003ec,Hisano:2004ds,Cirelli:2007xd}).

Relatedly, even if two-body processes dominate during freezeout, $n$-body interactions might come to dominate at late times if the velocity scaling is sufficiently strong. Therefore one question we would like to address is what limits can we place on such enhanced interactions.

\subsection{The unitarity bound}

For two-body scattering in the non-relativistic limit, partial-wave unitarity imposes an upper bound on the cross section for annihilation in the $l$th partial wave, which for distinguishable particles in the initial state is given by:
\begin{equation}\label{eq:optical_theorem} \sigma_l \le 4 \pi (2l+1)/|\vec{k}|^2,\end{equation}
where $\vec{k}$ is the momentum of one of the incoming particles in the center-of-mass frame (or alternatively, the momentum of an incoming particle scattering in a stationary potential). For $n > 2$, a similar bound can be set for non-relativistic scattering in hyperradial potentials \cite{2009PhRvL.103o3201M}; if $k$ is now the magnitude of the hyperradial momentum coordinate, the $1/k^2$ scaling is replaced by $1/k^{3 n - 4}$, and the prefactor is also modified. The ``event rate constant'' of \cite{2009PhRvL.103o3201M}, equivalent to our $\langle\sigma v^{n-1}\rangle$, scales with an extra power of $k/\mu_n$, where $\mu_n$ is the $n$-body reduced mass, giving an overall momentum scaling for the $l$th partial wave contribution of $\langle \sigma v^{n-1}\rangle_l \propto k^{5 - 3 n}$ in the unitarity-saturating limit.

The same unitarity scaling can be obtained from the optical theorem, as demonstrated in \cite{Kuflik:2017iqs}. Let us focus on $n\rightarrow 2$ processes, with an initial state denoted $i$ and a final state denoted $f$. The matrix element for such a process is the same as for the inverse $2 \rightarrow n$ process, $f \rightarrow i$. Considering the $f \rightarrow f$ forward scattering, we can use the optical limit to set an upper bound on this matrix element:
\begin{equation} \sum_i \frac{1}{S_i} \int d\Pi_n |\mathcal{M}_{f \rightarrow i}|^2 \le 2 \text{Im} \mathcal{M}(f \rightarrow f), \label{eq:optical1}\end{equation}
where the sum over initial states corresponds to a sum over spin configurations (and any other relevant quantum numbers), and $d\Pi_n$ is the standard Lorentz-invariant $n$-body phase space (as in Eq.~\eqref{eq:pils}, but with the $i=n+1\dots n+m$ labels replaced with $i=1\dots n$). 
Using this inequality and taking the non-relativistic limit to simplify Eq.~\eqref{eq:temp_average} for the specific case of $m=2$, one can obtain an upper bound on $\langle \sigma v^{n-1}\rangle$.
We leave the details of the computations to Appendix \ref{app:unitarity} and only report the main result here:

\begin{align}
\label{unitarity_bound}
\langle \sigma v^{n-1}\rangle & \le 2^{-\frac{1}{2} + \frac{3}{2}n} \left(\frac{T}{\pi} \right)^{-(3n - 5)/2} S_i \frac{g_4 g_5}{g_1\cdots g_n}   \left( \frac{m_1 + \cdots + m_n}{m_1 \cdots m_n} \right)^{3/2},
\end{align}
where, as before, $g_i$ counts the degrees of freedom in each state $i$. 

We see the expected scaling with $T^{-(3n-5)/2}$ ($\sim k^{-(3n -5)}$ in the non-relativistic limit). For $n=2$, in the simplified case where $m_i=m_\chi$ for every $i$, we obtain:
\begin{align}\langle \sigma v\rangle & \le \frac{16 \sqrt{2 \pi}}{m_\chi^2 \left(2 T/m_\chi\right)^{1/2}} S_i \frac{g_4 g_5}{g_1 g_2}, \end{align}
and for $n=3$, we obtain:
\begin{align}\langle \sigma v^2\rangle & \le \frac{48 \sqrt{3} \pi^2}{m_\chi^3 T^2} S_i \frac{g_4 g_5}{g_1 g_2 g_3}, \end{align}
in agreement with  \cite{Kuflik:2017iqs} up to the $g_i$ factors (set to 1 in that work) and the $S_i$ factor (absorbed into the definition of $\langle \sigma v^{n-1}\rangle$).

The $1/T^2$ unitarity scaling for the rate coefficient for three-body processes has been studied in the context of ultracold bosons (e.g. \cite{Braaten:2004rn,Braaten:2006vd} and references therein). For example, this scaling has been observed both theoretically and experimentally in dimer formation processes, where the initial state consists of three free particles and the final state contains a dimer plus a free particle \cite{2013PhRvL.110p3202R,2013PhRvL.111l5303F}. This scaling corresponds to a very large $s$-wave scattering length for the two-body $s$-wave interactions; at sufficiently low velocity/temperature, the cross section saturates at a scale set by the scattering length.

\subsection{Implications for cosmology}

For a $n$-body DM annihilation process, let us consider the scaling of the annihilation rate with redshift before the onset of structure formation, in the regime where the unitarity limit is saturated. While the DM remains at the same temperature as the SM radiation bath, the DM temperature redshifts as $(1+z)$, the typical velocity redshifts as $(1+z)^{1/2}$, and so the unitarity scaling $\langle \sigma v^{n-1}\rangle \propto k^{-(3n - 5)}$ corresponds to $\langle \sigma v^{n-1}\rangle \propto (1+z)^{\frac{5}{2} - \frac{3}{2} n}$, and the annihilation rate per particle $\Gamma_\chi \propto (1+z)^{\frac{5}{2} - \frac{3}{2} n} (1+z)^{3(n-1)} = (1+z)^{-\frac{1}{2} + \frac{3}{2} n}$. Once the DM temperature decouples from the cosmic microwave background (CMB) temperature, we instead have $k \propto 1+z$, and consequently $\Gamma_\chi \propto (1+z)^{5 - 3 n} (1+z)^{3(n-1)} = (1+z)^2$. Note that $H$ scales as $(1+z)^2$ in the radiation-dominated regime and $(1+z)^{3/2}$ in the matter-dominated regime, so $\Gamma$ decreases at least as rapidly as $H(z)$ with decreasing $z$, and (at least once the unitarity-scaling regime is entered) there is no possibility of an early freezeout followed by a later recoupling. After kinetic decoupling of the DM, the scaling in the unitarity regime is independent of $n$, and thus matches that of Sommerfeld-enhanced two-body annihilation.

In general, we expect that below some saturation velocity, the cross section will fall below the unitarity limit, with a weaker velocity scaling (or no velocity scaling, or even a low-velocity suppression). If $\langle \sigma v^{n-1}\rangle$ saturates (becoming constant) at some velocity, then below that velocity, the signal will only scale with $\rho^n$ rather than with velocity. In this case, for indirect searches probing the low-velocity regime, the effect of the velocity scaling at intermediate temperatures (above the saturation point) will be to greatly increase the normalization of the annihilation cross section $\langle \sigma v^2\rangle$ relevant for such searches, compared to the naively expected thermal value. This will be our standard assumption -- that the cross section is saturated -- in the indirect-detection constraints that follow. If saturation does not occur, or has not occurred for the velocities of interest for indirect detection, then the constraints we present in this work can still be used as estimates if the typical velocity of DM particles in the system of interest is known, but a detailed calculation would require inclusion of the full velocity dependence (as done for two-body annihilation in e.g. \cite{Boddy:2017vpe}). 

\section{Calculating signals from dark matter halos}
\label{sec:boost} 
A crucial ingredient in determining annihilation signatures is the contribution from low-mass halos, either isolated halos or subhalos within a larger system. This contribution is generally parametrized by the ``boost factor'', i.e. the ratio of the true signal to that expected from the smooth background DM density (either the cosmological DM density, for the case of probes of cosmological volumes, or the smooth density profile of the main halo). In the CDM paradigm, the smallest halos are also the oldest and most dense (having formed when the universe was denser); for few-body annihilation, the additional density scaling relative to two-body annihilation further enhances the signal from these small halos. 

Unfortunately, this contribution is difficult to model precisely, as the small halos in question are below the resolution of cosmological simulations. Thus we will first describe a benchmark estimate for the boost factor, and then discuss its possible uncertainties.

\subsection{Calculation of the average annihilation boost from isolated halos}

In order to estimate the enhancement of the signal due to the clumpiness of the universe, we first need to obtain the expected number density of structures over a wide range of masses, for which we use the standard Press-Schechter (PS) formalism \cite{Press:1973iz}. Let us start by introducing the matter power spectrum at redshift zero, which is given by:
\ba 
P(k) =  {\cal T}(k)^2 \, {\cal T}_{\chi}(k)^2 \, P_0(k).
\ea 
Here $k$ is the comoving wavenumber, $P_0(k)$ is the primordial power spectrum, and ${\cal T}(k)$ is the standard matter transfer function taken from CAMB \cite{2000ApJ...538..473L} with the fit for small scales provided by Ref.~\cite{Eisenstein:1997ik}. We use cosmological parameters $\Omega_{DM}h^2 =0.12$, $\Omega_bh^2=0.022$ and $H_0=67.27\text{ km}\,s^{-1}\text{Mpc}^{-1}$, consistent with Planck results \cite{Ade:2015xua}.

 ${\cal T}_{\chi}(k)$ encodes the suppression of power at small scales due to the free-streaming of DM particles. ${\cal T}_\chi(k)$ is in general model-dependent, and in particular depends on the temperature of the DM during cosmic history. For standard WIMP-like DM particles one can compute ${\cal T}_{\chi}(k)$ analytically (see e.g. \cite{Green:2005fa}). However, for the dark-sector models that we consider, we will simply approximate the suppression factor by an exponential cutoff at a characteristic wavenumber:
\ba
\label{Free_stream_T}
 {\cal T}_{\chi}(k) =e^{-k^2/2k_c^2}.
\ea
The cutoff $k_c$ depends on the details of the kinetic decoupling between the DM and the SM. In what follows we will study the effects of varying $k_c$ between $10  h/$Mpc and $10^7h/$Mpc. 
Direct observations of the Lyman-$\alpha$ forest at $k \lesssim 10 h/$Mpc, by HIRES/MIKE \cite{Viel:2013apy} and XQ-100 \cite{2016A&A...594A..91L}, have been used to set constraints on models of warm DM (WDM) \cite{Irsic:2017ixq} and fuzzy DM (FDM) \cite{Irsic:2017yje} via their suppression of the matter power spectrum at small scales. The WDM mass scale is currently constrained to be (conservatively) greater than $3.5$ keV, whereas the conservative lower mass bound in the FDM case is $20\times 10^{-22}$ eV (both limits are at $2\sigma$). We note that $k_c \sim k_{1/2}$, where $ k_{1/2}$ is defined as the scale where the power drops by a factor of 2 relative to the CDM case; for the FDM model, $k_{1/2}$ is estimated to satisfy $k_{1/2} \approx 4.5 (m_\chi/10^{-22} \text{eV})^{4/9}$/Mpc \cite{Hu:2000ke}. Taking the lower mass bound quoted above, we obtain $k_c \gtrsim 17/$Mpc $\approx 24 h/$Mpc, for the FDM model. Since we are relying on analytic estimates to translate between mass and $k_c$, and in any case, the cutoff need not have precisely the exponential form assumed for our transfer function, we consider $k_c \gtrsim 10 h/$Mpc. The largest cutoff we consider, $k_c \sim 10^{7}h/$Mpc, is comparable to the cutoff for a conventional WIMP scenario  \cite{Green:2005fa}.

The primordial power spectrum is given by:
\ba
P_0(k)={\cal A}_s \left(\dfrac{k}{k_0} \right)^{n_s-1},
\ea
where ${\cal A}_s \simeq 2.21 \times 10^{-9}$ is the amplitude of the primordial power spectrum, $k_0=0.05 \, {\text{Mpc}}^{-1}$ is the pivot scale and $n_s \simeq 0.969$ is the primordial scalar spectral index; we have taken the Planck 2015 best-fit values for the parameters \cite{Ade:2015xua}.

The variance of matter density perturbations for each halo mass can be obtained by smoothing the power spectrum over scales shorter than the size of the halo, i.e.:
\ba
\label{variance}
\hat \sigma^2(R) = \int_{k_{\rm{min}}}^{\infty} \dfrac{d^3 k}{(2\pi)^3} P(k)\, W^2(kR) 
\ea
where $k_{\rm{min}} \sim 10^{-4} \, h/{\text{Mpc}}$ corresponds to the largest observable scale, and $W(kR)$ is a window function that describes the size of the halo. The simplest window function would be a top-hat function in Fourier space. In  \cite{Schneider:2013ria} it has been argued that this choice of window function is indeed (at least marginally) better than other possibilities when the PS formalism is compared to the simulations.  Hence we only consider the sharp-$k$ window function:
\ba
\label{window}
W(kR) = \Theta(1-kR)  
\ea 
where $\Theta$ is the Heaviside step function and $R$ is a comoving length-scale associated with the halo. The disadvantage of this window function, however, is that the mass assignment to halos is not transparent, and a free parameter is needed that has to be fixed by calibration against simulation. Ref.~\cite{Schneider:2013ria} finds that the halo mass is related to the length-scale $R$ by:
\ba 
\label{M_R}
M = \dfrac{4 \pi}{3} \rho_{m,0} (CR)^3
\ea 
 with $C=2.7$, where $\rho_{m,0}$ is the matter energy density today.
This mass assignment indicates that the range over which we change the power spectrum cutoff  $k_c=( 10-10^7)h{\text{Mpc}}^{-1}$, roughly corresponds to halo masses in the  range $M_c \sim (3\times 10^{-8} - 3\times 10^{10} )M_\odot$; below which the structures are expected to be washed out due to the free streaming.
 
 The mass function of halos ({\it comoving} number density of halos  per logarithmic mass scale at each redshift) can then be written as 
 \ba 
 \label{mass_function}
 \dfrac{dn_h(M,z)}{d\ln M} =\dfrac{1}{2} {\cal F}(\nu(M,z)) \dfrac{\rho_{m,0}}{M} \dfrac{d\ln \nu(M,z)}{d\ln M}.
 \ea 
 Here $\nu(M,z)=\delta_c^2(z)/\hat \sigma^2(M)$, where $\delta_c(z) = 1.686/D(z)$ is the time-dependent critical density contrast, and $D(z)$ is the linear growth factor normalized to 1 at the present time. For the growth factor we do the following integral numerically \cite{1977MNRAS.179..351H,1980lssu.book.....P}:
 \ba 
 {\cal I}(z)=\dfrac{5}{2} \Omega_m \dfrac{H(z)}{H_0} \int_z^\infty \dfrac{(1+z')}{(H(z')/H_0)^3}\, dz'
 \ea 
 where $H(z)$ is the Hubble parameter, $H_0$ is the Hubble parameter evaluated today, and $\Omega_m$ is the present-day ratio of the matter energy density to the total energy density. The growth factor, in our convention, is then defined by $D(z)={\cal I}(z)/{\cal I}(0)$.
 For ellipsoidal collapse, the function ${\cal F}\left(\nu(M,z)\right)$ is given by:
 \ba
   {\cal F }\left(\nu(M,z)\right) =A_e \sqrt{\dfrac{2q_e\, \nu(M,z)}{\pi}}\left[ 1+(q_e\, \nu(M,z))^{-p_e}  \right] e^{-q_e \,\nu(M,z)/2},
 \ea
 with $A_e=0.3222, \, \, p_e=0.3$ and $q_e=1$ \cite{Schneider:2013ria}. 
Note that for the top-hat window function in Fourier space Eq.~\eqref{window} the halo mass function simplifies to 
\ba 
\label{mass_function_simp}
   \dfrac{dn_h(M,z)}{d\ln M} = \dfrac{\rho_{m,0}}{12 \pi^2 M \hat \sigma(M)^2 R(M)^3}{\cal F}(\nu(M,z))  \, P(1/R(M)),
\ea 
  where $R(M)$ is the comoving halo length-scale related to its mass via \eqref{M_R}.

 For computing the boost factor for DM interactions, we also need some information about the halo profile. It is easy to check that the NFW profile \cite{Navarro:1996gj} leads to a diverging integrated annihilation rate for $n$-body annihilation with $n>2$, due to the $1/r$ cusp in the DM density at the center of the halo. One can modify the profile around the center by assuming a smooth core, but the actual size of the core is quite uncertain. For the majority of our studies, we will instead consider the Einasto profile \cite{Einasto:1965czb}, which describes halo profiles in N-body simulations quite well, and for which the annihilation signal is well-defined:
 \ba 
 \label{Einasto}
 \rho_h(r) = \rho_{-2} \exp \left( \dfrac{-2}{\alpha_e} \left[ \left( \dfrac{r}{r_{-2}}\right)^{\alpha_e}-1 \right] \right).
 \ea 
 Here $\rho_{-2}$ and $r_{-2}$ are the density and radius at which the slope of the logarithmic density ($d \log \rho / d\log r$) is $-2$, and $\alpha_e$ is the Einasto shape parameter.  For a given halo mass, $\rho_{-2}$ has to be determined by requiring that the integral of density over the halo volume gives a consistent mass. However, note that  $\rho_{-2}$ is a prefactor which will be cancelled out in boost factor calculations, hence it is an irrelevant quantity there. We will discuss the other Einasto parameters below. 
 
Generalizing the results of \cite{Taylor:2002zd}, the boost factor for an individual halo in the presence of a general interaction (which need not be number-conserving), with $n$ initial-state particles, is given by{\footnote{Note that the boost factor depends on $n$ which has been made explicit in \eqref{Boost_halo} as a superscript. However, to simplify the notation, we will omit it hereafter.}}:
\ba
\label{Boost_halo}
B_h^{(n)}(M,z) = \dfrac{\langle \rho_h^n(r,M,z) \rangle}{\, \langle \rho_h(r,M,z) \rangle^n},
\ea
in which we have defined the following averaging of an arbitrary function ${\cal G}(r,M)$ over the halo profile
\ba
\langle {\cal G}(r,M) \rangle =\dfrac{4\pi \int_0^{r_{200}} r^2 {\cal G}(r,M) dr  }{V_h(M)},
\ea
with $V_h = \langle 1 \rangle = 4 \pi r_{200}^3(M,z)/3 $. Here $r_{200}$ is the approximate boundary of the halo, the virial radius, which is defined as the radius at which the average energy density of the halo (within that radius) becomes $\Delta_h=200$ times larger than the total background energy density, i.e. 
\ba
  \langle \rho_h(r,M,z) \rangle = \dfrac{M }{ V_h(M,z) }= \Delta_h \, \rho_c(z).
\ea 
Note that $\rho_h$ depends on redshift through the Einasto parameters. The parameters $r_{200}$ and $\alpha_e$ both need to be found by calibrating to either simulations or observations, the former of which is usually reported as the {\it concentration} parameter defined by $c(M,z)=r_{200}/r_{-2}$. We will discuss their explicit form later but, for now, it suffices to notice that both parameters depend on the halo mass as well as redshift.

Substituting the Einasto profile (Eq.~\eqref{Einasto}) into Eq.~\eqref{Boost_halo}, we obtain the following simple analytic expression for the halo's boost factor:
\ba
\label{B_halo}
B_h(M,z) = \left( \dfrac{3}{2 c^3} \right)^{1-n} \left(\dfrac{\alpha_e}{2} \right)^{(1-n)(3-\alpha_e)/\alpha_e} n^{-3/\alpha_e} \dfrac{\gamma(3/\alpha_e, 2 n\, c^{\alpha_e}/\alpha_e)}{\gamma^n(3/\alpha_e, 2\, c^{\alpha_e}/\alpha_e)},
\ea
in which $\gamma$ is the lower incomplete gamma function defined by
\ba
\gamma(x,y) =  \Gamma(x) - \Gamma(x,y),
\ea
where $ \Gamma(x)$ and $ \Gamma(x,y)$ are the gamma and the incomplete gamma functions, respectively. 

The universal boost factor due to the non-linear structures can then be written as:
\ba
\label{B_total}
B(z) = \dfrac{\langle \rho_h \rangle^{n-1}}{\rho_{m}^{n}(z)} (1+z)^3 \int_{M_{\rm{min}}}^{\infty} B_h(M,z)\dfrac{dn_h(M,z)}{d\ln M} dM,
\ea
where the factor $(1+z)^3$ appears because the mass function Eq.~\eqref{mass_function} is the comoving number density of halos. $M_{\rm{min}}$ is determined by the cutoff in the matter transfer function $k_c$; using the sharp-$k$ filter we have $M_\text{min} \lesssim M_c= \dfrac{4 \pi}{3} \rho_{m,0} (C/k_c)^3$ with $C=2.7$, following Eq.~\eqref{M_R}. In principle we could integrate down to $M_\text{min} = 0$ without large errors, as the exponential suppression in the mass function removes halos below $M_\text{min}$ from the integral in any case. In practice, however, we will take $M_\text{min} = 10^{-11} M_\text{sun}$ as a somewhat arbitrary and conservative choice. Note that the integral Eq.~\eqref{B_total} quickly saturates as a function of $M_{\text{min}}$ so that any choice of $M_\text{min}$ (provided that $M_{\text{min}} \lesssim 0.1 M_c$) is equally appropriate. Note that the above equations use the matter energy density, $\rho_m$, not the DM energy density, as baryonic matter would also contribute to the gravitational potential. 

The last integral in Eq.~\eqref{B_total} has to be performed numerically. Note that this boost factor neglects the contribution from the smooth background in the numerator, summing only over the contributions of collapsed halos. This is valid when most annihilation occurs in halos, but at high redshifts (prior to structure formation) or in regions where most halos have been disrupted, the boost factor instead approaches unity. To capture both limits, we define the effective CDM energy density as follows:
\ba
\rho_{\rm{eff}}(z) = (1+B(z))^{1/n} \, \rho_{\chi}(z).
\ea
Thus the $n$-body annihilation rate averaged over a region scales as $\rho_\text{eff}^n$, rather than $\rho^n$. Fig.~\ref{Fig:rho_eff} shows the results for $\rho_{\rm{eff}}$ as a function of redshift for $n=2, 3$ and $4$ for different parametrizations of the concentration parameter (which will be discussed below).
 
\subsection{The effect of substructures}

Within any halo there can be many subhalos that will enhance the boost factor. To estimate their impact, let us assume that all substructures are located around the virial radius of the host halo. This approximation yields a two-fold simplification: first, when integrating over the region of the halo where the subhalos are present, we can neglect the contribution from the main halo. Second, since the substructures are rather far from the center of the host halo, we can neglect tidal effects from the main halo, which are expected to truncate the substructures such that their true radius differs from naive estimates of the virialized radius \cite{Tormen:1997ik,Springel:2008cc}. 

Based on the above simplification, following the same steps that gave us the universal boost factor (Eq.~\eqref{B_total}), we can write the halo's boost factor modified by substructure ($\tilde B_h$) as follows:
\ba
\label{B_tilde}
 \tilde B_h(M,z) = B_h(M,z) + \int_{M_{\mathrm{min}}^{{\text{sub}}}}^{M_{\mathrm{max}}^{\text{sub}}}
 B_h(M_s,z) \dfrac{M_s}{M} \dfrac{dN_s(M,M_s)}{dM_s} dM_s.
\ea 
In the above relation we have assumed that $\langle \rho(r) \rangle = \Delta_h \, \rho_c(z)$, independent of the halo or substructure mass. $M_s$ is the subhalo mass, $M_{\mathrm{min}}^{{\text{sub}}}$ is the minimum halo mass set by the small-scale cutoff in the power spectrum (i.e. we set $M_\text{min}^\text{sub} = M_c = \dfrac{4 \pi}{3} \rho_{m,0} (C/k_c)^3$ as a natural choice), and $M_{\mathrm{max}}^\text{sub}$ is the maximum mass of subhalos (which is a function of the mass of the host halo). $N_s(M,M_s)$ is the {\it total} number of subhalos with mass larger than $M_s$ in a host halo with mass $M$. Ref.~\cite{Gao:2010tn} obtains a simple fitting formula for $N_s(M, M_s)$ from simulations, which we will follow here:
\ba
N_s(M,M_s) = \left(\dfrac{\mu }{\mu_1} \right)^{\kappa_1} e^{-(\mu/\mu_c)^{\kappa_c}},
\ea
where $\mu=M_s/M$, and we set $\mu_1=0.01$, $\mu_c= 0.1$, $\kappa_1=-0.94$ and $\kappa_c=1.2$. Note that these parameters in principle depend on both halo mass and redshift, but these dependences -- compared to other uncertainties -- have only a mild effect on the boost factor.
In particular, notice that since the boost factor is most sensitive to the smallest halos, changing the cutoff $\mu_c$ -- which only changes the number of the largest subhalos -- would not significantly modify the final results. The exponential cutoff in the subhalo number also makes the above integral insensitive to the actual value of $M_\text{max}^\text{sub}$, and we simply set $M_\text{max}^\text{sub}=10 M \mu_c$. On the other hand, the contribution of the substructures to the boost factor is more responsive to $M_\text{min}^\text{sub}$, although the sensitivity is still mild enough that the resulting uncertainty is negligible compared to the differences amongst the range of models we will consider. For example, if we take $M_\text{min}^\text{sub}=M_c/5$ rather than $M_\text{min}^\text{sub}=M_c$, $\rho_\text{eff}$ is enhanced by a factor smaller than $20\%$ in all cases (the actual value depends on the choice of $n$ as well as the concentration parameterization).

After computing the corrected boost factor $\tilde B_h$ we insert it into the universal boost factor \eqref{B_total} simply by replacing the {\it bare} halo boost factor $B_h$ with $\tilde B_h$.

\subsection{Einasto parameters and concentration uncertainties}
So far, we have used the Einasto profile and the associated free parameters (i.e. the concentration $c$ and the shape $\alpha_e$) to obtain the boost factor. We still need to discuss how the Einasto profile's parameters change as functions of halo mass and redshift. Note that for $n$-body annihilation the small dense halos become increasingly important as $n$ increases, and uncertainties in the concentration of the smallest halos have a large effect on the overall signal. Estimates for the concentration parameter often require extrapolation across a wide range of halo masses, as cosmological simulations cannot resolve the smallest DM halos. In contrast, the shape parameter seems to approach to a constant value for small halos so that the uncertainties in $\alpha_e$ are less important, although the large extrapolation is still necessary. 
Ref.~\cite{Klypin:2014kpa} finds simple redshift-dependent fits for $\alpha_e$, by calibrating to simulations\footnote{See also \cite{Gao:2007gh} where a slightly different coefficient for the second term is obtained whereas the first term (which is more important to us) is similar.}:
\ba
\label{alpha}
 \alpha_e(M,z) = 0.115 +0.0165 \nu
\ea 
where $\nu(M,z)=\delta_c^2(z)/\hat \sigma^2(M)$ is the universality function; small halos have $\nu \ll 1$ and so $\alpha_e$ is nearly independent of redshift and mass in this case. As a result of this observation and noticing that the smallest halos have a dominant contribution to the boost factor we simply fix $\alpha_e$ to follow the parameterization given in Eq.~\eqref{alpha}.

In order to estimate a plausible range for annihilation signals, we now consider several alternate parameterizations for the concentration parameter that have been developed in the literature. In the following discussions, and in the upcoming plots, we sort the concentrations that we are going to study according to the resulting typical size of the boost factor, starting from the one that has the largest effect.

According to Ref.~\cite{Comerford:2007xb}, observations of cluster-mass halos (with masses roughly $10^{13}-10^{15}M_\odot$ and redshift $z\lesssim 0.06$) suggest the following concentration-mass-redshift relation:
\ba 
\label{c_obs}
c_{\text{obs}}(M,z) = \left( \dfrac{14.5}{1+z} \right) \, \left(\dfrac{M}{1.3 \times 10^{13}h^{-1} M_\odot}\right)^{-0.15}.
\ea  

Ref.~\cite{Duffy:2008pz} finds a somewhat different concentration based on $N$-body simulations, taken from redshifts $z \lesssim 2$ and halo masses of $10^{11}-10^{15}M_\odot$:
\ba 
\label{c_sim1}
c_{\text{sim1}} (M,z) =\left( \dfrac{7.85}{(1+z)^{0.71}} \right) \, \left(\dfrac{M}{2 \times 10^{12}h^{-1}M_\odot}\right)^{-0.081}.
\ea 

Note that both these models have a power-law dependence on the halo mass, which is likely too optimistic in the sense of predicting an overly large concentration for small halos (with the degree of the overestimate being largest in the first parameterization above). Such small halos have similar collapse times over a wide range of masses, and therefore are expected to have similar natal concentrations; thus the concentration-mass relation is expected to flatten toward lower masses \cite{Sanchez-Conde:2013yxa, Ng:2013xha}. We include these power-law models for ease of comparison for earlier work, and to bracket uncertainties in the scaling of the concentration parameter for small halos.

Ref.~\cite{Klypin:2014kpa} avoids this power-law extrapolation issue by expressing the concentration in terms of the universality function $\nu$, as defined above:
\ba 
\label{c_sim2}
 c_{\text{sim2}} (M,z) = 6.5 \nu^{-0.8} (1+0.21 \nu).
\ea 
These results are obtained by fitting the Einasto profile to simulations of halos in the mass range $10^{11}-10^{15} M_\odot$, for redshifts $z \lesssim 5.5$.

Simulations discussed by Ref.~\cite{Duffy:2008pz} also show that the concentration becomes flatter as a function of mass at higher redshifts; based on this trend, Ref.~\cite{Mack:2013bja} proposes a flat concentration relation to serve as a maximally conservative estimate:
\ba 
\label{c_flat}
c_{\text{flat}} (M,z) =\dfrac{7.85}{(1+z)^{0.71}}.
\ea 
The parameterization of Ref.~\cite{Prada:2011jf} lies between this flat relation and the universality-function-based $ c_{\text{sim2}}$ estimate.

\begin{figure}
\includegraphics[scale=.42]{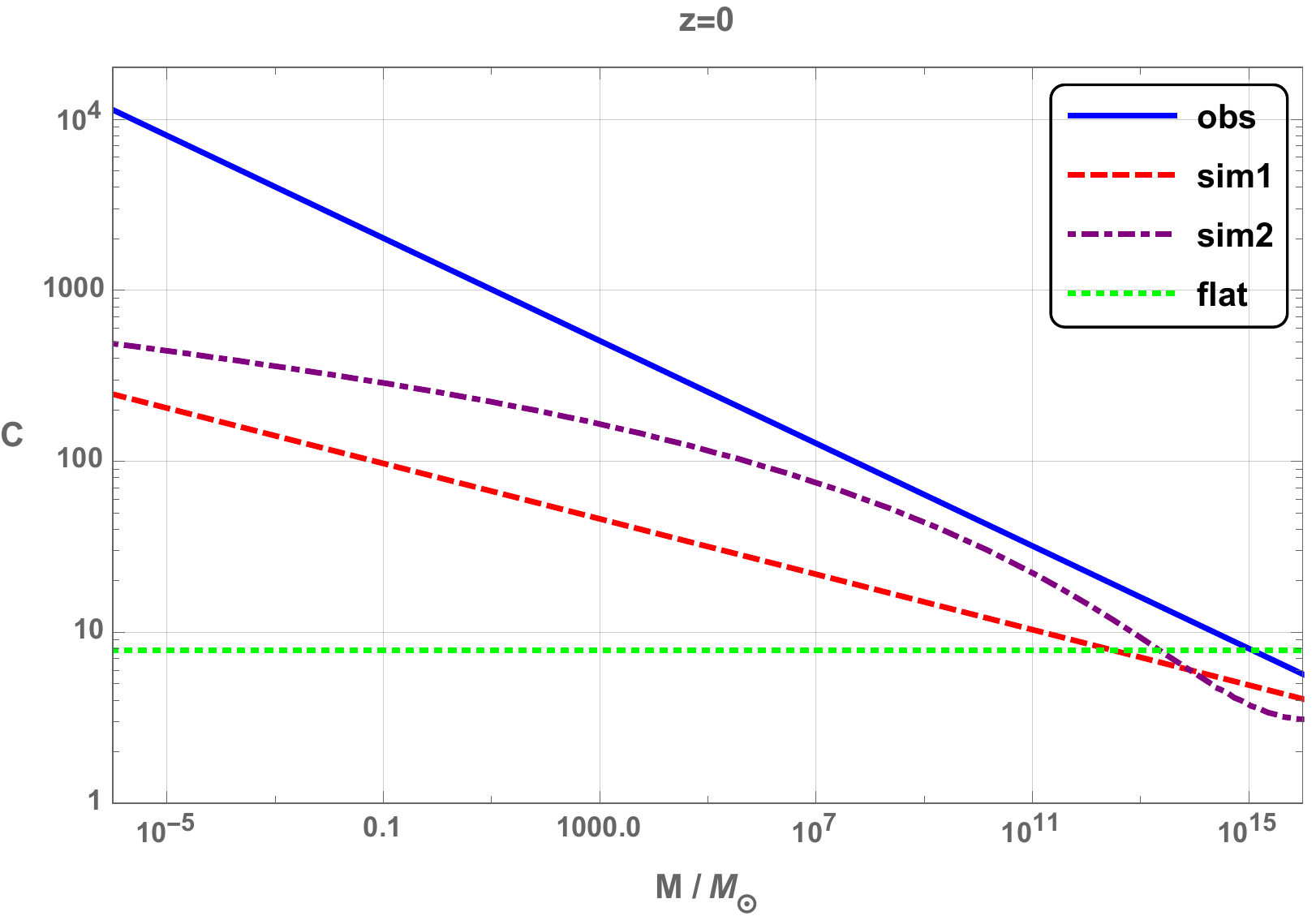}
\hspace{1cm}
\includegraphics[scale=.435]{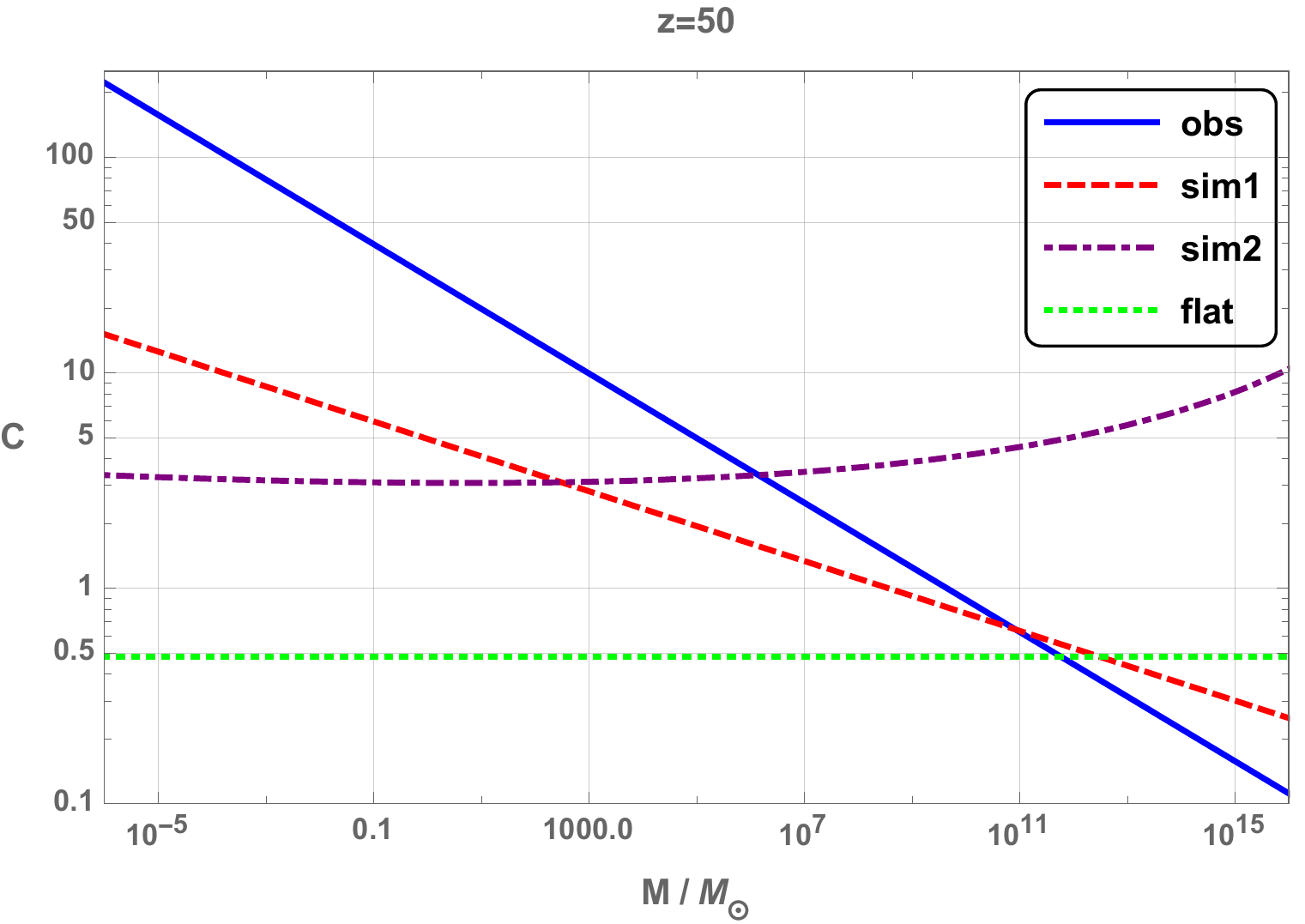}
\caption{Different concentrations as functions of mass at $z=0$ (left) and their extrapolation to redshift $z=50$ (right). For $c_{\text{sim1}}$ we used Eq.~\eqref{c_sim1} with $k_c=10^7h{\text{Mpc}}^{-1}$ to compute the variance and universality function $\nu$. Note that, for the smallest halos, which are the most important contributors to the boost factor, $c_{\text{sim2}}>c_{\text{sim1}}$ at low redshifts whereas $c_{\text{sim1}}>c_{\text{sim2}}$ at high redshifts.}
\label{concentrations}
\end{figure}

We will study all the above concentration parameterizations and compare the results; Fig.~\ref{concentrations} compares these parameterizations at $z=0$ and their extrapolation at $z=50$ as a function of halo mass. Note that often NFW profiles, rather than Einasto, are fitted to the observations/simulations to extract the concentration parameters; however, the NFW and Einasto profiles are quite similar except in the cores of halos. The differences are critical for annihilation signals, but not for the bulk halo shapes.

\begin{figure}
\hspace{3.4cm}
\includegraphics[scale=.334]{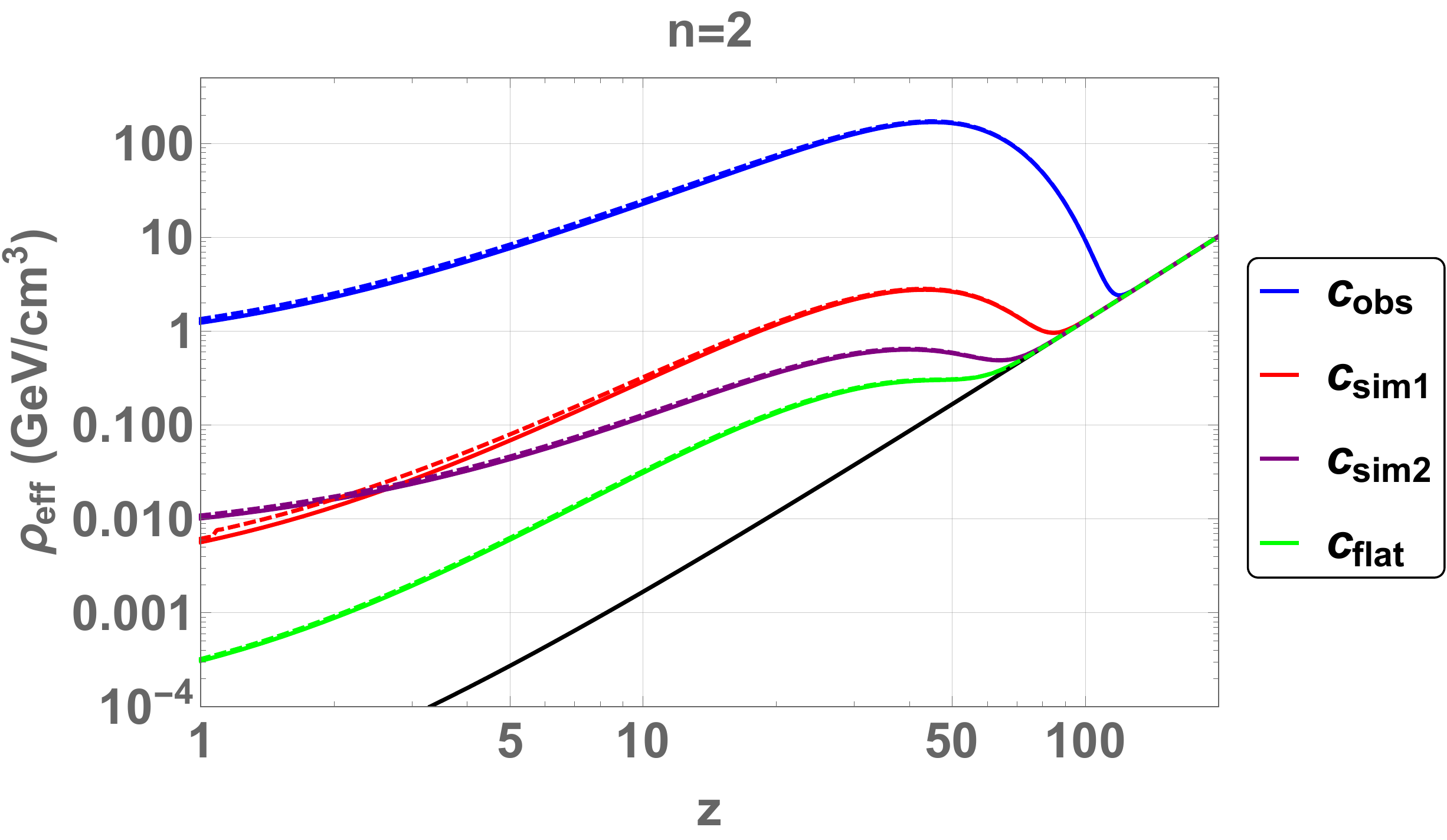}
\\ \vspace{.7cm}
\includegraphics[scale=.29]{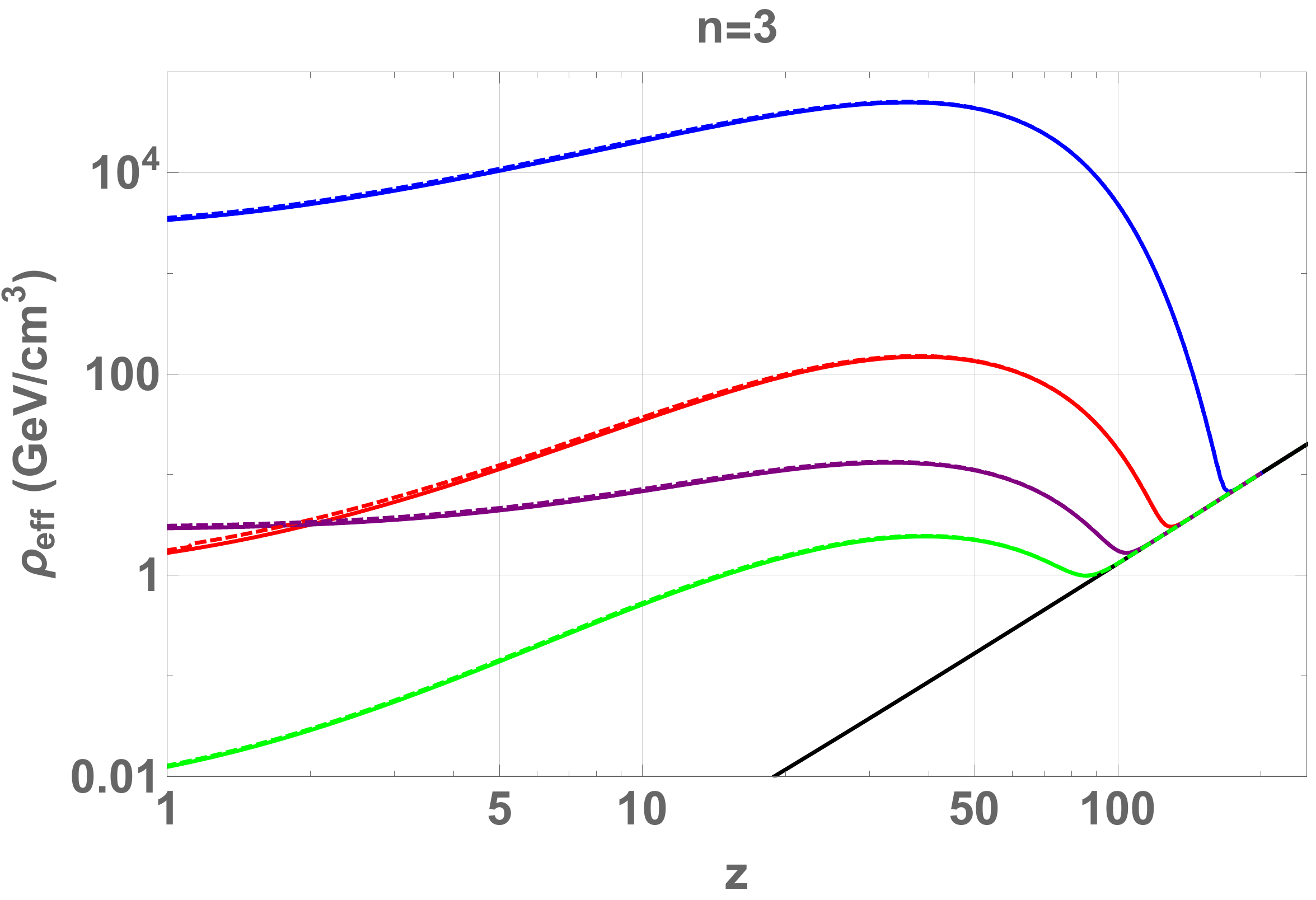}
\hspace{0.5cm}
\includegraphics[scale=.288]{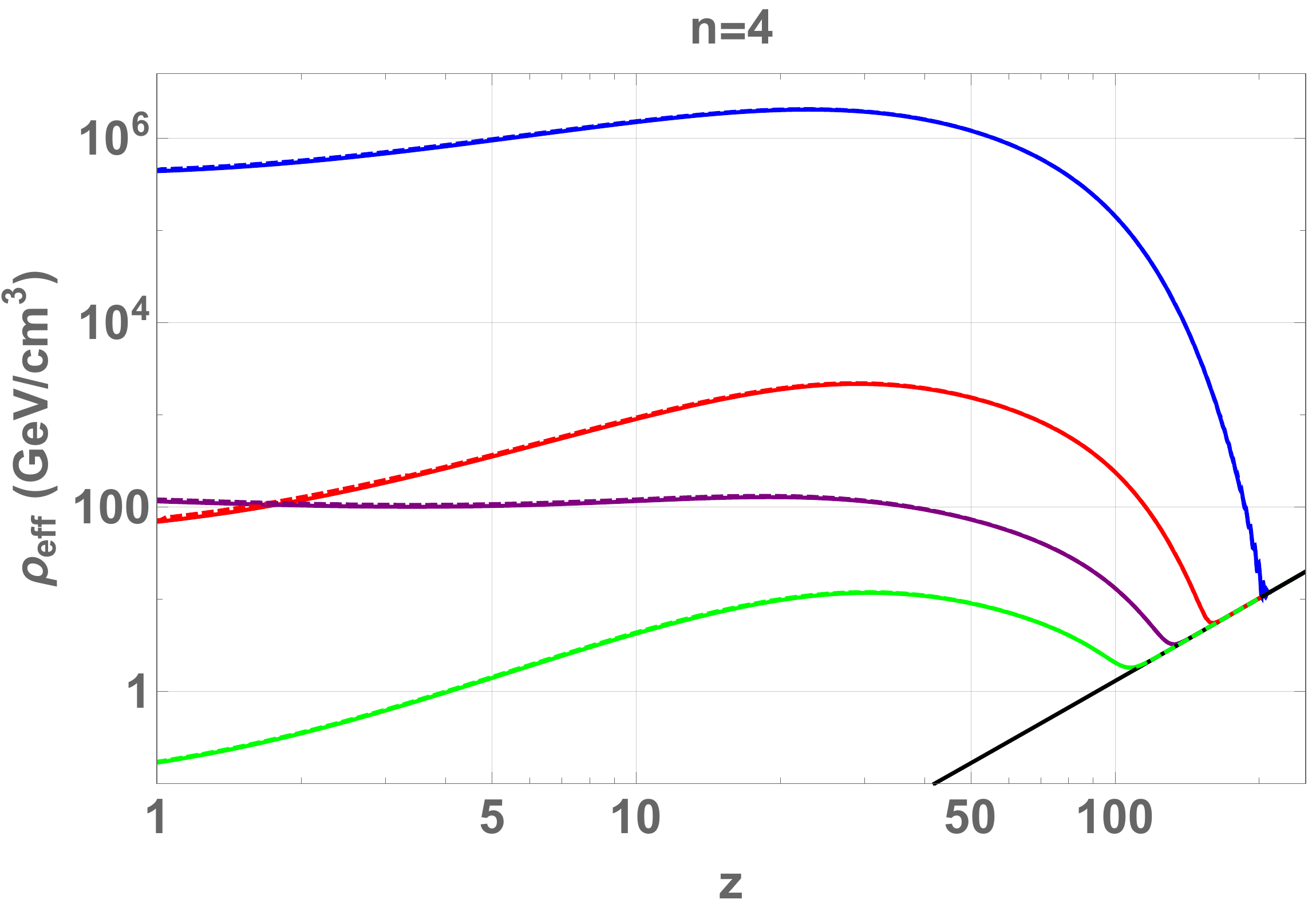}
\caption{The cosmological average $\rho_{\rm{eff}}$ as a function of redshift for $n=2$ (top), $n=3$ (bottom, left) and $n=4$ (bottom, right) for different concentration parametrizations. 
We have also plotted the smooth DM energy density $\rho_{\chi}(z)$ (black solid curve) for comparison. The solid curves are obtained by neglecting substructure whereas the dashed curves give the corrected $\rho_{\rm{eff}}$ including substructure. From top to bottom we have used Eq.~\eqref{c_obs} (blue), Eq.~\eqref{c_sim1} (red), Eq.~\eqref{c_sim2} (purple) and Eq.~\eqref{c_flat} (green) for the concentration parameter (note that the legend in the first plot applies to all plots). For $c_\text{sim2}$, given by Eq.~\eqref{c_sim2}, we have used the power spectrum with cutoff $k_c = 10^7h {\text{Mpc}}^{-1}$. }
\label{Fig:rho_eff}
\end{figure}

As we discussed earlier, the boost factor is dominated by the smallest halos. This fact allows us to derive an approximate scaling behaviour of the differential boost factor as a function of mass. First of all, notice that the variance in the low mass limit is almost a constant as a function of mass because a change in the halo mass, in this limit, only effectively changes the very small-scale contribution of the power spectrum in the variance integral, Eq.~\eqref{variance}, which is already quite small. This, in turn, implies that $\nu$ is almost a constant in the low-mass limit. Furthermore, notice that one can parametrize the matter power spectrum at short scales as $P(k \sim k_c) \simeq {\cal{P}} \left( \dfrac{k}{k_c} \right)^{-\tau_p} e^{-(k/k_c)^2}$ where ${\cal{P}}$ is an overall amplitude, the power-law factor comes from the transfer function with $\tau_p \simeq 2.9$ and the exponential cutoff is just ${\cal T}_\chi^2$ with ${\cal T}_\chi$ defined in Eq.~\eqref{Free_stream_T}. 

On the other hand, the halo boost factor Eq.~\eqref{B_halo}, in the low mass limit, depends on halo mass only through the concentration by a power-law relation $B_h \propto c^{3(n-1)}$. This is because $\gamma(x,y\gg 1) \sim \Gamma(x)$, and for small halo masses $c$ is large and $\alpha_e$ is small (and almost a constant).  
Finally, from Eq.~\eqref{B_total}, 
\ba 
\dfrac{dB}{dM}(M,z) = \dfrac{\langle \rho_h \rangle^{n-1}}{\rho_{m}^{n}(z)} (1+z)^3  B_h(M,z)\dfrac{dn_h(M,z)}{d\ln M}.
\ea 
Putting all mass dependent factors together and using Eq.~\eqref{mass_function_simp} for the mass function and noticing that $k \propto M^{-1/3}$ as well as $k_c \propto M_c^{-1/3}$ we get 
\ba 
\dfrac{dB}{dM}(M,z) \propto  \left(\dfrac{M}{M_c} \right)^{\frac{\tau_p}{3}-3\tau_c (n-1)-2} e^{-(M/M_c)^{-2/3}},
\ea 
where we have omitted the redshift dependence as well as the overall amplitude, and we have assumed that the concentration scales as a power law with respect to mass, with some index $\tau_c$, i.e. $c \propto (M/M_c)^{-\tau_c}$. Note that all the different concentrations discussed above satisfy the condition $\tau_c \geq 0$. Clearly, the above approximate differential boost factor peaks around $M \sim M_c$, showing that the boost factor is mostly sensitive to the halos with the lowest mass. 

In the above estimation, we neglected the substructures. As is evident in Fig.~\ref{Fig:rho_eff}, the substructures would be less important in the universal boost factor. This is because, in the small mass limit, the number density of isolated halos is typically larger than that contained in larger halos, although they are of the same order of magnitude. 

\begin{figure}
\hspace{3.4cm}
\includegraphics[scale=.29]{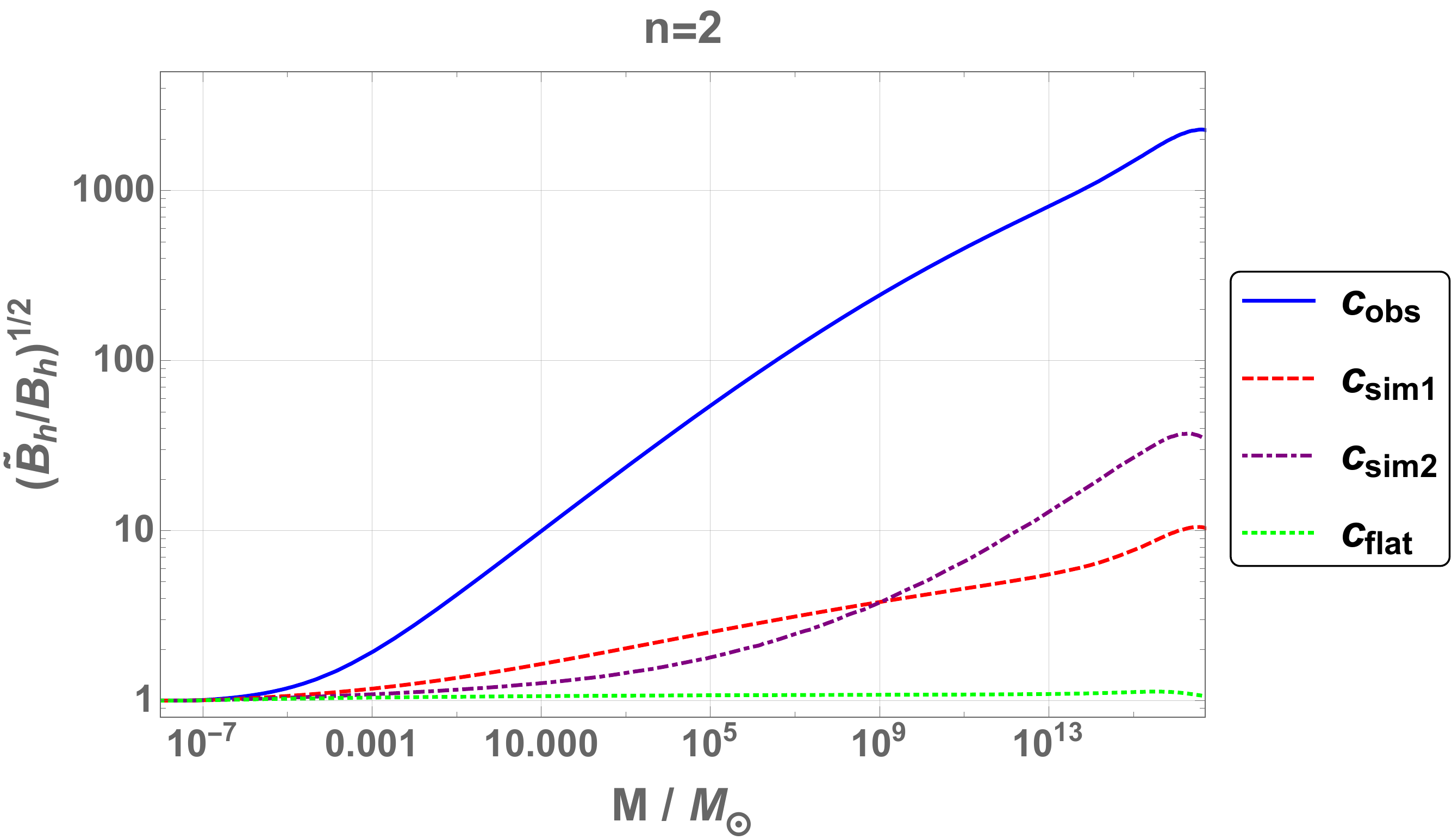}
\\ \vspace{.8cm}
\includegraphics[scale=.33]{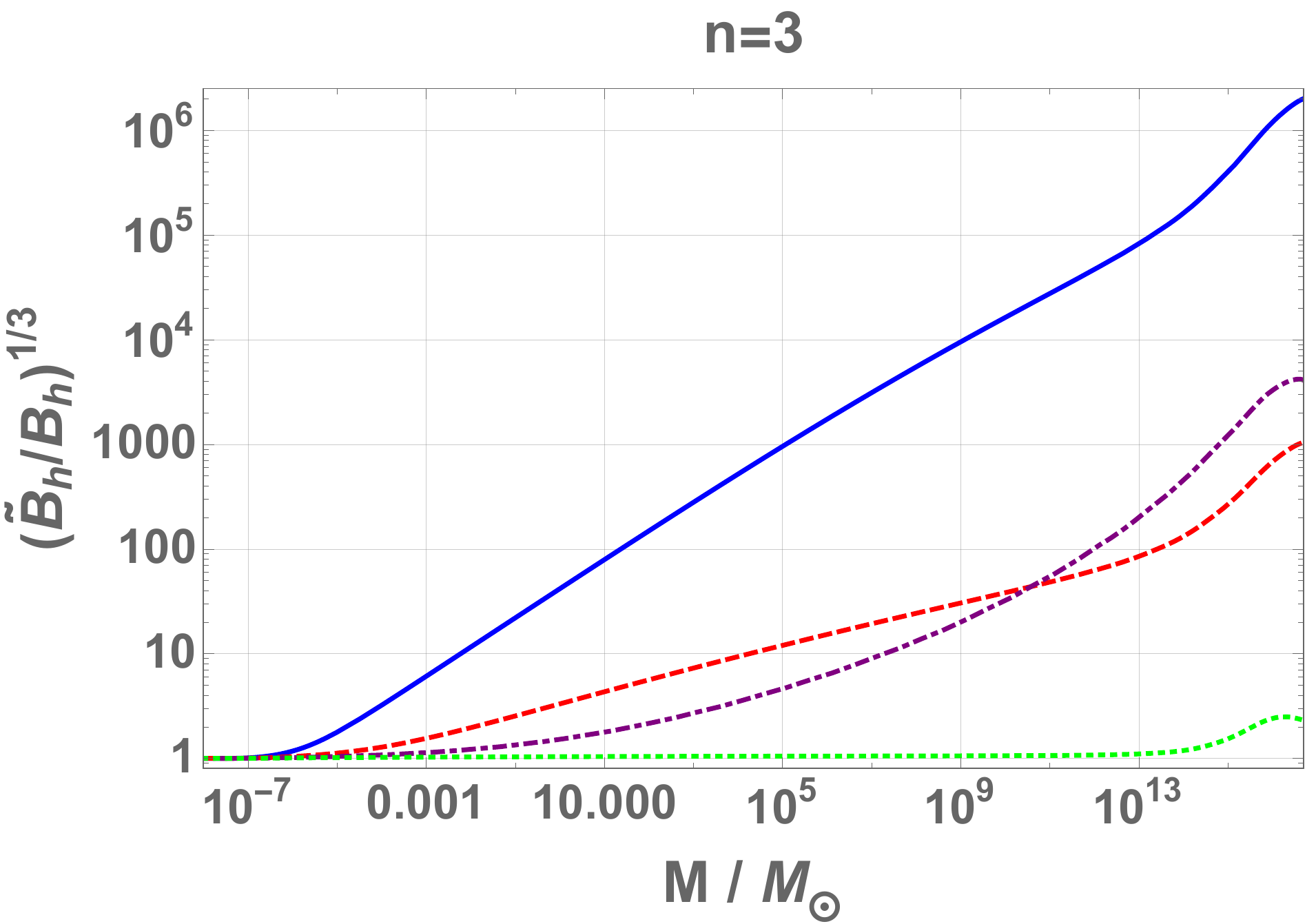}
\hspace{.3cm}
\includegraphics[scale=.33]{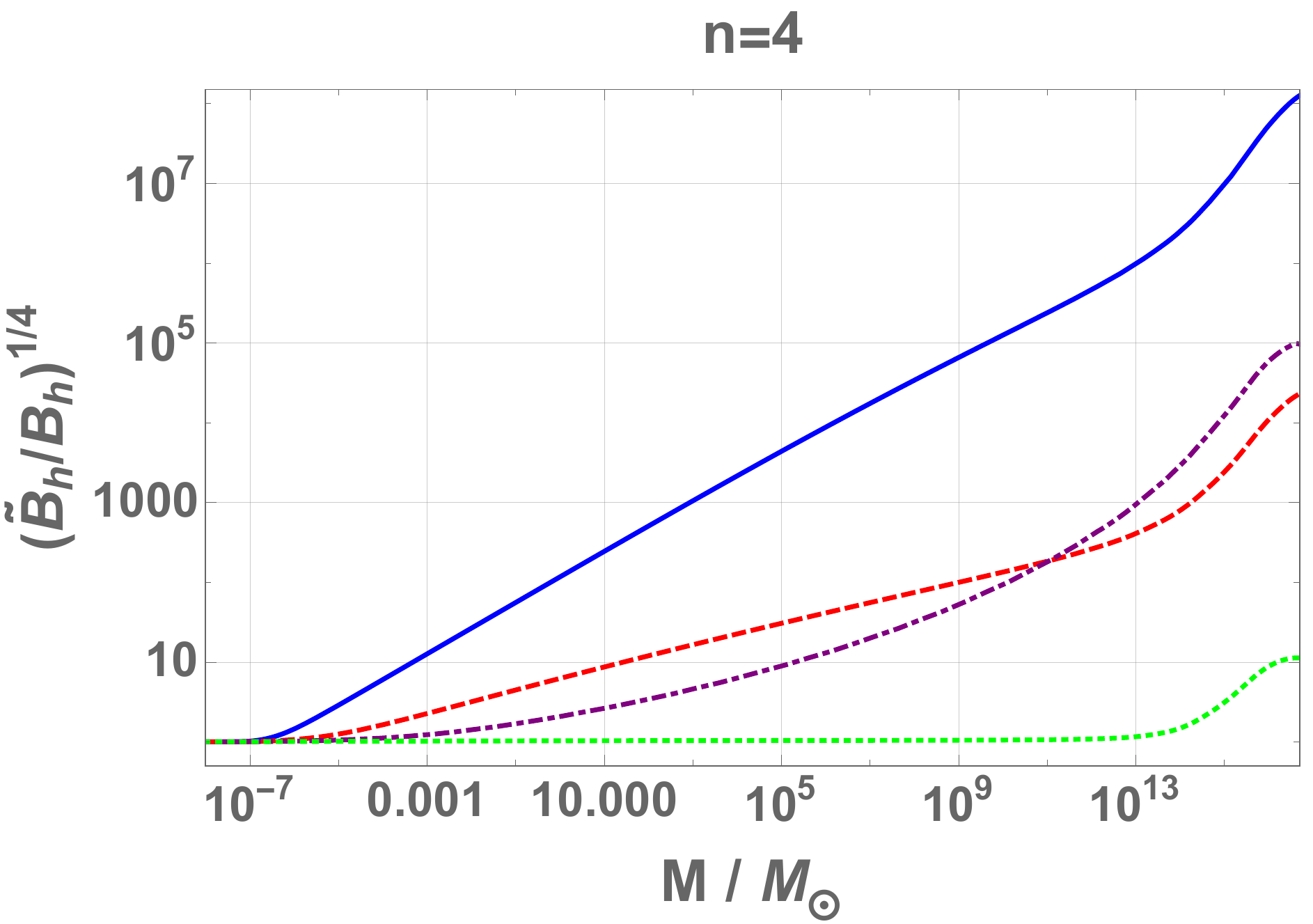}
\caption{The ratio $(\tilde B_h/B_h)^{1/n}$ with $n=2$ (up), $n=3$ (bottom, left) and $n=4$ (bottom, right) at redshift $z=0$ for different concentration parametrizations (see the legend, which applies to all panels). We have set $k_c=10^7 h {\text{Mpc}}^{-1}$.}
\label{Fig:B_ratio}
\end{figure}

However, for studying individual clusters, the effect of substructures can be significant. Note that the signal from a cluster is proportional to $B_h$ when the substructures are neglected, and is proportional to $\tilde B_h$ when they are taken into account. To illustrate the significance of substructures, it is suitable to plot the ratio $( \tilde B_h/B_h)^{1/n}$ as a function of cluster mass which indicates how important the effect of substructures is on the effective DM energy density. See Fig. \ref{Fig:B_ratio} where this ratio has been depicted for $n=2,3$ and $4$. It is evident that the effect of substructure in the total signal is significant.
\section{Diffuse signals}
\label{sec:constraints} 

In this section we consider various indirect-detection searches for DM annihilation in halos and the intergalactic medium, and their capacity to constrain $n$-body annihilation. We will summarize how to estimate the size of the DM signals in each of these channels, and then discuss the resulting limits. Because we expect the signal to rise steeply at lower masses, we will focus on searches relevant to constraining sub-GeV DM, from the cosmic microwave background and from telescopes measuring X-rays and soft gamma rays.

\subsection{Parametrization of the expected signal}
We will characterize the strength of the indirect-detection signals by a normalization parameter:
\begin{equation} \xi \equiv (dE/dV dt)/\rho_{\chi,\text{loc}}^n, \label{eq:xi} \end{equation} 
where $dE/dV dt$ is the {\it local} rate of energy injection into SM particles (or unstable particles which decay purely to the SM) per unit volume per unit time, and $\rho_{\chi,\text{loc}}$ is the {\it local} DM mass density. For example, for DM in the intergalactic medium far away from any structures, the local DM density could be well approximated by the cosmological average DM density, whereas for DM annihilating within a halo, the local DM density would be determined by the halo density profile, and whether the annihilation is occurring within a denser substructure. In both cases, the annihilation signal scales with the appropriate power of the local DM density, and the ratio $\xi$ is determined by the particle physics of the theory. This parameter $\xi$ controls the normalization of the signals to which our indirect-detection observations are sensitive, so in the following sections, we will formulate our constraints in terms of $\xi$.

The dependence of $dE/dV dt$ on redshift, position etc due to variations in the DM density is explicitly canceled out in the quantity $\xi$. However, $\xi$ could still possess redshift- or position-dependence; for example, if the annihilation cross section has a non-trivial velocity dependence.

Note that $\xi$ has mass dimension $5 - 4 n$, so from naive dimensional analysis, we expect it to be a rapidly varying function of the DM mass for $n > 2$. If a $\mathcal{O}(1)$ fraction of the power per annihilation goes into production of SM particles, then $\xi$ and $\langle \sigma v^{n-1}\rangle$ are related by $\xi \sim \langle \sigma v^{n-1}\rangle m_\chi^{1-n}$. The exact relationship must be calculated for each model, and depends on combinatoric factors (i.e. how many identical particles are present in the initial state) as well as the fraction of the annihilation power that goes into SM particles. In the unitarity saturating limit, as discussed in Sec.~\ref{sec:sommerfeld}, the annihilation cross-section scales as  $\langle \sigma v^{n-1}\rangle \propto k^{5 - 3 n}$. The resulting scaling of the parameter $\xi$ is then expected to be $\xi \propto m_\chi^{5-4n} v^{5 - 3n}$, where $v$ is the typical relative velocity between DM particles.

\subsection{Cosmic microwave background limits}
\label{sec:CMB}

\subsubsection{Methodology}

DM annihilation during the cosmic dark ages and the epoch of reionization can modify the cosmic microwave background (CMB), and is thus tightly constrained (e.g.~\cite{Madhavacheril:2013cna}). These constraints depend mostly on the amount of power injected in electromagnetic channels, and to a lesser degree on the spectrum of the injected particles; for decay \cite{Slatyer:2016qyl} and (two-body) $s$-wave annihilation, they provide broadly-applicable and nearly model-independent limits on the decay lifetime / annihilation rate, for DM masses in the keV to multi-TeV range. These limits are particularly stringent for light DM, excluding the thermal relic cross section for $s$-wave annihilation for DM masses below $\sim 10$ GeV (assuming there is no substantial branching fraction into neutrinos). In this section we extend these limits to 3- and 4-body annihilation. While in the case of 2-body annihilation the dominant signal comes from $z\sim 600$, from Fig. \ref{Fig:rho_eff} we can infer that signals from the epoch of structure formation will be relatively more important for $n > 2$.

The CMB signal is largely controlled by the rate of energy going into hydrogen ionization at redshift $z$ from DM annihilation, which can be written in terms of the parameter $\xi$ (Eq.~\eqref{eq:xi}) as:
\begin{eqnarray}
\left(\dfrac{dE}{dVdt}\right)_\text{ionH} & = & \xi \,  g_\text{ionH}(z) \Omega_{DM}^n \rho_{c,0}^n (1 + z)^{3n} \,.
\end{eqnarray}     
Here $\rho_{c,0}$ is the critical density in the present day, $\Omega_{DM}$ is the DM mass density as a fraction of the critical density.\footnote{If the annihilating DM is a small fraction of the total DM, then the power injection parameter $\xi$ will be suppressed accordingly.} Note that the quantity $\left(dE/dVdt\right)_\text{ionH}$ is the energy injection rate per unit volume averaged over the whole universe (in contrast with $(dE/dVdt)$ in the definition of $\xi$ (Eq.~\eqref{eq:xi}), which is defined locally). The function $g_\text{ionH}(z)$ captures the non-trivial rescaling of the injected energy by the boost factor $(\rho_\text{eff}/\Omega_{DM} \rho_c)^n$, and also the amount of energy deposited into ionization at redshift $z$, as opposed to other channels. We will also include the effects of extra excitations and heating from the secondary products of DM annihilation, which are controlled by similar rescaling functions $g_c(z)$, where $c$ is a channel label.

The functions $g_c(z)$ are calculated, using the results of \cite{Slatyer2015a}, from an integral over energy injection at all previous redshifts, taking into account the time needed for the injected particles to cool and lose their energy. These functions thus depend on the boost factor evaluated at all redshifts higher than the redshift of interest $z$. See \cite{Slatyer2015a} for more details; the parameter here called $g_c(z)$ is there labeled $f_c(z)$.
We will follow the methodology described in Ref.~\cite{Slatyer:2016qyl} to determine the CMB signatures of $n$-body annihilation; that is, we consider injection of photons and $e^+ e^-$ pairs with injection energies ranging from keV to multi-TeV scales, and redshift dependence appropriate to $n$-body annihilation (i.e. $dE/dV dt$ scales as $\rho_\text{eff}(z)^n$). We compute the impact on the CMB using the \texttt{CLASS} public code \cite{Lesgourgues:2011re}, and then perform a principal component analysis (PCA) to estimate the variance in the impact on the CMB anisotropy spectrum for injections of particles at different energies. We marginalize over the standard six cosmological parameters in the PCA (see Ref.~\cite{Slatyer:2012yq} for details). We then validate the limits by a Markov Chain Monte Carlo (MCMC) analysis, using the likelihoods from Planck 2015 (TT + TE + EE, low-$\ell$ and high-$\ell$, and the lensed $C_\ell$) \cite{Ade:2015xua}, and the \text{MontePython} public code \cite{Audren:2012wb}.

We note that limits can also be placed by requiring that the increased ionization level not overproduce the total optical depth, and that heating of the gas by DM annihilation products not violate constraints on the gas temperature \cite{Diamanti2014,Liu:2016cnk}. We have checked these limits following the methodology of \cite{Liu:2016cnk} and using  \texttt{CLASS} to compute the modifications to the temperature and ionization histories. In particular, we require that the high-redshift optical depth to recombination from DM annihilation satisfy \cite{Adam:2016hgk}:
\begin{equation}
\delta\tau = - \int_{6}^{z_{CMB}} dz n_e(z) \sigma_T \dfrac{dt}{dz}  \leq 0.044,
\end{equation}
and that the temperature history satisfy \cite{Bolton2010,Becker2011,Bolton2011}:
\begin{eqnarray}
\text{log}_{10}\left(\dfrac{T_{IGM}\left(z=6.08\right)}{K}\right)=4.21 \substack{+0.06 \\ -0.07}, \quad \text{log}_{10}\left(\dfrac{T_{IGM}\left(z=4.8\right)}{K}\right)=3.9\pm 0.1.
\end{eqnarray}
We take the conservative approach of ignoring non-DM sources of heating, and requiring that the heating from DM does not exceed these limits.

However, under these assumptions, we find that these limits are always weaker than the CMB anisotropy bounds, so we will focus on the latter in our discussion. We note that measurements of the 21cm signal from neutral hydrogen during the cosmic dark ages could potentially set tighter bounds on the thermal history and hence on the contribution to heating from DM annihilation \cite{Valdes:2007cu,Valdes:2012zv}, but at present the only claimed detection of such a signal \cite{bowman2018absorption} suggests a gas temperature lower than in the standard scenario with no DM annihilation, making the extraction of limits on annihilation somewhat scenario-dependent \cite{Liu:2018uzy}. 

\subsubsection{Principal components and weighting functions}

\begin{figure}
 \includegraphics[width=7.5cm]{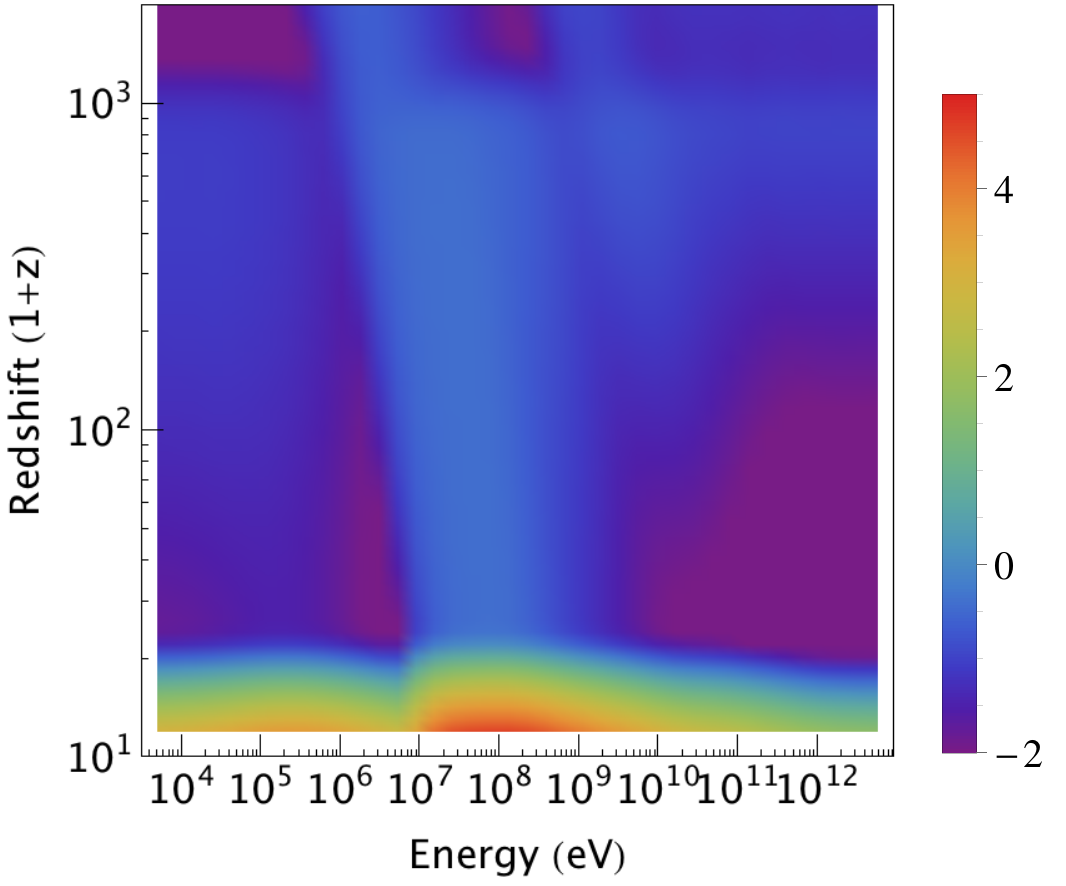}
 \hspace{.7cm}
 \includegraphics[width=7.5cm]{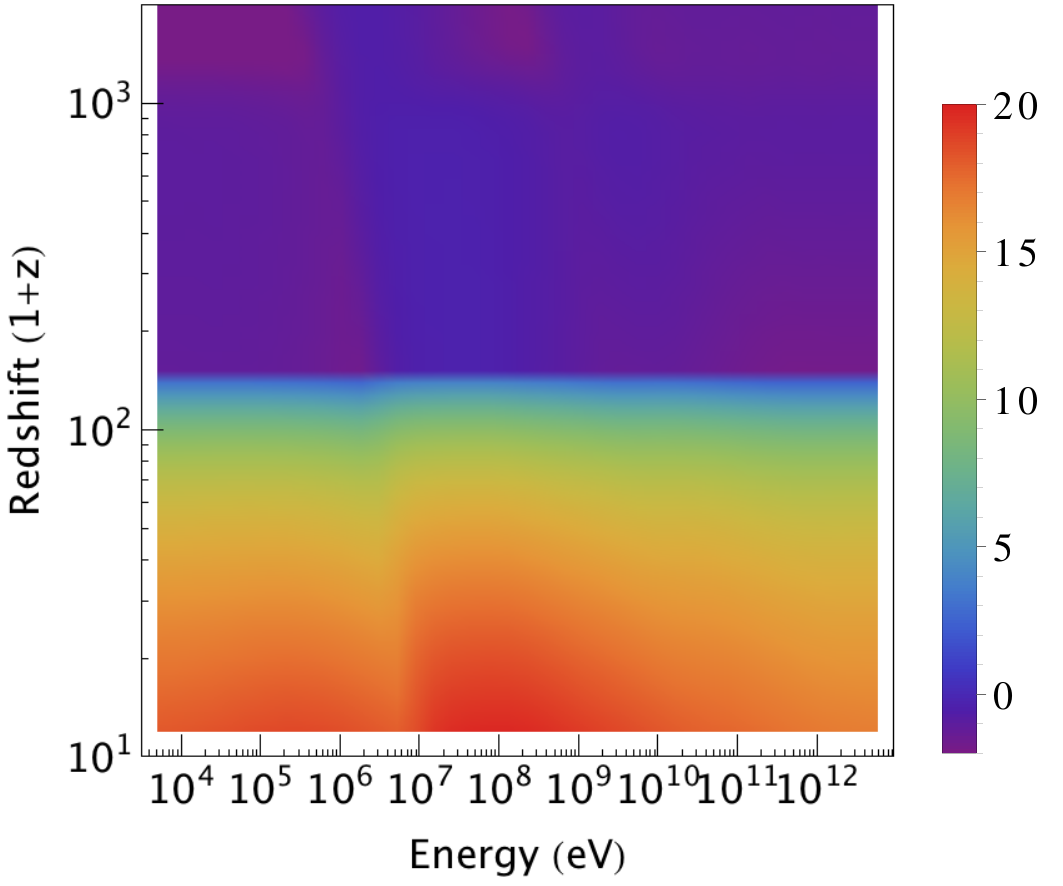}
 \\ \vspace{.7cm}
 \includegraphics[width=7.5cm]{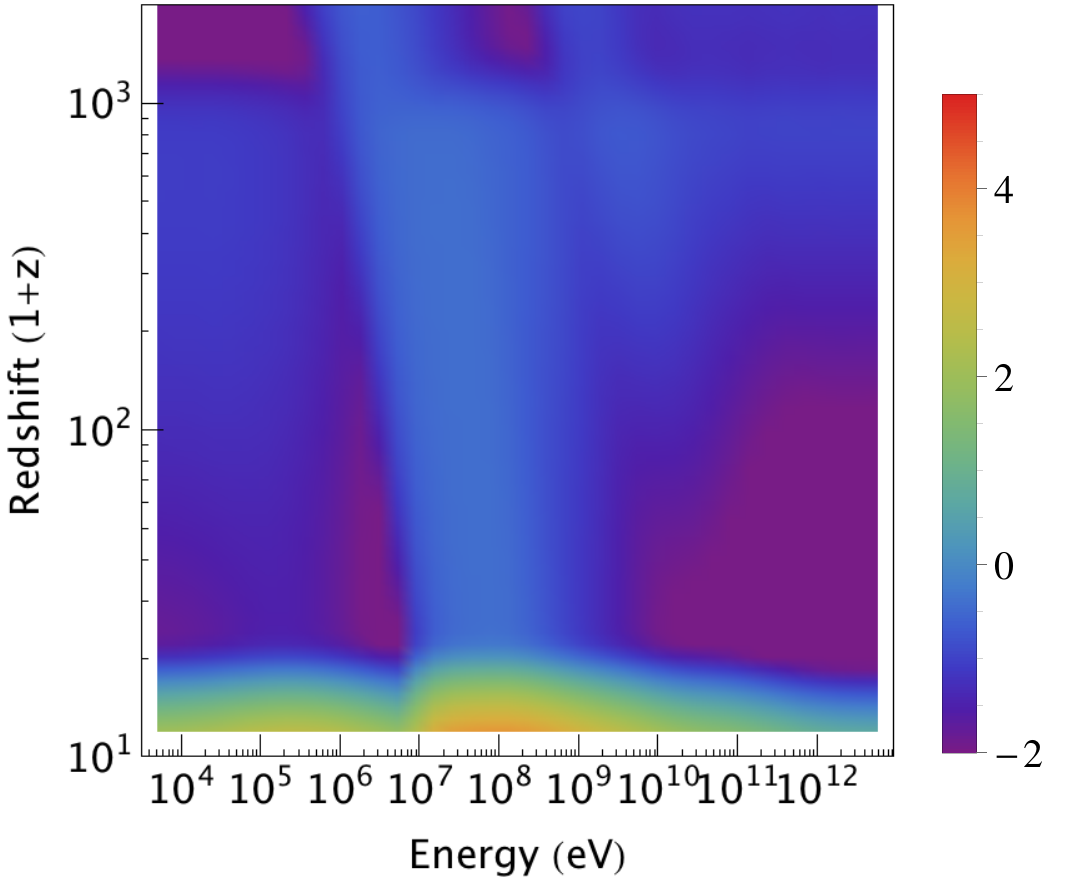}
 \hspace{.7cm}
 \includegraphics[width=7.5cm]{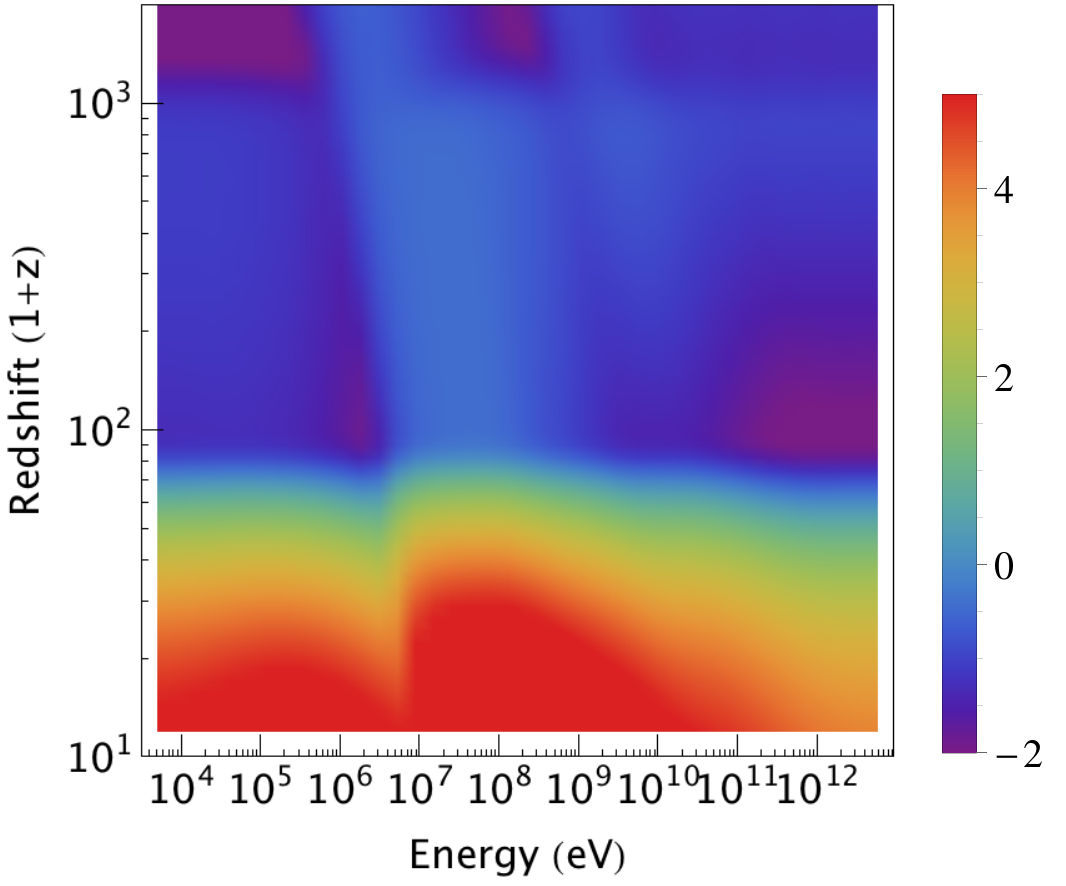}
 \caption{
The log of the energy deposition efficiency function for hydrogen ionization, $\log_{10}(g_\text{ionH}(z))$, for injection of electron-positron pairs by 3-body annihilation over a range of energies, for concentration models $c_{\text{obs}}$ (upper panel) and $c_{\text{flat}}$ (lower panel) with cutoff scale $k_c$ of $10 \,h {\text{Mpc}}^{-1}$ (left panel) and $10^7 \,h {\text{Mpc}}^{-1} $ (right panel).}
\label{efficiency2}
\end{figure}

As an illustrative example to build intuition, let us first consider the signals from 3-body annihilation with two extreme concentration models, $c_{\text{obs}}$ and $c_{\text{flat}}$, and cutoff scales of $10  \,h {\text{Mpc}}^{-1}$ and $10^7 \, h {\text{Mpc}}^{-1}$. These choices span the full range of $\rho_{\text{eff}}$ among the structure formation models we considered. Fig. \ref{efficiency2} shows the function $g_\text{ionH}(z)$ and its dependence on injection energy, and redshift, for these model choices. The onset of structure formation at late times spurs a steep rise in $g_{\text{ionH}}(z)$ at low redshifts; this effect is more pronounced when $c_{\text{obs}}$ is used rather than $c_{\text{flat}}$, and when the cutoff momentum scale is increased from $10  \,h {\text{Mpc}}^{-1}$ to $10^7 \, h {\text{Mpc}}^{-1} $, corresponding to lower-mass halos being allowed. This is expected, as these choices give rise to larger $\rho_{\text{eff}}$ at low redshift, as shown in Fig.~\ref{Fig:rho_eff}.

\begin{figure}
\hspace{2cm}
 \includegraphics[width=12.cm]{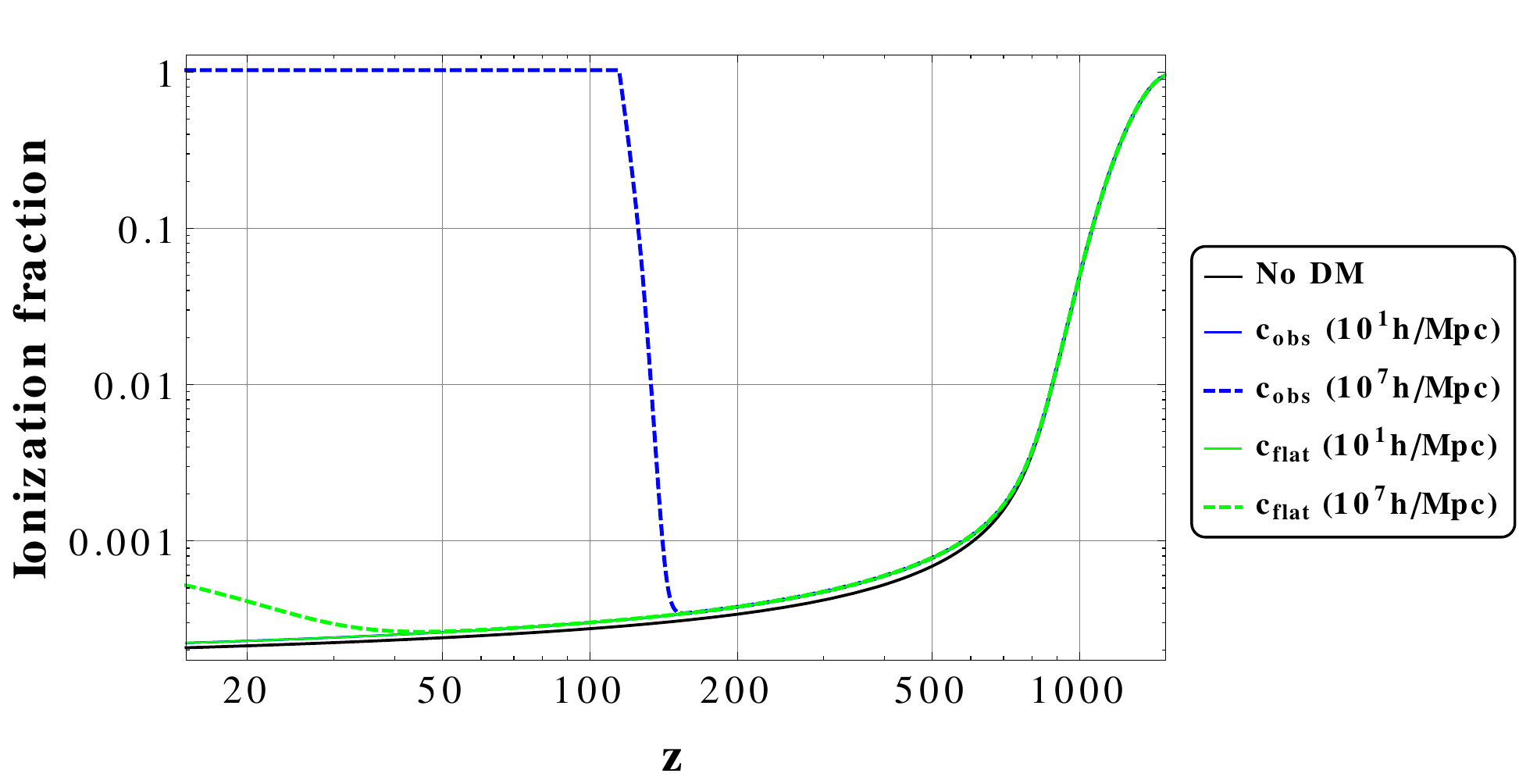}
 \caption{
Modifications to the hydrogen ionization fraction by 3-body annihilation producing 100 MeV $e^+ e^-$ pairs (that is, each particle has 100 MeV of kinetic energy) from a 3-body interaction setting $\xi =1.2 \times 10^{7}$ MeV$^{-7}$, for different concentration models and cutoff scales. Note that the blue and green solid curves overlap with each other.}
\label{ionizationtest}
\end{figure}

The changes to the ionization history from a large energy injection (to make the effect pronounced and easily visible to the eye) are shown in Fig.~\ref{ionizationtest}, for the various combinations of cutoff scale and concentration model. As a benchmark for this example, we consider annihilation that produces electrons and positrons (in equal numbers) with energies of 100 MeV; in the case of a decaying non-relativistic mediator (as we will discuss for a specific model in Sec.~\ref{sec:models}), this would correspond to a mediator mass of 200 MeV.

We note that in all cases there is an appreciable cutoff- and concentration-independent modification to the ionization history at high redshift, corresponding to the signal from the smooth DM density in the epoch prior to structure formation; the choice of cutoff and concentration affects the dominant contribution from halos at $z \lesssim 200$.

To understand the impact of changing the energy of injected particles, or injecting photons rather than electrons and positrons, we perform a principal component analysis with respect to injection energy and species. As for 2-body annihilation \cite{Finkbeiner:2011dx} and decay \cite{Slatyer:2016qyl}, we find that the first principal component consistently dominates the variance in the 3-body case, capturing over 90\% of the variance; this is true for all the different concentration models and cutoffs we tested. We can thus characterize the effects of 3-body DM annihilation (for a given structure formation model) by a largely DM-model-independent pattern of perturbations to the CMB anisotropy spectrum, together with a model-dependent normalization factor. The same approach can be applied for 4-body annihilation, and likewise we find that the first principal component consistently contributes over 95\% of the variance.

\begin{figure}
  \includegraphics[width=7.cm]{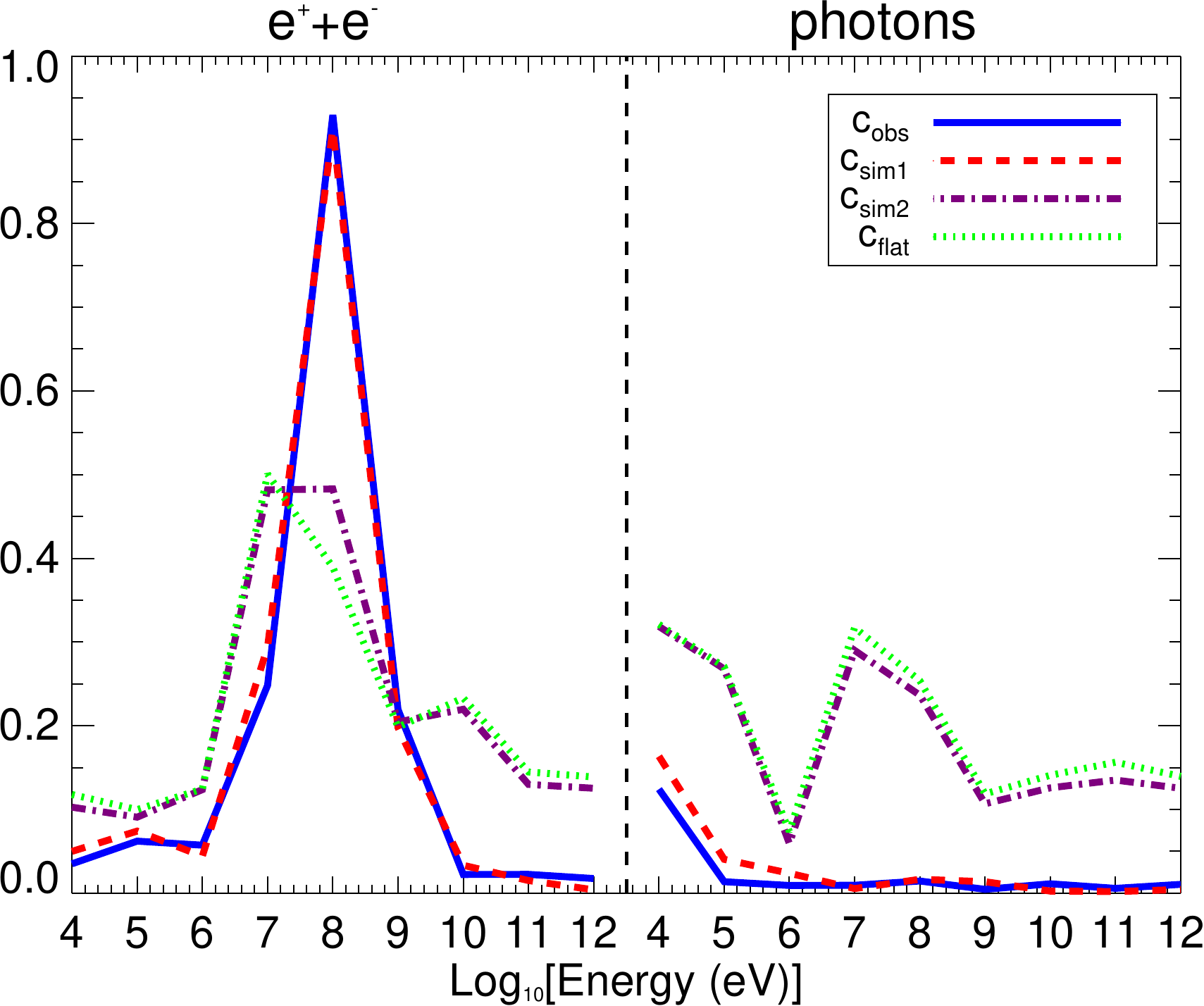}
  \hspace{1cm}
  \includegraphics[width=7.cm]{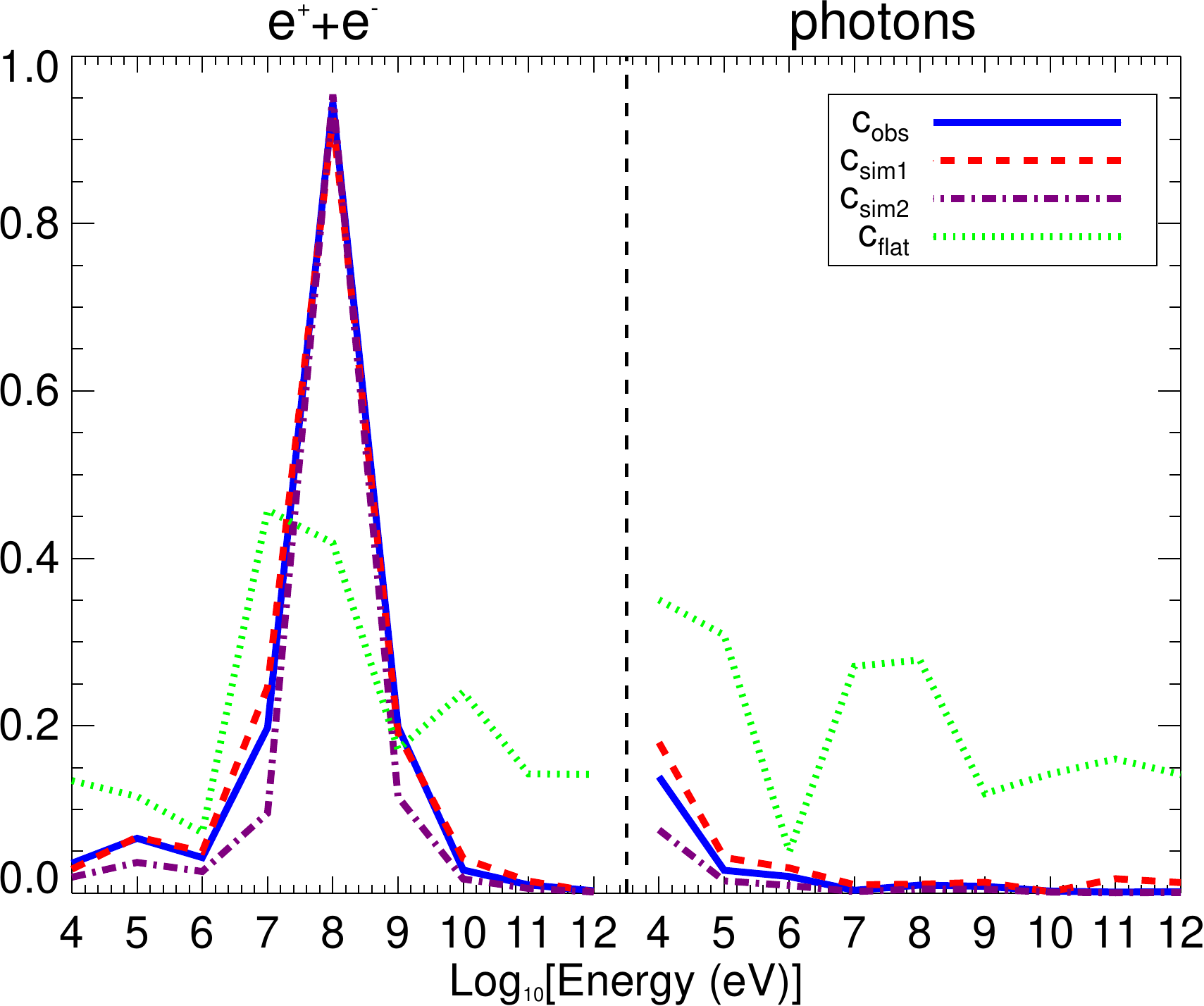}
 \caption{
First principal component, with respect to injection energy and injected particle species, for $k_c = 10^7 \, h {\text{Mpc}}^{-1} $ and different concentration models (see text for model details). We show results for 3-body annihilation (left) and 4-body annihilation (right).}
\label{PCA}
\end{figure}

The first principal component (PC), evaluated for 3-body and 4-body annihilation with different concentration models but a high $k_c = 10^7 \, h{\text{Mpc}}^{-1} $, is shown in Fig.~\ref{PCA}; this curve describes the significance of the changes to the CMB as a function of injection energy and species, holding $\xi$ (and the structure formation model) fixed. The model-dependent normalization factor is given by $\xi$ multiplied by an injection-spectrum-dependent efficiency factor, with the latter obtained by taking the integral of the spectrum against the first PC (see \cite{Slatyer:2016qyl} for further discussion). 

We see that the shape of the first PC differs noticeably between different concentration models; this is due to the differing redshift dependence of the signals in these cases. At low redshifts, the universe is more transparent, and the variation in deposition efficiency between injected particles of different energies is more pronounced. For example, in the 3-body case we see that the two concentration models with a weaker signal at low redshifts have very similar first PCs, as do the two models with a stronger signal at low redshifts, but the two sets of first PCs are quite different from each other. This is because in the models with smaller concentrations, the signal is dominated by high redshifts, whereas in the higher-concentration models the signal peaks near the end of the cosmic dark ages, as suggested by Fig.~\ref{ionizationtest}. In the 4-body case, structure formation is relatively more important, and only in the case of the $c_\text{flat}$ concentration model is the first PC very different (suggesting that only in this case does the high-redshift signal dominate). The choice of a high $k_c$ accentuates these differences; with $k_c = 10\, h {\text{Mpc}}^{-1} $, the PCs for all concentration models resemble those for the $c_\text{flat}$ concentration model, as the signal is dominated by high redshifts prior to the onset of structure formation.

We see that the estimated $2\sigma$ constraint on $\xi$ is generically strongest for injections of 100 MeV $e^+ e^-$ pairs (each particle has 100 MeV of kinetic energy). Cross-checking our PCA results, using the first PC only, against \texttt{MontePython}, for 3-body annihilation with our bracketing choices of concentration model and cutoff scale, we find fairly good agreement, with the PCA typically giving a constraint too strong by $\sim 30-40\%$. Results are given in Table  \uppercase\expandafter{\romannumeral1}. We also test the effect of including additional PCs and find it to be small, as expected.

\begin{table}
\def\arraystretch{1.3}
\begin{tabular}{ m{4.cm}|m{4.cm} m{4cm} c}
\hline
 Injection histories  & PCA (1PC) & PCA (3PCs) & \texttt{MontePython}    \\  
 \hline 
 $c_{obs}$ $(10^1 h {\text{Mpc}}^{-1})$  & 1.2$\times10^{7}$ & 1.1$\times10^{7}$ &  1.3$\times10^{7}$   \\
 $c_{obs}$ $(10^7 h {\text{Mpc}}^{-1})$  & 3.3$\times10^{-4}$ & 3.3$\times10^{-4}$  &   5.1$\times10^{-4}$  \\
 $c_{flat}$ $(10^1 h {\text{Mpc}}^{-1})$   & 1.2$\times10^{7}$ & 1.1$\times10^{7}$  &  1.4$\times10^{7}$ \\
 $c_{flat}$ $(10^7 h {\text{Mpc}}^{-1})$   & 1.2$\times10^{7}$ & 1.1$\times10^{7}$ &  1.4$\times10^{7}$  \\
 \hline
\end{tabular}
\caption{2-$\sigma$ constraints on the DM annihilation parameter $\xi$ for 3-body annihilation, for a range of cutoff scales and concentration parameter models, as estimated by PCA  by only 1 PC(first column), and 3 PCs (second column) and cross-checked using \texttt{MontePython} (third column). Units are MeV$^{-7}$.}
\end{table}

\begin{figure}
 \includegraphics[width=8cm]{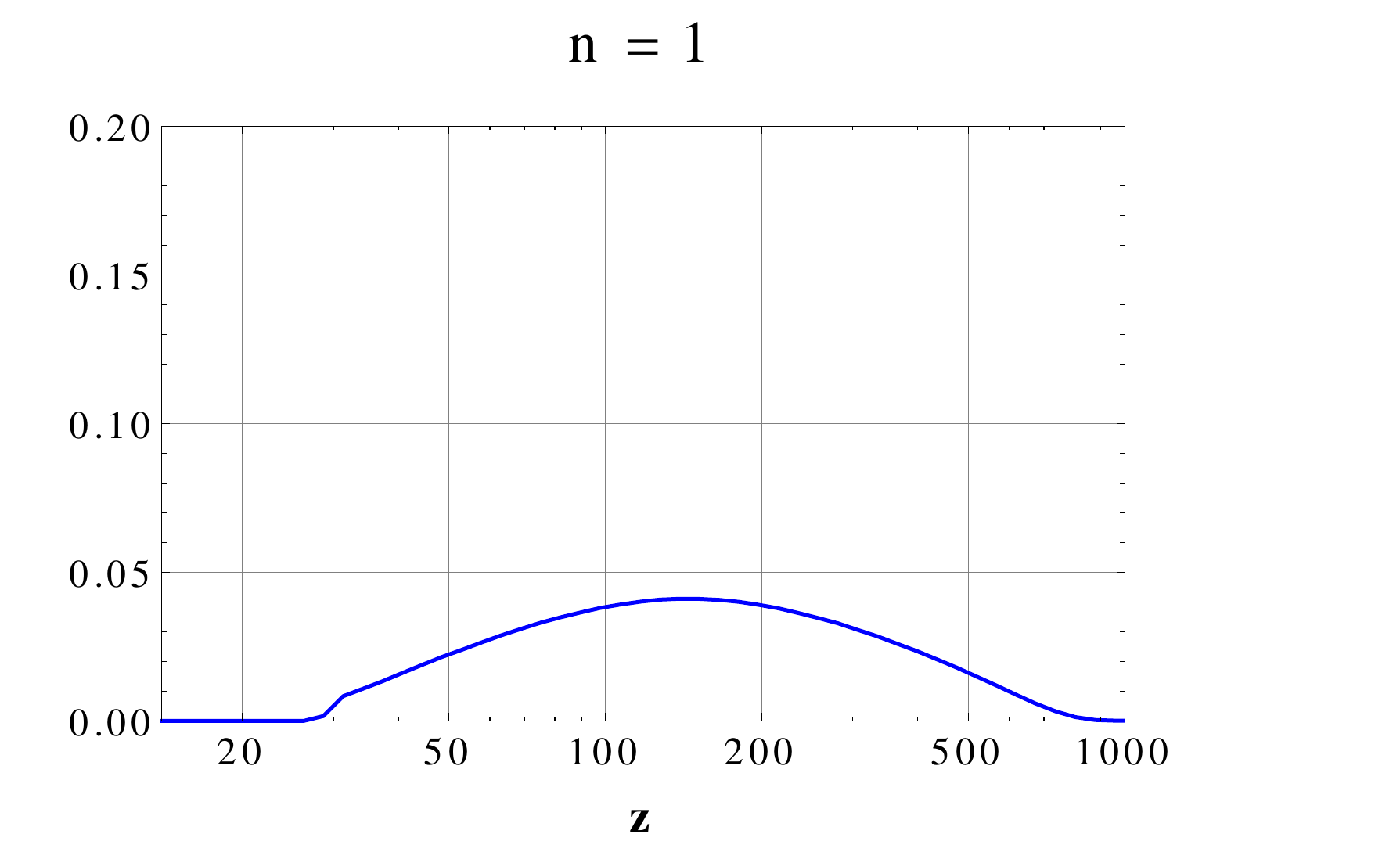}
 \includegraphics[width=8cm]{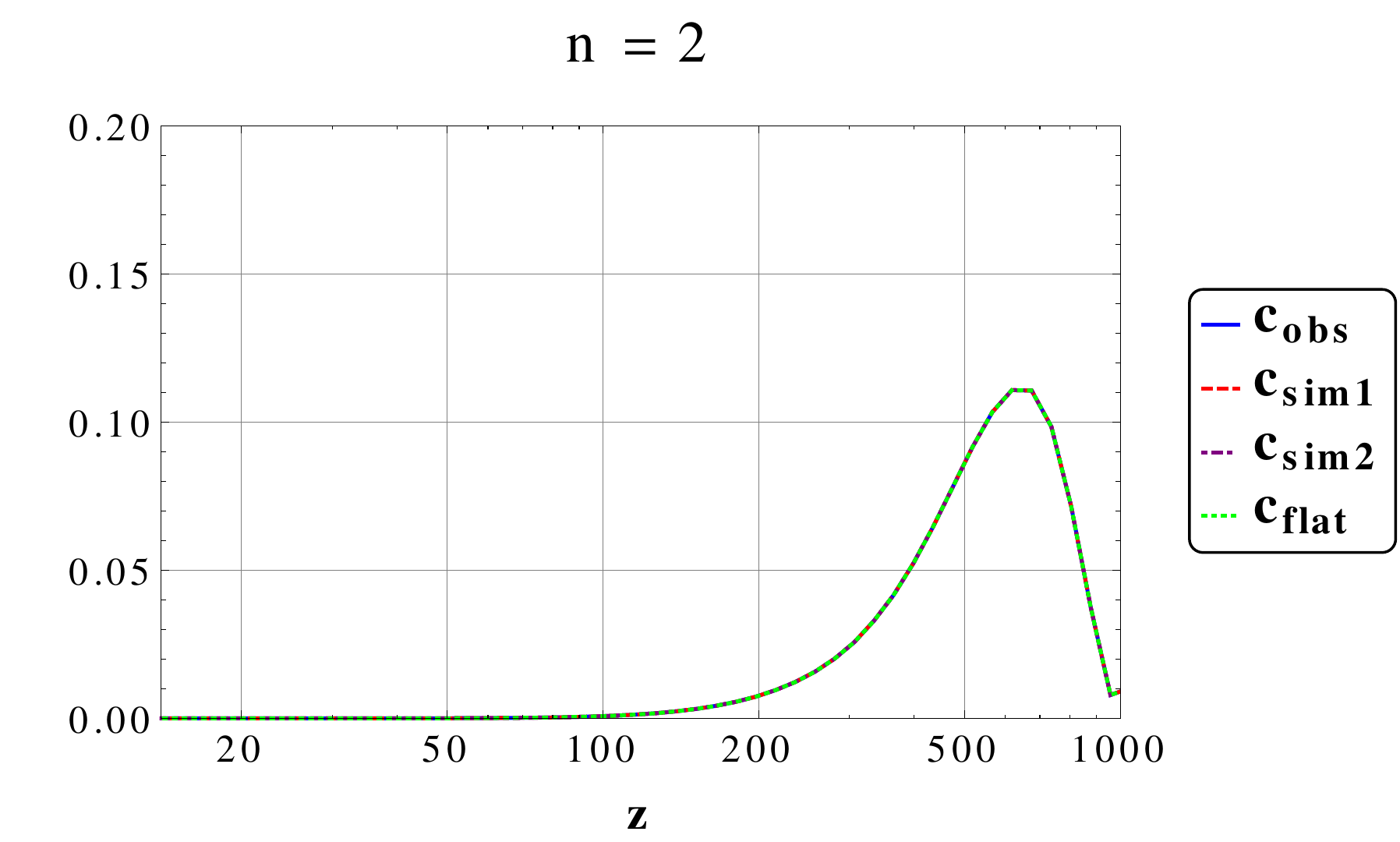}
  \\ \vspace{.7cm}
 \includegraphics[width=8cm]{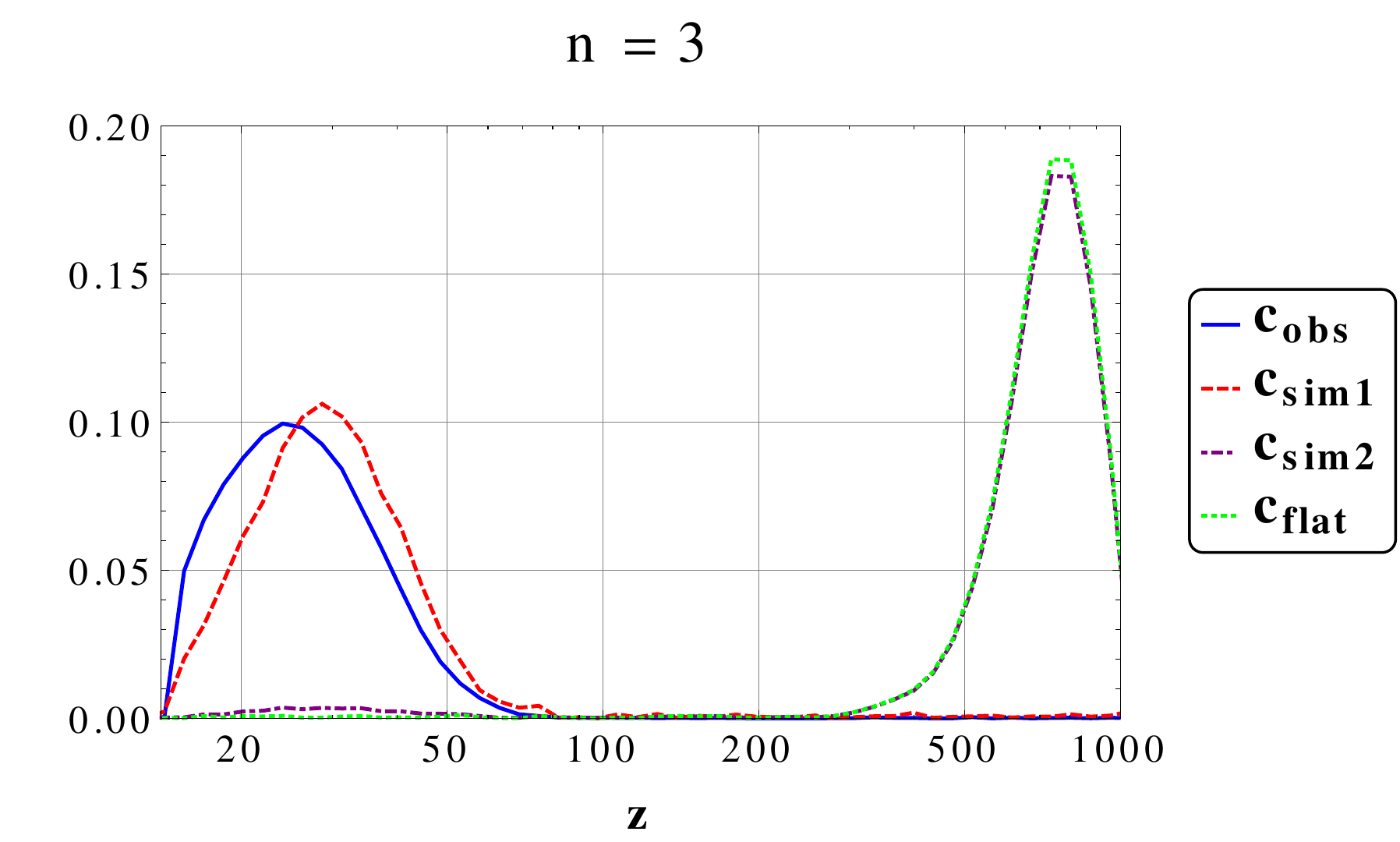}
  \includegraphics[width=8cm]{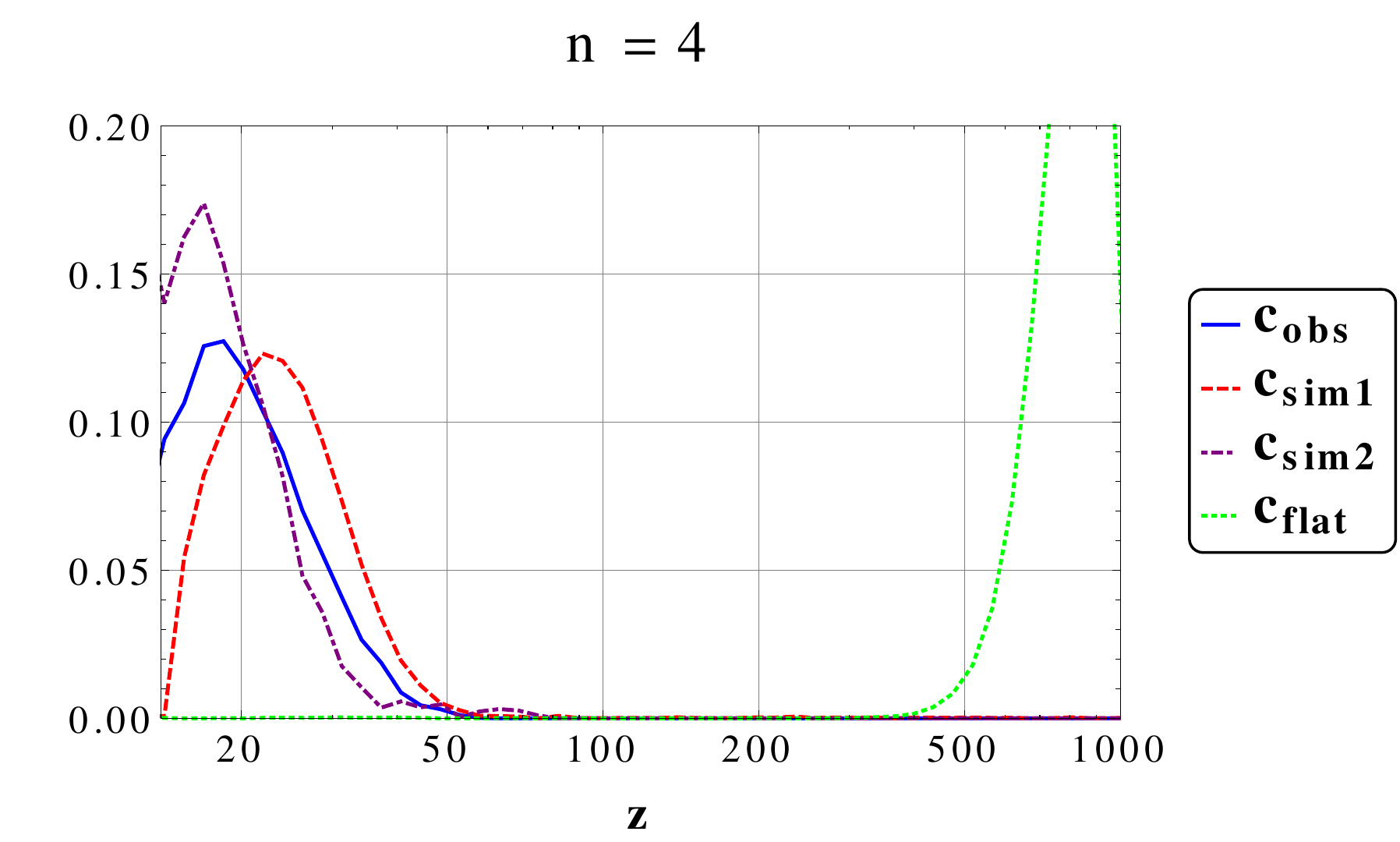}
 \caption{
Weighting function (see text) for decaying DM (upper left), 2-body DM annihilation (upper right), 3-body DM annihilation (lower left) and 4-body DM annihilation (lower right). Where relevant, the cutoff scale is taken to be $10^7h/$Mpc.}
\label{weightfn}
\end{figure}

To indicate the importance of annihilation during different epochs, in Fig.~\ref{weightfn} we plot the ``weighting function'' ${\cal W}(z)$, as derived in \cite{Slatyer2015a}. This function is defined so that  $\int d\ln(1+z) \xi g_\text{ionH}(z) {\cal W}(z)$ governs the size of the CMB signal for arbitrary $g_\text{ionH}(z)$, in the approximation where the analysis is truncated to the first principal component. This weighting function is extracted from the Fisher matrix governing the effect on the CMB of energy injections at different redshifts \cite{Finkbeiner:2011dx}. 

In the cases of 3-body and 4-body annihilation, where $g_\text{ionH}(z)$ generically becomes very large at low redshifts due to the effects of substructure and the small energy injection from the homogeneous dark matter component, the weighting function approximation is generally not adequate for accurate estimates of the constraints, because small errors in ${\cal W}(z)$ at low redshifts are amplified by the large $g_\text{ionH}(z)$ prefactor in the integral. However, it still provides a valid qualitative picture of which redshifts dominate the signal under different assumptions for structure formation.

As expected from  previous work \cite{Galli2009, Finkbeiner:2011dx,Slatyer:2016qyl} the function peaks at redshift $\sim$ 600 for $n=2$, and at lower redshift ($z \sim 100-200$) for DM decay ($n=1$). 
We see that for $n=4$ with a cutoff of $10^7\, h {\text{Mpc}}^{-1} $, for all but the ``$c_\text{flat}$'' concentration parameter evolution, the signal is strongly peaked at low redshifts, and small halos are expected to be very important. For 3-body annihilation, on the other hand, these assumptions on the cutoff and concentration lead to an interesting bimodal structure for the weighting function, where two peaks are potentially present; one at redshift $\sim 30$ driven by structure formation, and one at redshift $\sim 800$ driven by the high smooth DM density at early times. From this figure, we expect the former low-redshift peak to dominate for the $c_\text{obs}$ and $c_\text{sim1}$ prescriptions, for a cutoff scale of $k_c = 10^7 h$/Mpc; for the $c_\text{sim2}$ and $c_\text{flat}$ prescriptions, in contrast, the low-redshift peak becomes negligible and the high-redshift peak dominates. Smaller $k_c$ cutoffs (i.e. more depletion of small halos) enhance the high-redshift peak relative to the low-redshift peak. These results are consistent with our inferences from Fig.~\ref{PCA}.

\begin{figure}[h]
 \includegraphics[width=7.7cm]{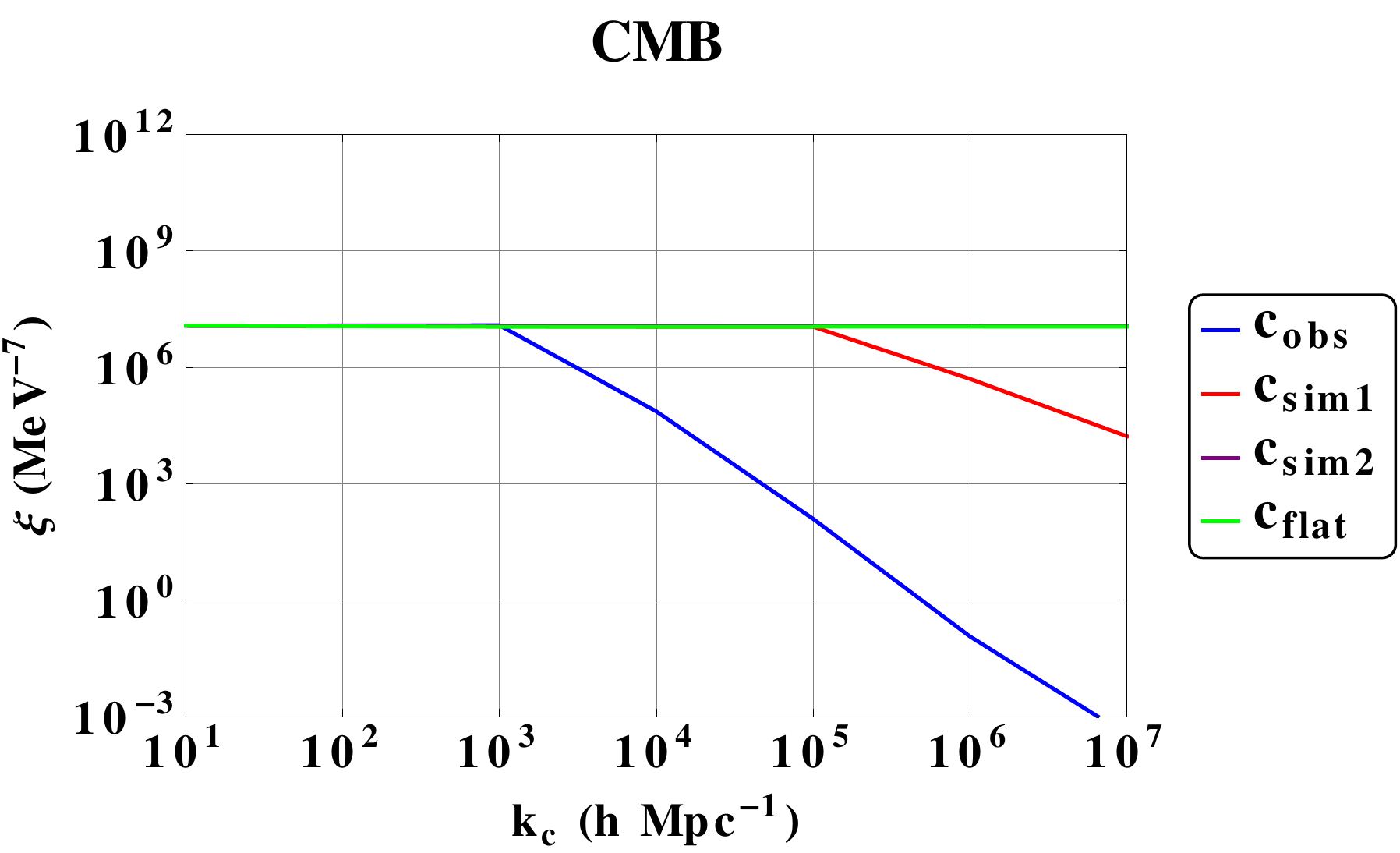}
   \hspace{.4cm}
    \includegraphics[width=7.7cm]{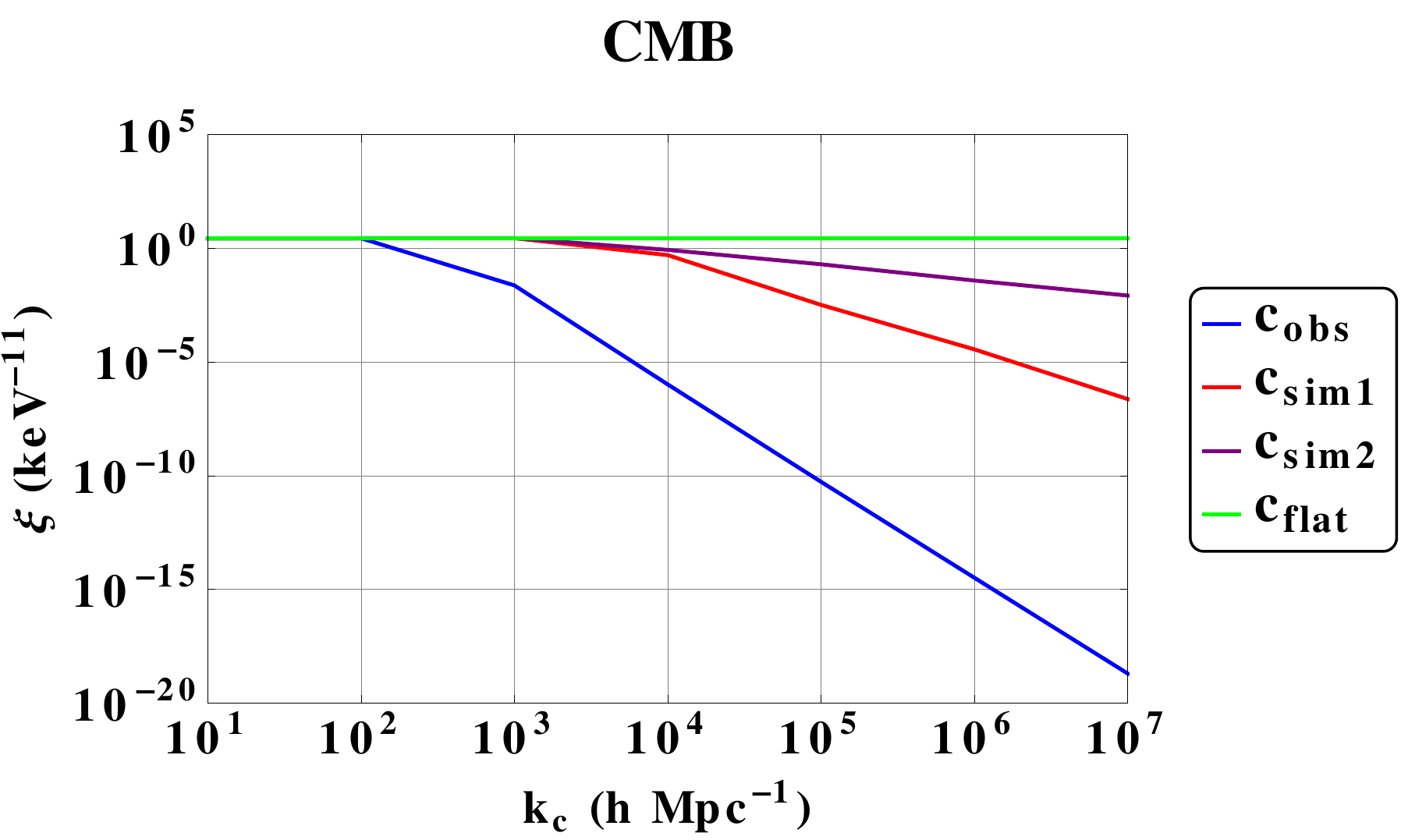}
 \\  \\ 
     \includegraphics[width=7.9cm]{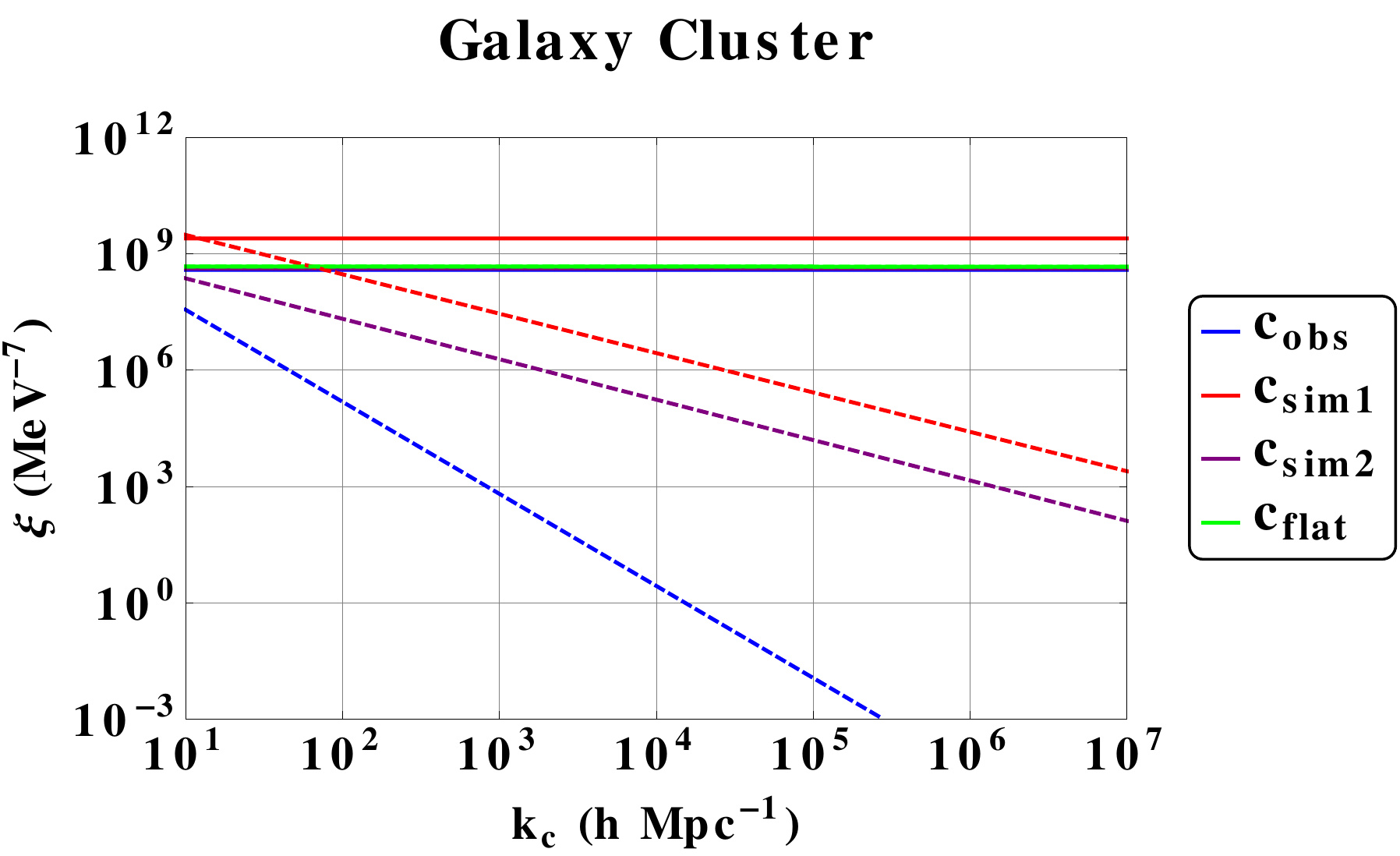}
        \hspace{.4cm}
 \includegraphics[width=7.9cm]{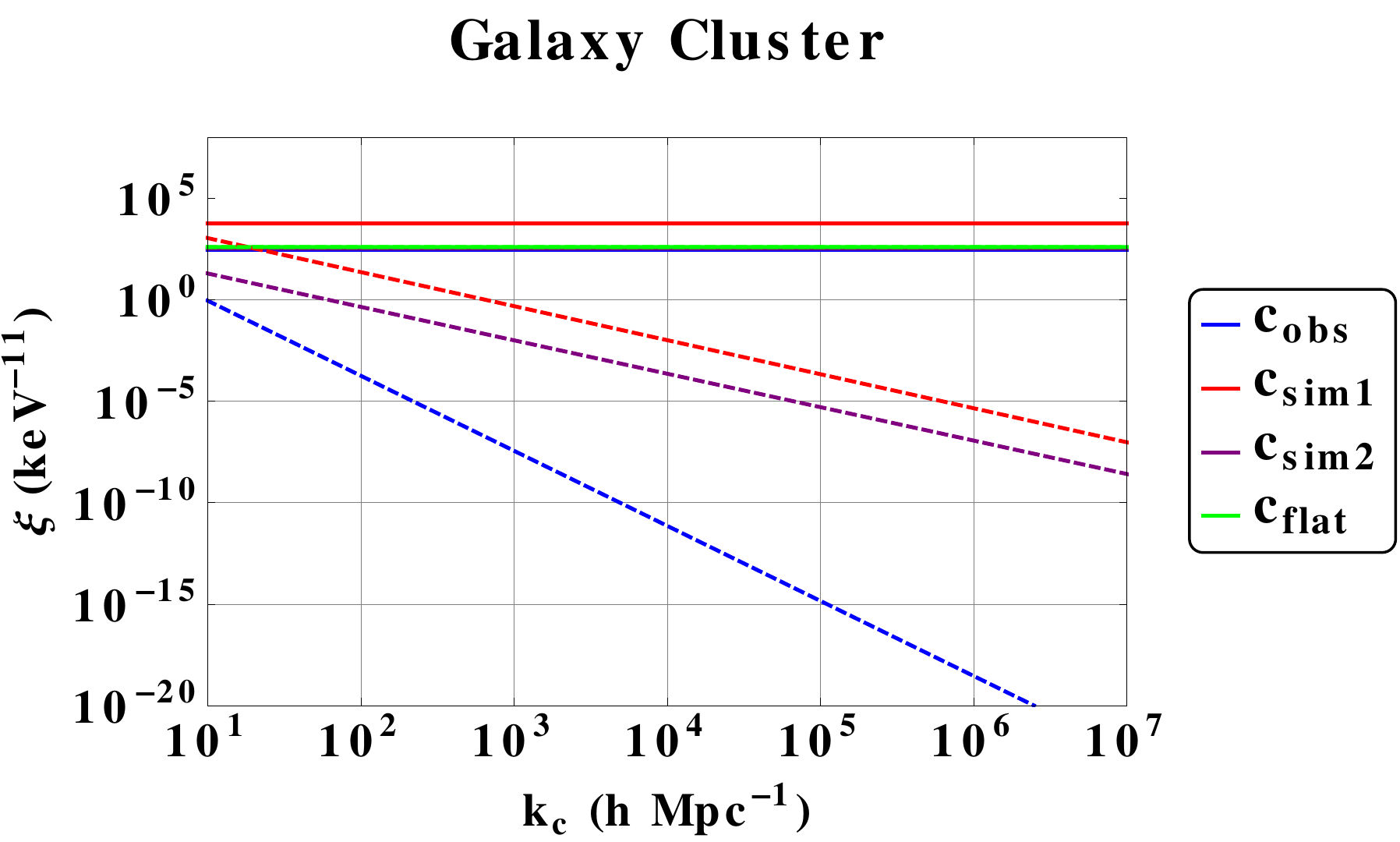}
 \caption{
2-$\sigma$ indirect-detection limits on the annihilation parameter $\xi$ for 3-body ({\bf left}) and 4-body ({\bf right}) annihilation as a function of small-scale cutoff. {\bf Upper panel:} CMB limits for different concentration models assuming annihilation to 100 MeV $e^+e^-$ pair. 
{\bf Lower panel:} Estimated limits from the Perseus cluster for 10 keV photons, for different concentration models; dashed (solid) lines show the constraints with (without) including substructure.}
\label{limits}
\end{figure}

Repeating the constraint calculation for different cutoffs and concentration parameters, we obtain the constraints shown in (the upper panel of) Fig.~\ref{limits}. For the purpose of demonstrating the dependence on the structure formation parameters, we assume an injection of $e^+ e^-$ pairs such that each particle has 100 MeV of (kinetic) energy, corresponding to the peak of the first PC; the limits may be translated to any other energy (or to photons) by rescaling the limit by the first PC.

\subsection{Constraints from the present-day universe: general points}

The early universe has high DM density and the CMB provides a sensitive observational probe, but for decay and conventional 2-body annihilation, limits from dwarf galaxies (e.g. \cite{Fermi-LAT:2016uux}), the Milky Way (e.g. \cite{Essig2013, Chang:2018bpt}) and galaxy clusters (e.g. \cite{Lisanti:2017qlb}) are often considerably more stringent. However, these bounds depend on the energy and spectrum of the annihilation products to a greater degree than the CMB limits.

As discussed in Section \ref{sec:parametric}, the mass range of greatest interest for $n > 2$ is 1 GeV and lower, which is below the optimal energy range for the Fermi Gamma-Ray Space Telescope and well below energy thresholds for ground-based gamma-ray telescopes. For keV-scale photons, X-ray telescopes provide sensitive constraints: Chandra \cite{1538-3873-114-791-1}, Suzaku \cite{Mitsuda:2007mq}, and XMM-Newton \cite{Jansen:2001bi} cover the range from 0.1 to 10 keV, and NuSTAR \cite{Harrison:2013md} is sensitive in the 3-80 keV range. INTEGRAL and EGRET data provide sensitivity to somewhat higher-energy photons, in the hard X-ray and soft gamma-ray band.

For DM below the $\sim 100$ keV mass scale, the available SM annihilation channels are rather limited at low DM velocities; only neutrinos and photons are kinematically allowed. In the X-ray range, therefore, we will focus on the limits on photon lines. If the DM annihilates to relativistic intermediate particles which subsequently decay into photons, then the velocity of the intermediate particles will broaden the resulting photon spectra (to ``boxes'' rather than lines), likely making them somewhat less detectable. 10 MeV-GeV electrons can inverse-Compton-scatter on starlight photons (energy $\sim 1$ eV), producing a spectrum of photons peaked in the keV-MeV range, and so X-ray telescopes may also have sensitivity to DM in this higher mass range.

A detailed study of X-ray sensitivity to such spectra is beyond the scope of this paper; the line limits we show should be viewed as an upper limit on the potential sensitivity (with that sensitivity being attained when the spectrum is very sharply peaked). Our principal goal here is to estimate the approximate sensitivity of various searches to multi-particle annihilation, and to compare the sensitivity of various local observations and the CMB bounds of the previous section.

\subsection{Constraints from X-ray observations of the Galactic Center}

Let us first consider the possible signal from the center of the Milky Way, where the DM density is expected to be large. We expect that substructure should be tidally disrupted in the Galactic Center (GC), so we do not include any boost factor from small halos. If we assume that the energy injection from the $n$-body annihilation is dominated by photons of energy $\sim E_\gamma$, then if the local rate of energy injection into SM particles per volume per time is given by $\xi \rho_{\chi,\text{loc}}^n$, the number of photons produced per volume per time is given by $\xi \rho_{\chi,\text{loc}}^n / E_\gamma$. The integrated photon number flux from a given field of view, adjusted for the detector efficiency, is then given by:
\begin{equation}
F_\gamma^{GC}  = \frac{1}{E_\gamma} \xi \Delta \Omega \mathcal J, 
\label{eq:nustar}
\end{equation}
where $\Delta \Omega = \int_{\text{FOV}} d\Omega \mathcal E$ accounts for the size of the field of view and the energy-independent detector efficiency $\mathcal E$, and $\mathcal J$ is the averaged J-factor given by,
\begin{equation}
\mathcal J = \dfrac{1}{ \Delta { \Omega}}  \int_{\text{FOV}} d\Omega \mathcal E  \int_{\text{los}} d  \ell \, \rho_{_{MW}}(r)^n,
\label{eq:jfac} \end{equation}
where $n$ labels the number of initial-state particles, $\rho_{_{MW}}(r)$ is the DM density profile as a function of the galactocentric radius $r$, and $\ell$ is the line of sight distance from the observer, which is related to $r$ and the angle from the GC $\psi$ by $r=\left(R_\odot^2+\ell^2 - 2 R_\odot \ell \cos\psi \right)^{1/2}$.

We test two Einasto density profiles (see Eq.~\eqref{Einasto}) with shape parameters $\alpha_e^{^{MW}} = 0.3$ and $0.17$ (following Ref.~\cite{Perez:2016tcq}) and $r_{-2}^{^{MW}} = 20$ kpc, as well as a NFW profile with a large 1.5 kpc core:
\begin{equation}
\rho_{_{MW}}(r)  = \dfrac{\rho_s}{ \left(\dfrac{r}{r_s}\right)^{\gamma_s} \left(1+\dfrac{r}{r_s}\right)^{3-\gamma_s}},
\end{equation}
where $r_s = $ 20 kpc, $\gamma_s = 0.7$, and $\rho (\text{r} < \text{1.5 kpc}) = \rho (\text{1.5 kpc})$. In both cases the density is normalized to $\rho (R_\odot) = $ 0.4 GeV, where $R_\odot = 8$ kpc. We will refer to the Einasto profile with $\alpha_e^{^{MW}} = 0.3$ as ``Einasto'' and $\alpha_e^{^{MW}} = 0.17$ as ``Shallow Einasto''.

NuSTAR has performed pointed observations of a $1.4^\circ \times 0.6 ^\circ$ region around the Galactic Center. In Ref.~\cite{Perez:2016tcq}, the total $\Delta \Omega$ (including the efficiency factor) is approximately 3.8 $\text{deg}^2$; this result is obtained from a weighted sum of 12 observations. We use the efficiency map due to vignetting effects for two observations, as shown in Ref.~\cite{Perez:2016tcq}, to compute the efficiency-corrected J-factor. For 3-body annihilation, for the cored-NFW,  Shallow Einasto, and Einasto profiles, we obtain respectively $\mathcal J = 206.3$, 876.5, and $2.22\times10^4$ $\text{GeV}^3  \,\text{cm}^{-9} \,\text{kpc}\,\text{sr}^{-1}$, whereas for 4-body annihilation the corresponding values are 684.0,  $7.05\times 10^3$, and $7.83\times10^5$ $\text{GeV}^4\,\text{cm}^{-12}\,\text{kpc}\,\text{sr}^{-1}$.  

The model-independent constraint on the decay rate shown in \cite{Perez:2016tcq} fluctuates between $ \Gamma \lesssim 3\times10^{-27} \text{s}^{-1}$ and $2\times10^{-29} \text{s}^{-1}$ for the 3 - 79 keV energy range, with no strong trend with energy. For 10 keV photons, the limit is approximately $10^{-28} \text{s}^{-1}$, representative of this region; we convert this bound to a limiting photon flux using Eq.~\eqref{eq:nustar}, yielding $F_\gamma^{GC} \approx 8.3\times10^{-5} \, \text{cm}^{-2} \,\text{s}^{-1}$ for a decay J-factor of $\mathcal{J} = 29 \, \text{GeV} \,\text{cm}^{-3} \,\text{kpc} \,\text{sr}^{-1}$. Using the 3- and 4-body J-factors given above, we convert this flux limit into a limit on $\xi$, and display the results in Fig.~\ref{limitsmass}. 

\subsection{Constraints from the Galactic halo in soft gamma rays}
\label{sec:GH}
For 100 keV-GeV DM, observations of the diffuse gamma-ray emission from our Galactic halo by hard X-ray and soft gamma-ray observatories can provide stringent constraints \cite{Essig2013}. Ref.~\cite{Essig2013} shows conservative limits on annihilation and decay by requiring that photon signals not overproduce the total observed diffuse photons by more than $2\sigma$ in any energy bin. Since no spatial information is employed, it is again straightforward to convert the limits to the 3-body and 4-body cases by simply rescaling the bounds by the $\cal J$-factor within the relevant region of interest (ROI), which controls the observed photon flux. (Note that here we include only the signal from the smooth Galactic halo, and assume that the limits given in Ref.~\cite{Essig2013} are likewise dominated by the Galactic emission.) In the following observations, $\cal J$-factors are quoted in units of $\text{GeV}^n\,\text{cm}^{-3 n}\,\text{kpc}\,\text{sr}^{-1}$ for $n$-body processes (we assume ${\cal E}$ is constant over the relevant regions of interest throughout, and so can be set to 1 in Eq.~\eqref{eq:jfac}; a constant but non-unity efficiency can be absorbed into the total exposure / observation time, which cancels out in our comparison between constraints on processes with different $n$).

The observations employed are:
\begin{itemize}
\item The HEAO-1 \cite{Gruber:1999yr,Marshall:1980} A2 detector sets stringent limits for photon energies between 3 - 50 keV, with a ROI covering two longitude ranges between $238^{\circ}$ and $289^{\circ}$ and $58^{\circ}$ and $109^{\circ}$, with a latitude range $20^{\circ} < | b | < 90^{\circ}$. In the cored-NFW, shallow Einasto and Einasto profiles respectively, the $\cal J$-factors for decay are 5.38, 5.36, and 5.20; for 3-body annihilation 0.38, 0.38, and 0.36; for 4-body annihilation 0.13, 0.13, and 0.12.
\item INTEGRAL \cite{Bouchet:2008rp} observations of the ROI with $| \ell | < 30^{\circ}$ and  $| b | < 15^{\circ}$, in the energy range 20 keV to 8 MeV. Beyond 2 MeV, the signal to noise ratio is relatively large, so data up to 2 MeV energy was used to constrain DM. In the cored-NFW, shallow Einasto and Einasto profiles respectively, the $\cal J$-factors for decay are 20.5, 19.6, 24.5; for 3-body annihilation they are  62.2, 67.8, and 879.9; for 4-body annihilation they are 167.5,  297.7, and $5.32\times10^4$.
\item COMPTEL \cite{Kappadath:1996} observed 0.8 - 30 MeV photons by averaging over $| \ell | < 60^{\circ}$ and  $| b | < 20^{\circ}$.  Data from 1-15 MeV was used to set the constraints. In the cored-NFW, shallow Einasto and Einasto profiles respectively, the $\cal J$-factors for decay are 14.87, 14.35, and 16.57; for 3-body annihilation 27.2, 28.74, and 338.1; for 4-body annihilation 67.63,115.81, and  $2.01\times10^4$.
\item EGRET \cite{Strong:2004de} data consists of a set of regions sampling the whole sky. The intermediate latitude region, labeled E in the analysis of Ref.~\cite{Strong:2004de}, observes the energy range between 20 MeV and 10 GeV, in a ROI with $0^{\circ} < \ell < 360^{\circ}$ and  $20^{\circ} < | b | < 60^{\circ}$. In the cored-NFW, shallow Einasto and Einasto profiles respectively, the $\cal J$-factors for decay are 6.38, 6.32, and 6.30; for 3-body annihilation 1.17, 1.08, and 1.38; for 4-body annihilation 0.85, 0.72, and 1.19.
\end{itemize}

\subsection{Constraints from galaxy clusters}

It is also possible we can detect a signal from galaxy clusters. However, in galaxy clusters the signal is strongly dependent on the degree of small-scale substructure. Accordingly, we will show results for a range of concentration parameter estimates and cutoff scales.
 
 The integrated photon flux from $n$-body annihilation in a cluster, if the signal is dominated by photons of energy $E_\gamma$, can be estimated as: 
\begin{eqnarray}
F_\gamma^c  &=& \frac{\xi}{E_\gamma} \tilde B_h(M) \dfrac{R_{200}^3}{ 3 d^2} \left(\Delta_h \rho_c(z) \right)^n
\end{eqnarray}  
where $\tilde B_h(M)$ is the boost factor given by Eq.~\eqref{B_tilde}, $d$ is our distance from the cluster, and $M$ is the mass of the galaxy cluster. For comparison, we also study the signal expected from a cluster with no substructure by replacing $\tilde B_h$ above with $B_h$, given by Eq.~\eqref{B_halo}.
XMM-Newton and Chandra have observed a number of galaxy clusters and performed searches for photon lines in the few-keV range. 
 
 Here we use Perseus as an example target. As shown in \cite{Tamura:2014mta}, the flux limit observed by Suzaku is $\sim 10^{-1} \,\text{photons}  \,\text{cm}^{-2}  \,\ \text{s}^{-1}  \,\text{sr}^{-1}$, for photon energies in the range from 1 keV to 10 keV, with field of view $\sim 320 \, \text{arcmin}^2$ centered on the cluster center. Similar flux limits, corresponding to $\sim 10^{-5}$ photons $\text{cm}^{-2}$ $\text{s}^{-1}$, have been set for a 3.5 keV line in several clusters and galaxies. For example, there have been claims of a 3.5 keV line flux in the core of Perseus at a flux of $5.2\times 10^{-5}$ photons $\text{cm}^{-2}$ $\text{s}^{-1}$ \cite{Bulbul:2014sua}; the Chandra ACIS-S and ACIS-I observations yielded fluxes of $1.02\times 10^{-5}$ photons $\text{cm}^{-2}$ $\text{s}^{-1}$ \cite{Boyarsky:2014jta}; for M31 there is a claim of a signal with a flux of $0.49\times 10^{-5}$ photons $\text{cm}^{-2}$ $\text{s}^{-1}$ \cite{Boyarsky:2014jta}; for Virgo there is an upper limit $0.91\times 10^{-5}$ photons $\text{cm}^{-2}$ $\text{s}^{-1}$ \cite{Bulbul:2014sua}. We will not in this work seek to explain the claims of a 3.5 keV line, but they demonstrate that a line at this flux level is potentially detectable.
 
 We use $E_\gamma$ =10 keV as an example to estimate the constraints on $\xi$. The cluster parameters for Perseus are taken to be $R_{200} = 1.90 \, \text{Mpc}, M_{200} = 7.71 \, M_\odot$, $z = 0.0183$ and $d=77.7$Mpc (luminosity distance), following \cite{SanchezConde:2011ap} (note that Ref. \cite{SanchezConde:2011ap} assumes a NFW profile for Perseus, but with respect to the total mass and virial radius, we expect the difference between the Einasto and NFW profiles to be small) with $\Delta_h = 200$ as before. The resulting bound is shown in Fig.~\ref{limits} (lower panel) for a range of different choices for the substructure models. We see that in the absence of substructure, the cluster bounds are always weaker than the CMB limits, but they rapidly outpace the CMB bounds as the amount of substructure increases.

\subsection{Discussion of diffuse constraints}

\begin{figure}[h]
\centering
 \includegraphics[width=11.5cm]{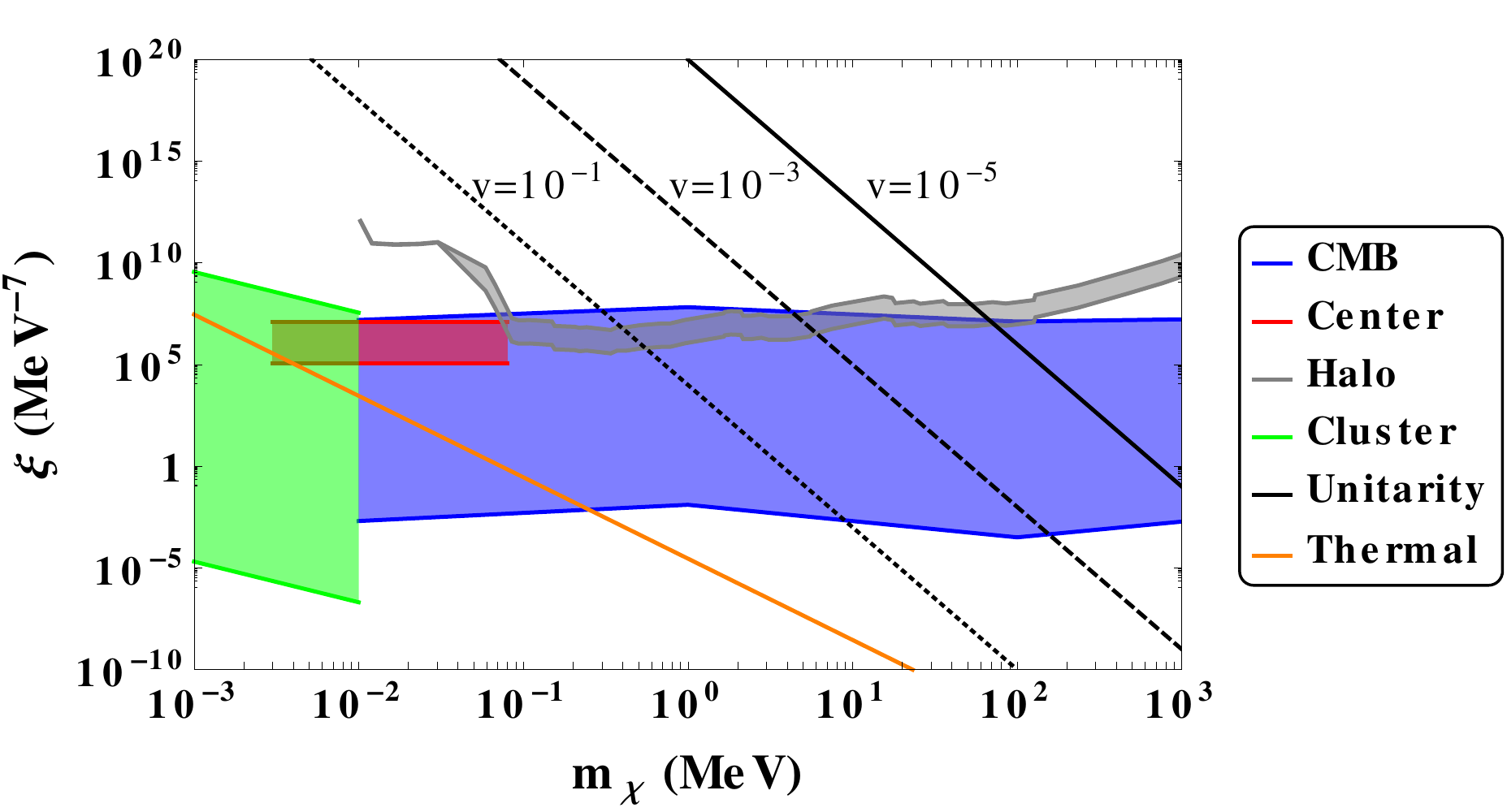} \\
 \vspace{.4cm}
 \includegraphics[width=11.5cm]{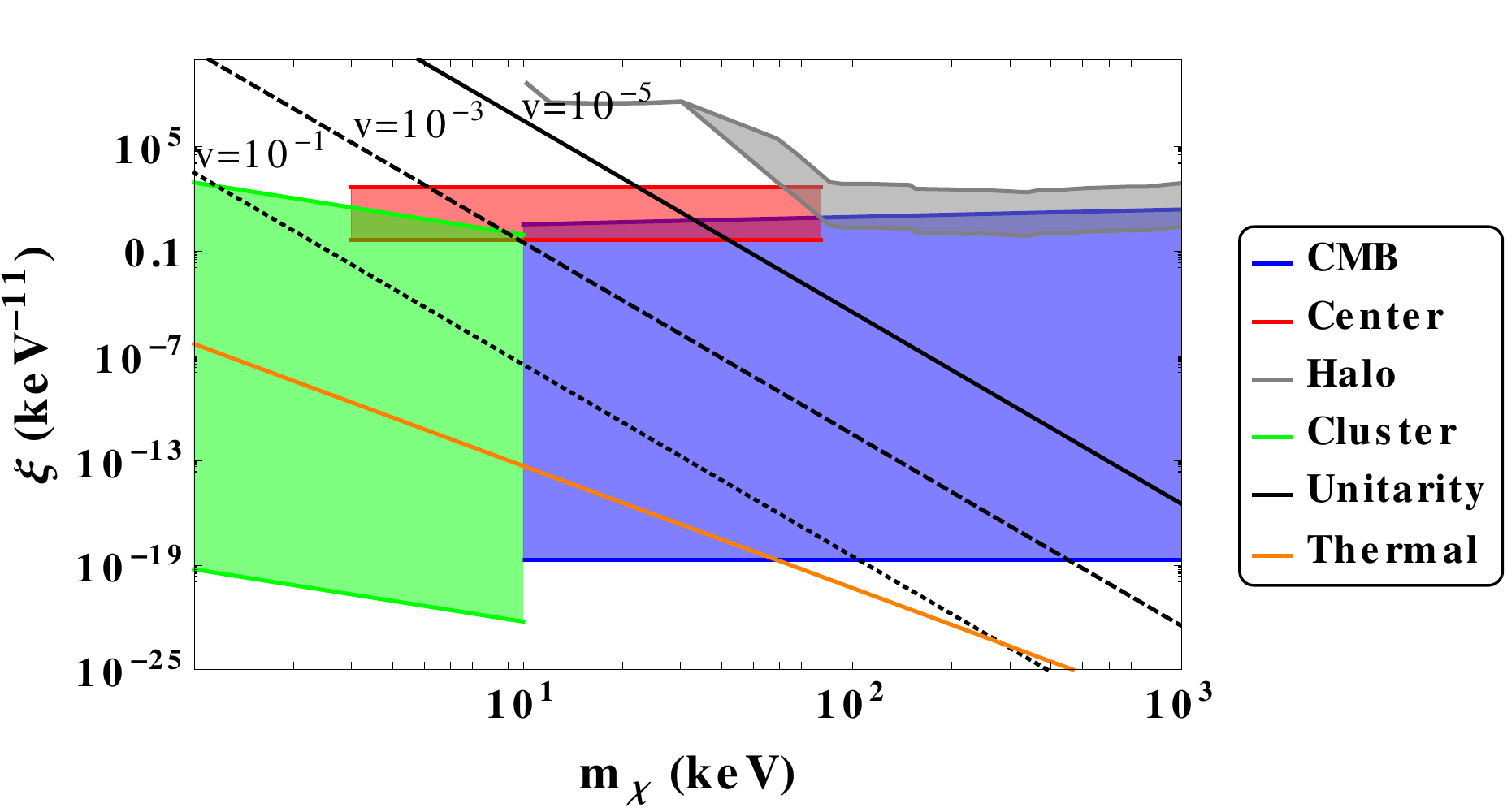}
 \caption{Limits on $\xi$ for 3-body (upper panel) and 4-body (lower panel) annihilation as a function of $m_\chi$, comparing with limits from unitarity and estimates for a thermal relic scenario. The {\bf blue} band shows the range of CMB constraints (estimated from principal component analysis) for different small-scale structure models, with the edges being determined by two extreme cases: $c_{flat}$ concentration with $k_c=10 \,h\text{Mpc}^{-1}$, and $c_{obs}$ concentration model with $k_c=10^7  \,h\text{Mpc}^{-1}$. Below $m_\chi =$ 1 MeV we use the efficiency for injection of photons, assuming the photon energy traces the DM mass; for masses above 1 MeV we use the efficiency for injection of electron-positron pairs instead, again assuming the electron/positron energy is given by the DM mass. The {\bf{green}} band shows the estimated constraint from X-ray observations of Perseus, for the same range of small-scale structure models. The {\bf{red}} bands show the estimated range of constraints from NuSTAR observations of the Milky Way, spanning the cored-NFW, shallow Einasto and Einasto profiles. The  {\bf{gray}} bands show the constraints from diffuse X-ray and gamma-ray observations of the Galactic halo, summarized in Sec.~\ref{sec:GH}, for different DM profiles (Einasto, shallow-Einasto and cored-NFW). The {\bf  black} lines show the approximate unitarity limit from Eq.~\eqref{eq:m_unitarity} for different typical DM velocities $v$. The {\bf orange} line is an estimate of $\xi$ for a thermal relic as a function of mass, taken from Eq.~\eqref{eq:thermal}  setting $g_* = 10 $ and $x_f = 50$, and ignoring combinatoric factors.}
\label{limitsmass}
\end{figure}

The indirect-detection limits on 3-body and 4-body DM annihilation from the CMB, the Galactic center, the Galactic halo and galaxy clusters are summarized in Fig.~\ref{limitsmass}. The late-time constraints from the CMB and galaxy clusters are highly uncertain, due to our insufficient knowledge about the non-linear structures and substructures of the universe; the major uncertainties can be quantified by the small-scale cutoff and the halo concentration, as shown in Fig.~\ref{limits}. These uncertainties are translated to the width of the bands in Fig.~\ref{limitsmass}. For limits from the Galactic center and halo, where no substructure is assumed, the width of the bands reflects the different density profiles considered. Taking into account the effect of substructures would result in stronger constraints (i.e. pushing down the Galactic halo / center bands in Fig.~\ref{limitsmass}); however, this is a complex problem, with numerical errors in simulations potentially modifying the inferred subhalo distribution in non-trivial ways (e.g. \cite{Kravtsov:2003sg,Giocoli:2009ie,vandenBosch:2018tyt}). Such a study is beyond the scope of this paper.

The relation between the energy of the annihilation products and the DM mass is model-dependent in general -- for example, if the annihilation produces an unstable mediator which subsequently decays, the energy of the decay products depends on the mass of the mediator, as well as the number of particles in the final state of its dominant decays -- but for the purposes of this figure, we assume the annihilation products are injected with an energy equal to the DM mass. Where photon spectral information is required, we assume either that the photon energy traces the DM mass, or that (for DM masses above 1 MeV) the photon spectrum is that produced by final-state radiation from electrons and positrons; we choose whichever spectrum yields the stronger limit. Likewise, for the CMB constraints, we choose whichever channel ($e^+ e^-$ or photons) yields the stronger constraint, assuming that the energy of injected particles traces the DM mass. These constraints are in general not highly sensitive to the exact energy of the injected particles, so we do not expect these approximations to strongly affect the results (although models that produce X-ray photons with a similar total energy but a much broader spectrum are likely less constrained than models that yield photon lines).

In the case of Galactic center observations by NuSTAR, the constraints fluctuate rapidly with energy, as discussed above; we do not show these rapid fluctuations, but instead take the lower limit on the decay lifetime to be $\sim 10^{28}$ s within the NuSTAR energy range, which corresponds to a fixed energy-independent limit on $\xi$. Including the fluctuations in the bound would modify the limits by up to one order of magnitude in either direction in very narrow energy ranges, which does not qualitatively change the results.

Fig.~\ref{limitsmass} also shows how the limits compare to the unitarity bound for different velocities, and to the estimated size of a thermal relic signal. We see that in general, constraining thermal scenarios requires a model with more than the minimal amount of small-scale structure. However, in the presence of velocity enhancements, even the most pessimistic (and hence robust) constraints can potentially probe the parameter space for $n$-body annihilation. For example, the CMB signal from high redshifts can be predicted very reliably, and generically dominates the overall CMB signal in models with little small-scale structure (see Fig.~\ref{weightfn}). Future improvements in simulations of small-scale structure may help narrow down the uncertainties and allow robust constraints to be placed on scenarios that are not near the unitarity limit.

These limits in turn let us estimate the DM mass range that could potentially be probed by indirect searches. As discussed in Sec.~\ref{sec:sommerfeld}, theoretical consistency requires the annihilation cross-section to satisfy the unitarity bound, Eq.~\eqref{unitarity_bound} (unless there is a long-range interaction present and many partial waves contribute to the annihilation). As long as Eq.~\eqref{unitarity_bound} is satisfied, neglecting numerical prefactors, we roughly have $\xi \lesssim T^{(5-3n)/2} \, m_\chi^{5(1-n)/2}$. In the non-relativistic limit we may estimate the temperature as $T \sim m_\chi \langle v^2 \rangle$ -- where $v$ is the velocity of the particles and the average is over its distribution --  implying $\xi \lesssim \langle v^2 \rangle^{(5-3n)/2} \, m_\chi^{5-4n}$. If an observational constraint $\xi_{\text{obs}}$ is well below the unitarity bound, it can probe the theoretically allowed region of parameter space. This happens if
\ba 
m_\chi \lesssim \langle v^2 \rangle^{\frac{5-3n}{2(4n-5)}}  \, \xi_{\text{obs}}^{1/(5-4n)}.
\label{eq:m_unitarity}
\ea 
Note that for $n>2$ this upper bound on mass is only weakly sensitive to the actual observational constraint $\xi_{\text{obs}}$. For example, for $\sqrt{\langle v^2 \rangle} \sim 10^{-3}$ (corresponding to the typical virialized velocity of particles within halos), and $n=3$, an observational constraint in the range $ \xi_{\text{obs}} \sim (10^{-10} - 10^{10}) \text{MeV}^{-7}$, can constrain models with $m_\chi \lesssim 1-1000 \text{MeV}$ (with tighter limits having the potential to test higher masses), whereas for $n=4$ and observational constraints in the range $ \xi_{\text{obs}} \sim (10^{-25} - 10^{5}) \text{keV}^{-11}$ we could potentially test models with  $m_\chi \lesssim 3-10^3 \text{keV}$.

\section{Individual sources: ultracompact minihalos (UCMHs)}
\label{sec:ucmh} 
As we saw earlier, the $n$-body annihilation signal is more sensitive to small dense halos for higher $n$, and in general it will be more competitive with 2-body signals in regions of high density. Thus we now study the potential observability of $n$-body annihilation in particularly dense regions of DM that could exist in the present-day universe; we focus here on the possibility of dense ultracompact minihalos (UCMHs).

\subsection{Formation and density profile of UCMHs}

A UCMH is a (hypothetical) dense object that was formed at recombination or earlier; after matter-radiation equality, the UCMH develops through the accretion of DM onto the primordial seed. This contrasts with regular hierarchical structure formation where halos form at much later times. Below we briefly review the UCMH formation and the resulting density profile, following the notation and conventions of \cite{Bringmann:2011ut}.

The initial seed for the UCMH formation is the DM contained within the distance scale of a mode at horizon crossing, with initial mass: 
\ba 
M_i \simeq \left[ \dfrac{4\pi}{3}\rho_\chi H^{-3} \right]_{aH=1/R_i}  = \dfrac{1}{2} H_0^2 m_\text{Pl}^2 \Omega_{DM} R_i^3 \simeq 1.4 \times 10^{11} \, \left(\dfrac{\Omega_{DM} h^2}{0.12} \right) \, \left(\dfrac{R_i}{{\text{Mpc}}} \right)^3 \, M_{\odot},\,\,\,\,
\label{eq:rdef}
\ea 
where $R_i$ is the comoving radius of the mode that has just entered the horizon at the UCMH formation redshift, and $H_0$ and $\Omega_{DM}$ are respectively today's Hubble parameter and the DM abundance as a fraction of the critical energy density. 

The mass of the resulting object remains almost unchanged until matter-radiation equality, after which structures start to grow. Since the density perturbation contrast grows linearly with scale factor we have:
\begin{align}
M_{\text{UCMH}}(z<z_{\text{eq}}) &= \dfrac{1+z_{\text{eq}}}{1+z} M_i \nonumber \\
&\simeq  4.7 \times 10^{13} \left(\dfrac{1+z_{\text{eq}}}{3382} \right) \, \left(\dfrac{1+z}{10}\right)^{-1} \, \left(\dfrac{\Omega_\chi h^2}{0.12}\right)\, \left(\dfrac{R_i}{{\text{Mpc}}} \right)^3 M_{\odot}.
\label{eq:rimass}
\end{align}
The accretion of the ambient DM onto the UCMH ceases to be efficient when hierarchical structure formation begins. Consequently, the final mass of the UCMH is roughly $M_{\text{UCMH}}(z\sim 10)$. 

The density profile of UCMHs can be analytically estimated based on a spherical collapse model for accretion of DM layers onto the primordial seed \cite{Fillmore:1984wk,Bertschinger:1985pd}. The self-similar solution yields the profile $\rho(r)= A(z)/r^{9/4}$, where \cite{Bringmann:2011ut}
 \ba 
 A(z)=\dfrac{3\omega_\chi  M_{\mathrm{UCMH}}(z)}{16 \pi R_{\mathrm{UCMH}}(z)^\frac34} \label{eq:az}
 \ea 
 and $\omega_\chi=\Omega_{DM}/\Omega_m$ is the fractional DM contribution to the total matter today. This profile is consistent with simulations performed in \cite{Vogelsberger:2009bn,Ludlow:2010sy}, in contrast with earlier simulations suggesting even steeper profiles \cite{Mack:2006gz}. 
 
 To obtain the normalization factor $A(z)$, we need to know how the mass and radius of the UCMH evolves with redshift. Based on N-body simulations, Refs.~\cite{Ricotti:2007au,Ricotti:2007jk} find the radius of the UCMH to obey:
 \ba 
 \label{eq:rmass}
 R_{\text{UCMH}}(z) = 0.019 \, \text{pc} \, \left( \dfrac{1000}{z+1}  \right) \, \left(\dfrac{M_{\text{UCMH}}(z)}{M_\odot}  \right)^{1/3}  
 \ea 
 As before, since the UCMH stops growing roughly when hierarchical structure formation starts to be effective, the final radius of the UCMH is given roughly by its value at $z\sim 10$.\footnote{As a caution, these results were derived assuming best-fit values from WMAP3 for the cosmological parameters.} Note that inserting the redshift scaling of $R_\text{UCMH}(z)$ and $M_\text{UCMH}(z)$ into Eq.~\eqref{eq:az} leads to a density normalization $A(z)$ that is approximately constant with redshift. 
 
  After the UCMH has formed with this steep density profile, various physical factors (to be discussed below) are expected to reduce the density in the core region; we will model the resulting profile as a simple density cutoff:
 \ba 
 \rho_{_{_{\mathrm{UCMH}}}}(r)=
  \begin{cases}
  \rho_\text{cut} \hspace{2.2cm}\,\text{for $r<r_c$}\\
  \rho_\text{cut} \left(\dfrac{r}{r_c} \right)^{-9/4}  \hspace{.53cm}\text{for $r_c < r <R_{\mathrm{UCMH}}$} \\
  0 \hspace{2.4cm}\,\,\,\,\,\, \text{for $R_{\mathrm{UCMH}}<r$}.\\
  \end{cases}
 \ea 
 Here $r_c$ is the core radius, $\rho_\text{cut} =A(z)/r_c^{9/4}$ is the core density and $R_{\mathrm{UCMH}}$ is the virialized radius of the UCMH. The core mass is $M_\text{cut} = 4\pi \rho_\text{cut} r_c^3/3$.
 
Note however that more recent works \cite{Gosenca:2017ybi,Delos:2017thv} claim that the steep $r^{-9/4}$ density profile should be disrupted during structure formation, or alternatively is impossible to produce from an initial Gaussian random field, with realistic minihalos being better described by shallower NFW or $r^{-1.5}$ density profiles.

We will focus in this section on the optimistic case where the DM density rises steeply with decreasing $r$, with the self-similar $r^{-9/4}$ profile, until saturating at some core radius. Where relevant, we will briefly discuss how the calculation would change for a power-law profile with a different slope. We will also test a specific example of a minihalo population with density scaling as $r^{-1.5}$ at small $r$, using as our model the halo simulated in \cite{Delos:2017thv}:
 \begin{equation}
\rho(r) = \dfrac{1.3\times10^{12}}{\left(r/r_s\right)^{3/2}\left(1+r/r_s\right)^{3/2}}\,\left(M_\odot/\text{kpc}^3\right).
\label{eq:density1-5} \end{equation}
with $r_s = 10^{-4}$ kpc. In this example, the halos are sourced by a peak in the primordial power spectrum at a scale of 7 $\text{kpc}^{-1}$. As with the $r^{-9/4}$ profile, we will also truncate this profile to a flat-density core at an appropriate radius.

In addition to the uncertainties in the density profile, the UCMH abundance is not known a priori, so the results we obtain should not be considered as robust limits on the annihilation rate. Instead, we will provide estimates for the circumstances under which UCMHs could give rise to potentially observable signals, in the presence of 3-body or 4-body annihilation.

\subsection{Core radius estimation }
There are several distinct physical processes that may contribute to the UCMH core formation. We shall discuss the most significant ones below. In this work, we shall ignore any interplay between these effects, and estimate the actual core size as the largest cutoff radius induced by the various processes, for the $r^{-9/4}$ density profile. Where relevant, we will discuss what changes to the calculation or additional ingredients are needed to compute the core sizes for other density profiles.

\subsubsection*{$r_v$ from velocity of DM}
The nonzero velocity of DM particles after decoupling washes out small scale perturbations as well as the inner cusp of the UCMH. To estimate the core radius due to the peculiar velocity of particles, we employ a velocity dispersion estimated from N-body simulations, as a function of comoving radius $R_i$ \cite{Ricotti:2007au,Ricotti:2009bs}:
\ba 
\sigma_v (z) \simeq 1.58 \, \left(\dfrac{R_i}{1 \text{Mpc}}\right)^{0.85}\left(\dfrac{1000}{z+1} \right)^{-1/2} \text{km}\,\text{s}^{-1}.
\label{eq:vdis}
\ea 
For $\rho \propto r^{-9/4}$ the velocity dispersion can then be related to the UCMH mass by:
\ba 
\sigma_v (z) \simeq 0.14 \, \left(\dfrac{1000}{z+1} \right)^{1/2} \, \left(\dfrac{M_{\text{UCMH}}(z)}{M_\odot}\right)^{0.28} \text{m}\,\text{s}^{-1}.
\ea 
From angular momentum conservation, the mean rotational velocity of particles at radius $r$ from this velocity dispersion is:
\ba 
v_{\text{rot}} (r,z)= \dfrac{\sigma_v(z)R_{\text{UCMH}}(z)}{r}.
\ea 
On the other hand, the Keplerian velocity of particles that orbit at the edge of a constant-density core is given by
\ba 
v_{\text{Kep}}(r_c,z) = \sqrt{\dfrac{GM_\text{cut}}{r_c}} = \sqrt{\dfrac{4 \pi G A(z)}{3 r_c^{1/4}}}.
\ea 
The core radius from the peculiar velocity of particles ($r_v$) can then be estimated as the radius where the typical size of the peculiar velocity becomes comparable to the velocity of infalling DM particles, i.e. $v_{\text{rot}}(r_v,z)\simeq v_{\text{Kep}}(r_v,z) $. This results in the estimate:
\ba 
r_v = \left(\sigma_v(z)^{2} R_{\text{UCMH}}(z)^{2} \dfrac{3}{4\pi G A(z)}\right)^{4/7}.
\label{eq:vcore}
\ea 
This radius should be evaluated at the redshift of collapse, as this is when the inner density profile of the UCMH is determined; as discussed in Ref.~\cite{Bringmann:2011ut}, subsequent evolution primarily modifies the outer regions of the minihalo. To obtain a conservative (i.e. maximally large) estimate of the core size, we take the redshift of collapse to be at recombination, $z=1000$.

For  $\rho \propto r^{-\beta}$ profile, the power 4/7 in Eq.~\eqref{eq:vcore} should be replaced by $1/\left(4-\beta\right)$, and the density normalization will need to be adjusted as appropriate. Eq.~\eqref{eq:vdis} depends on the initial conditions of the fluctuations prior to the collapse, and should be largely independent of the eventual halo density profile, but the relationship between the eventual UCMH mass and the comoving scale $R_i$ in Eq.~\eqref{eq:rimass} is dependent on the same formation history that affects the density profile.

For the specific case of the profile described in Eq.~\eqref{eq:density1-5}, from Ref.~\cite{Delos:2017thv} we have $M_{\text{UCMH}} = 8.1 \, M_\odot$ at the simulated redshift of $z = 400$. Integrating the density profile of Eq.~\eqref{eq:density1-5}, the radial limit of integration must be $R_{\text{UCMH}}(z=400) = 1.4\times10^{-4}$ kpc to obtain the correct mass. We take the collapse redshift to be $z=400$ since it is the largest redshift for which we have a mass normalization; this may overestimate the core size, but we will find that even with this conservative approach, $r_v$ is very small in this scenario and other coring effects dominate. For small radii ($r \ll r_s = 10^{-4}$ kpc, the scale radius), the profile can be approximated as a $r^{-3/2}$ profile with normalization:
\ba 
\rho = \dfrac{1.3\times10^{12}}{\left(r/r_s\right)^{3/2}}\,\left(M_\odot/\text{kpc}^3\right);
\label{eq:1-5norm}
\ea 
this approximation is adequate for $r \lesssim 3 \times 10^{-5}$ kpc. By replacing $A(z)$, $R_\text{UCMH}(z)$ and $\beta$ in Eq.~\eqref{eq:vcore} we obtain $r_v = 3.2 \times 10^{-14} \text{Mpc}$. This value is well inside the region where the density scales as $r^{-3/2}$, so this estimate is self-consistent.

\subsubsection*{$r_{\mathrm{ann}}$ from annihilation}
The DM annihilation inside the UCMH would wash out the self-similar profile, by depleting regions of sufficiently high DM density, as discussed in Sec.~\ref{sec:parametric}. Hence another cutoff radius is set by requiring that the core density be low enough that DM annihilations would not deplete a $\mathcal{O}(1)$ fraction of the DM over the age of the universe \cite{Bringmann:2011ut}:
\ba 
r_{\mathrm{ann}}& \simeq & \left(\dfrac{A(z) \sqrt[n-1]{\tau_{_{\text{UCMH}}} \langle \sigma v^{n-1}\rangle  } }{m_\chi} \right)^{4/9},\label{eq:anncore}
\ea 
where $\tau_{\text{UCMH}}$ is the age of the system. For UMCHs observed today, the relevant timescale is the age of the universe, $\tau_{\text{UCMH}}$ = 13.76 Gyr. For a density profile other than $\rho \propto r^{-9/4}$, of the form $\rho \propto r^{-\beta}$, the index of the bracketed term in Eq.~\eqref{eq:anncore} should be replaced by $1/\beta$. Likewise, $A(z)$ should be replaced by the correct normalization of the density profile.

For the profile of Eq.~\eqref{eq:density1-5}, we use the appropriate normalization (Eq.~\eqref{eq:1-5norm}) for the $r^{-3/2}$ profile, and check that the inferred core radius lies within the region where this approximation to the profile is valid.

\subsubsection*{$r_{L}$ from Liouville's theorem}

Liouville's theorem tells us that the DM phase-space distribution function $f_c$ must be less than (or equal to) the peak of the initial distribution $ f^i_{\text{max}}$. We can go beyond the earlier simple analysis for fermionic DM by specifying an initial condition at the redshift of kinetic decoupling, where the temperatures of the DM and the SM begin to diverge. Assuming the distribution is approximately Maxwellian at this epoch, we can write:
\ba 
f^i(q) d^3r \,  d^3q = \dfrac{\rho_\chi(T_{kd})}{m_\chi(2\pi m_\chi T_{kd})^{3/2}} e^{-\dfrac{q^2}{2m_\chi T_{kd}}} d^3r \,  d^3q
\ea 
Here $\rho_\chi$ is the background DM energy density and $T_{kd}$ is the DM temperature at kinetic decoupling.  $f^i(q=0) = \dfrac{\rho_\chi(T_{kd})}{m_\chi(2\pi m_\chi T_{kd})^{3/2}}$ is then the maximum of the initial distribution. 

Assuming the velocity distribution of the DM in the core is non-relativistic and approximately Maxwellian, the phase space density is given by:
\ba 
f(r,q) = \dfrac{\rho_{\text{cut}}}{m_\chi^4\left(2\pi \langle v^2\rangle \right)^{3/2}} e^{-\dfrac{q^2}{2 m_\chi^2 \langle v^2\rangle}},
\ea 
where $\langle v^2\rangle$ is the average velocity-squared of the particles in the core. Liouville's theorem then gives:
\ba
f(r,q=0) = \dfrac{A(z) r_L^{-9/4}}{m_\chi^4\left(2\pi \langle v^2\rangle \right)^{3/2}} \le \dfrac{\rho_\chi(T_{kd})}{m_\chi(2\pi m_\chi T_{kd})^{3/2}}.
\ea

If we assume that the particles in a core with radius $r_L$ are virialized, we have $\langle\mathcal K \rangle=-\langle \mathcal U\rangle/2$ where $\cal K$ and $\cal U$ are the kinetic energy and potential energy of the system respectively. For the kinetic energy we have $\langle \mathcal K \rangle=\frac{1}{2}N_{\text{cut}} m_\chi \langle v^2 \rangle=\frac{1}{2}M_\text{cut} \langle v^2 \rangle$, whereas the  potential energy is given by $\langle \mathcal U\rangle = - \int_0^{r_L} \dfrac{G M(r)}{r} \dfrac{dM(r)}{dr} dr$ (here $M(r)=4\pi \rho_{\text{cut}} r^3/3$ is the mass within a sphere with radius $r$). Taking the integral and using the virial theorem we find:
\ba 
\langle v^2\rangle
= \dfrac{4 \pi }{5 m_\text{Pl}^2} A(z) \, r_L^{-1/4}.
\ea 

We can now use these relations to obtain a lower bound for the core radius:
\ba 
r_L^{15/8} \ge \left(\dfrac{5 T_{kd}}{4 \pi m_\chi} \right)^{3/2} \, \dfrac{A(z)^{-1/2} m_\text{Pl}^3}{\rho_\chi(T_{kd})}.
\ea 
Provided the decoupling occurs when the DM is non-relativistic, $T_{kd} \le m_\chi$, and since $\rho_\chi(T_{kd}) \propto T_{kd}^{3}$, taking $T_{kd} \approx m_\chi$ yields a lower bound on the core size of:
\ba r_L \ge \left(\left(\dfrac{5}{4 \pi} \right)^{3/2} \, \dfrac{A(z)^{-1/2} m_\text{Pl}^3}{\rho_\chi(m_\chi)} \right)^{8/15}. \ea

For fermionic DM, as discussed previously, Pauli exclusion can also put a constraint on the maximum density (the so-called Tremaine-Gunn bound), and hence the core radius; the calculation is identical to that above, except that the maximum of the initial distribution $\rho_\chi(T_{kd})/(m_\chi (2 \pi m_\chi T_{kd})^{3/2})$ is replaced with 2. Thus the corresponding constraint can be obtained by replacing $\rho_\chi(m_\chi)$ by $2 m_\chi (2 \pi m_\chi^2)^{3/2}$, obtaining the limit:
\ba r_{_{TG}} \ge  \left(\left(\dfrac{5}{8 \pi^2} \right)^{3/2} \, \dfrac{A(z)^{-1/2} m_\text{Pl}^3}{2 m_\chi^4} \right)^{8/15}. \ea

We find that the core radius $r_{_{TG}}$ obtained by setting $f^i_\text{max} = 2$ is much smaller than $r_L$ even for a DM mass of $1$ keV. For  $T_{kd}$ smaller than $m_\chi$ this ratio is even smaller. This is to be expected since the DM cosmological energy density, once the DM becomes non-relativistic, should be much smaller than $m_\chi^4$. So we can safely ignore this constraint on the core radius in the thermally coupled scenario.

In both cases, if the density profile immediately outside the core is $\rho \propto r^{-\beta}$ rather than $r^{-9/4}$, then the only replacement in the bounds is that the $8/15$ index is replaced by $1/(3 - \beta/2)$, and the normalization of $A(z)$ must be modified. Again, we approximate the profile of Eq.~\eqref{eq:density1-5} as a $r^{-3/2}$ profile with appropriate normalization, since the core lies within the inner region.

\subsection{Benchmark models}

We consider two benchmarks to test for sensitivity to UCMHs. On one hand, we consider a generic thermal 3-body annihilation scenario with $x_f=50$, $T_d=m_\chi$ and coupling $\alpha =1$, for both annihilation and scattering processes, corresponding to $\xi = 2.5\times10^{-11}\text{MeV}^{-7}$.  In this case, the DM mass required to obtain the correct relic density is $m_\chi \sim 33\, \text{MeV}$, using the estimates of Sec.~\ref{sec:parametric}.

As a second benchmark, we use a parameter point for ``Not Forbidden Dark Matter'' (NFDM) \cite{Cline:2017tka}, for $m_\chi = 0.5$ GeV, $\varepsilon = 10^{-6}, \alpha^{\prime} = 4.3$ and $r_{NFDM} = 1.8$, corresponding to $\xi = 3.4\times10^{-15} \text{MeV}^{-7}$, which yields the correct relic abundance and is not ruled out by current experiments. We will discuss this class of models in more detail in Sec.~\ref{sec:models}. Both models, as expected for thermal relic DM with 3-body annihilations, cannot be tested by the diffuse signal searches discussed previously.

\begin{figure}[h]
 \includegraphics[width=8.2cm]{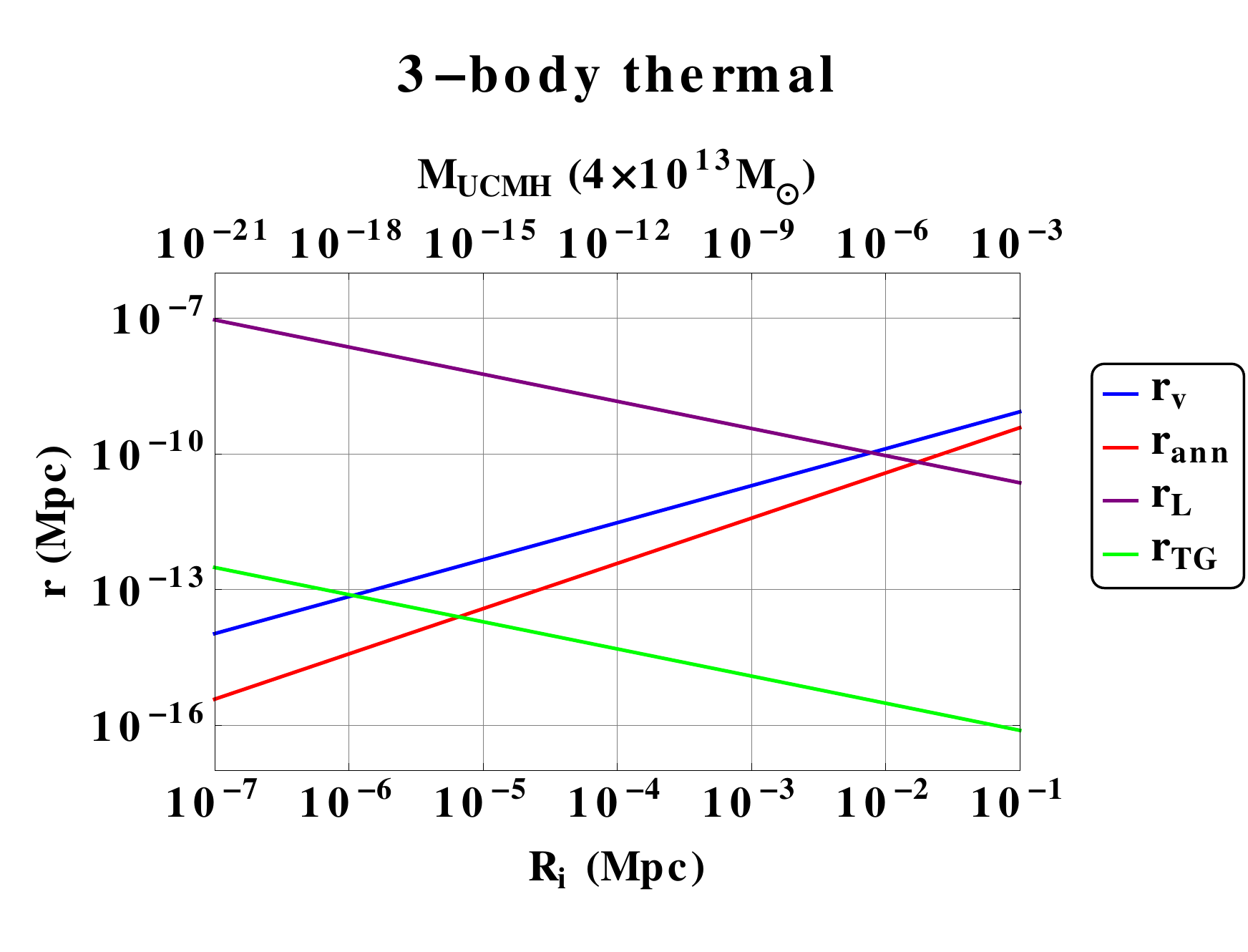}
 \includegraphics[width=8.2cm]{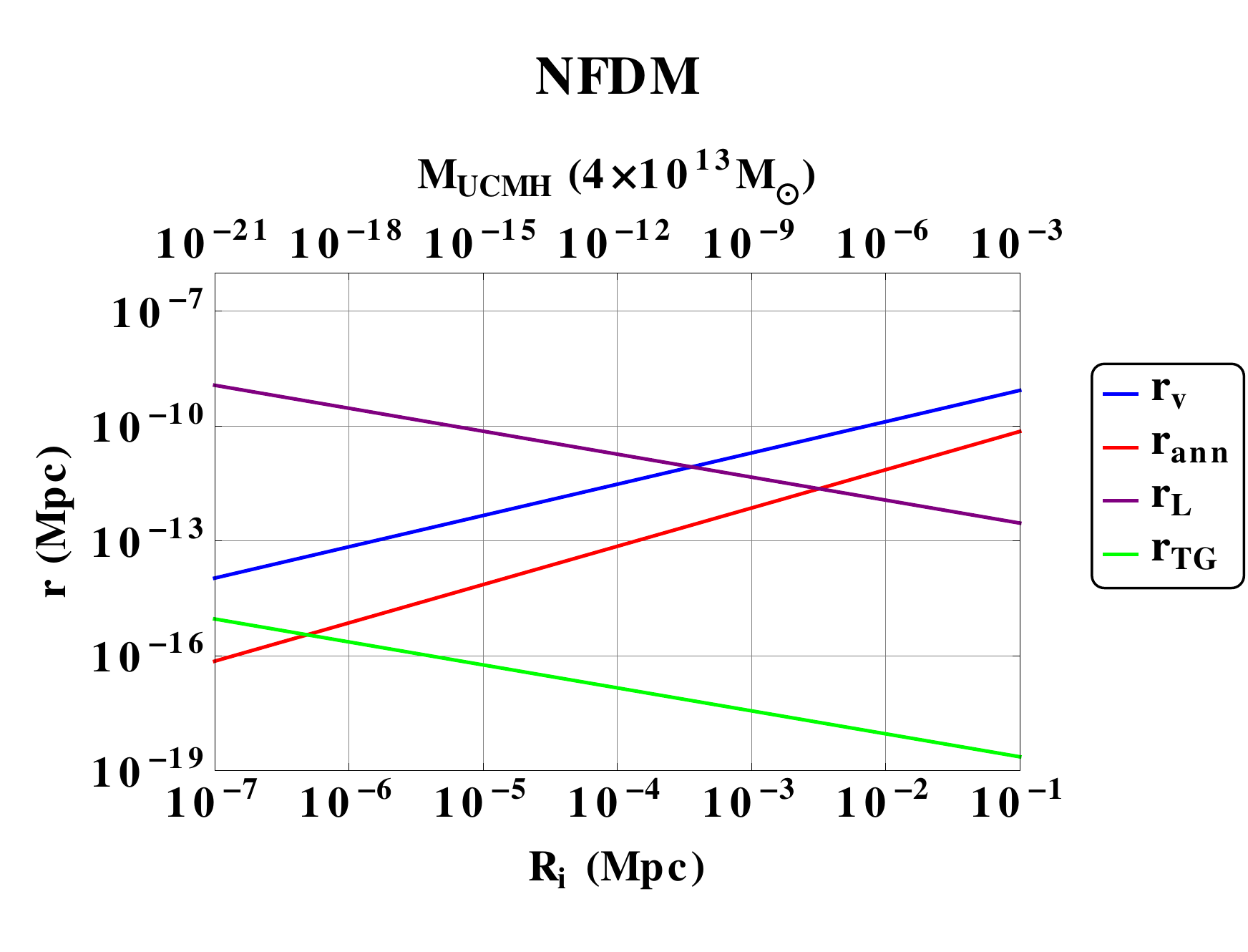}
 \caption{UCMH core sizes induced by various physical processes for a benchmark thermal relic model with 3-body annihilation  (left) and the NFDM benchmark (right) assuming density profiles $\rho \propto r^{-4/9}$ as a function of $R_i$ associated with the UCMH (defined in Eq.~\eqref{eq:rdef}) and its mass.}
\label{rcut}
\end{figure}

For these two model choices, in Fig.~\ref{rcut} we compare the core sizes originating from the mechanisms discussed above. We express the size of the UCMH by the comoving radius of the corresponding mode, $R_i$ and its mass; $R_i$ is related to the UCMH mass by $M_{\text{UCMH}}$ by Eq.~\eqref{eq:rmass}. For the profile in Eq.~\eqref{eq:density1-5} the maximum core size is set by $r_L$, with values of $r_L =  1.8 \times 10^{-9}$ Mpc for the generic thermal benchmark and $ 4.7 \times 10^{-11}$ Mpc for the NFDM benchmark.

\subsection{Point source signals from UCMHs}
The integrated annihilation power in each UCMH for $n$-body annihilation for $\rho \propto r^{-4/9}$ is given by:
\begin{align}\label{S_n} S_n & = 4\pi \xi \int_0^{R_\text{UCMH}} \rho_{_{_{\mathrm{UCMH}}}}^n r^2 dr \nonumber \\
& =  A(z)^n \xi \left[4 \pi \int_{r_c}^{R_\text{UCMH}} r^2 dr  r^{-(9/4) n} + (4/3)\pi r_c^{3 - (9/4) n} \right] \nonumber \\
& = \frac{4 \pi}{3}  A(z)^n \xi r_c^{3 - (9/4) n} \left(\frac{1}{1 - \frac{4}{3n}} \right), \end{align}
where in the last line we have assumed that $R_\text{UCMH} \gg r_c$.

UCMHs could appear as point sources, not associated with any known sources, in observations by telescopes across a range of energies \cite{Bringmann:2011ut}. The optimal telescope will depend (as previously) on the mass range of the DM; \emph{Fermi} may have sensitivity to relatively heavy DM candidates ($\sim 1$ GeV and above), whereas lighter DM could produce soft gamma-ray or X-ray point sources.

If the annihilation signal is dominated by photons with energies $\sim E_\gamma$, the predicted photon number flux from a single UMCH from $n$-body annihilation is:
\begin{equation}
F_\gamma^{\text{UCMH}}(d) = \frac{1}{4\pi d^2} \frac{S_n}{E_\gamma},
\end{equation}
where $d$ is the distance between Earth and the UCMH.
 By matching this flux to the sensitivity of a particular point source search, we can define $d_\text{obs}$ as the maximum distance for a UCMH to be detectable in that search. We assume that the distribution of UCMHs tracks the DM density profile of the Milky Way, which we take to be a NFW profile with parameters chosen to match \cite{Battaglia:2005rj}: $c = 18, M_{200} = 9.4 \times 10^{11} M_{\odot}, r_s = 17.0 \text{kpc}$. These parameters are consistent, within uncertainties, with more recent studies of the Galactic halo, e.g. \cite{Nesti:2013uwa}. We can then determine the expected number of detectable sources, for a UCMH spectrum dominated by a particular mass scale, by integrating the total DM mass within $d_{\text{obs}}$ and within the field of view of the experiment, multiplying by the fraction of DM in UCMHs (which we will denote $\eta_\text{UCMH}$), and dividing the result by the UCMH mass. For experiments such as \emph{Fermi}, with a large field of view where the sensitivity can vary depending on position on the sky, we can divide the sky into pixels within which the point source sensitivity (and hence $d_\text{obs}$) is approximately constant, repeat this analysis for each pixel, and then sum the total number of detectable sources.
 
The predicted distribution for the \emph{actual} number of observed sources from such a population will be a Poisson distribution; we will quote only the expectation values for the number of observed sources in the figures that follow, from which the probability of seeing any specific number of sources can be immediately derived. 
 
We consider the point source sensitivity of:
\begin{itemize}
\item The Chandra point source search \cite{Elvis:2009rr}, which has a field of view of $ 0.9 \, \text{deg}^2$ at $\left(\text{RA}, \text{DEC} \right) \approx \left( 150, 2\right) \text{deg}$, and a sensitivity of $5.7\times10^{-16} \,\text{erg} \,\text{s}^{-1}\, \text{cm}^{-2} $ between 0.5-10 keV for X-ray photons, modeling the photon count flux as a fairly hard power-law with a spectral index of $1.4$ (i.e $dN_\gamma/dE \propto E^{-1.4}$). Variations to the assumed spectrum will change the sensitivity by $\mathcal{O}(1)$ factors; for example, the sensitivity in the 0.5-2 keV band alone is $1.9\times 10^{-16}  \,\text{erg} \,\text{s}^{-1}\, \text{cm}^{-2} $, while in the 2-10 keV band it is $7.3 \times 10^{-16}  \,\text{erg} \,\text{s}^{-1}\, \text{cm}^{-2} $.

Note that for our 3-body benchmarks, the photons produced directly by annihilation will not dominantly lie in the Chandra energy band, suppressing detectability. However, these models would generically produce electrons and positrons; when $\mathcal{O}(10-100)$ MeV electrons and positrons inverse Compton scatter $\sim 1$ eV starlight, the resulting spectrum should peak in the X-ray band (the photons' initial energy being increased by a factor of $\mathcal{O}((E_{e^-}/m_e)^2) \sim 100-10^5$). These particles can also lose energy through (for example) bremsstrahlung and synchrotron; a full analysis of these secondary photon signals would take into account all energy loss mechanisms for the electrons/positrons, and the possibility that they might propagate sufficiently far before losing all their energy that the UCMHs would no longer appear as point sources. These effects will tend to reduce the detectability of a signal from inverse Compton scattering, but their treatment is beyond the scope of this work.

Consequently, we will show default results for two benchmark models if 1 photon in this band is produced per annihilation; note that this corresponds to a rather small fraction ($\sim$ keV/100 MeV $\sim 10^{-5}$) of the DM mass energy being converted into X-ray photons. We will also demonstrate and discuss how the limit changes if the number of photons per annihilation is increased or decreased, using the NFDM benchmark as our baseline.

The Chandra point source search is more directly applicable to lower-mass benchmarks, e.g. in the 4-body case. As we will discuss below, 4-body thermal benchmark scenarios are not detectable by Chandra, even with optimistic assumptions.

\item The survey mode of \emph{Fermi} over a 10-year period, which views the full sky, and has a sensitivity of order a few $\times 10^{-8} \,\text{photons} \,\text{s}^{-1}\, \text{cm}^{-2} $ for point sources producing photons at $\sim 100$ MeV energies, improving to a few $\times 10^{-9} \,\text{photons} \,\text{s}^{-1}\, \text{cm}^{-2}$ for power-law spectra extending up to higher energies. 

We obtain the relevant sensitivity maps for this case using the package \texttt{fermipy} \cite{2017arXiv170709551W}, requiring test statistic $> 25$ and $>10$ photons per energy bin for detected sources, and using source-class data with the \texttt{P8R2\_V6} instrument response functions. Note that for the 3-body thermal benchmark model we consider, the DM mass is $33$ MeV, and the resulting photon spectrum is expected to peak below the energy range to which \emph{Fermi} has sensitivity. For the NFDM model, the annihilation we consider is mediated by a 0.9 GeV dark photon, which is between the $\omega$ and $\phi$ meson resonances. The dominant photon-producing channels are from the production and decay of mesons, and from final state radiation from electrons and muons \cite{Tanabashi:2018oca, Hooper:2012cw}; the full photon spectrum is thus quite complicated to model. As an alternative to computing the full spectrum, we consider a broad spectrum with $dN/dE \propto E^{-2}$ between 30 MeV and 450 MeV. Such a spectrum yields about 10 photons per annihilation, so approximately 1 photon per annihilation if 10$\%$ of the energy goes into photons. From \cite{Hooper:2012cw}, annihilation mediated by a 1 GeV gauge boson produces on average 0.7 photons, and 5$\%$ of the energy goes into photons, similar to the result for our simplified spectrum. Therefore, by default we assume 1 photon per annihilation; we will also show the effect of changing the brightness of a source by setting different numbers of photons per annihilation.
\end{itemize}

\begin{figure}[h]
 \includegraphics[width=8.5cm]{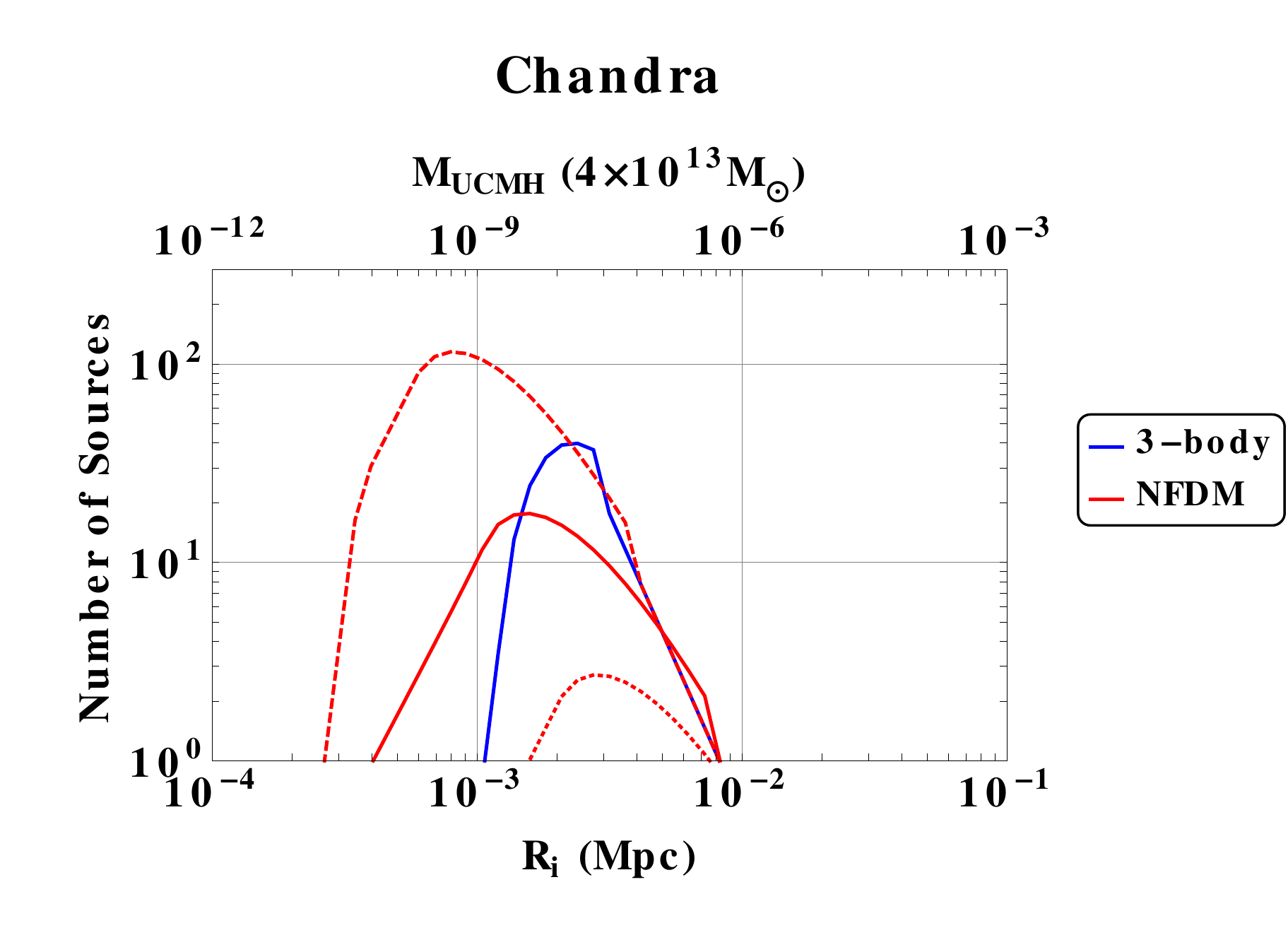}
 \includegraphics[width=8.5cm]{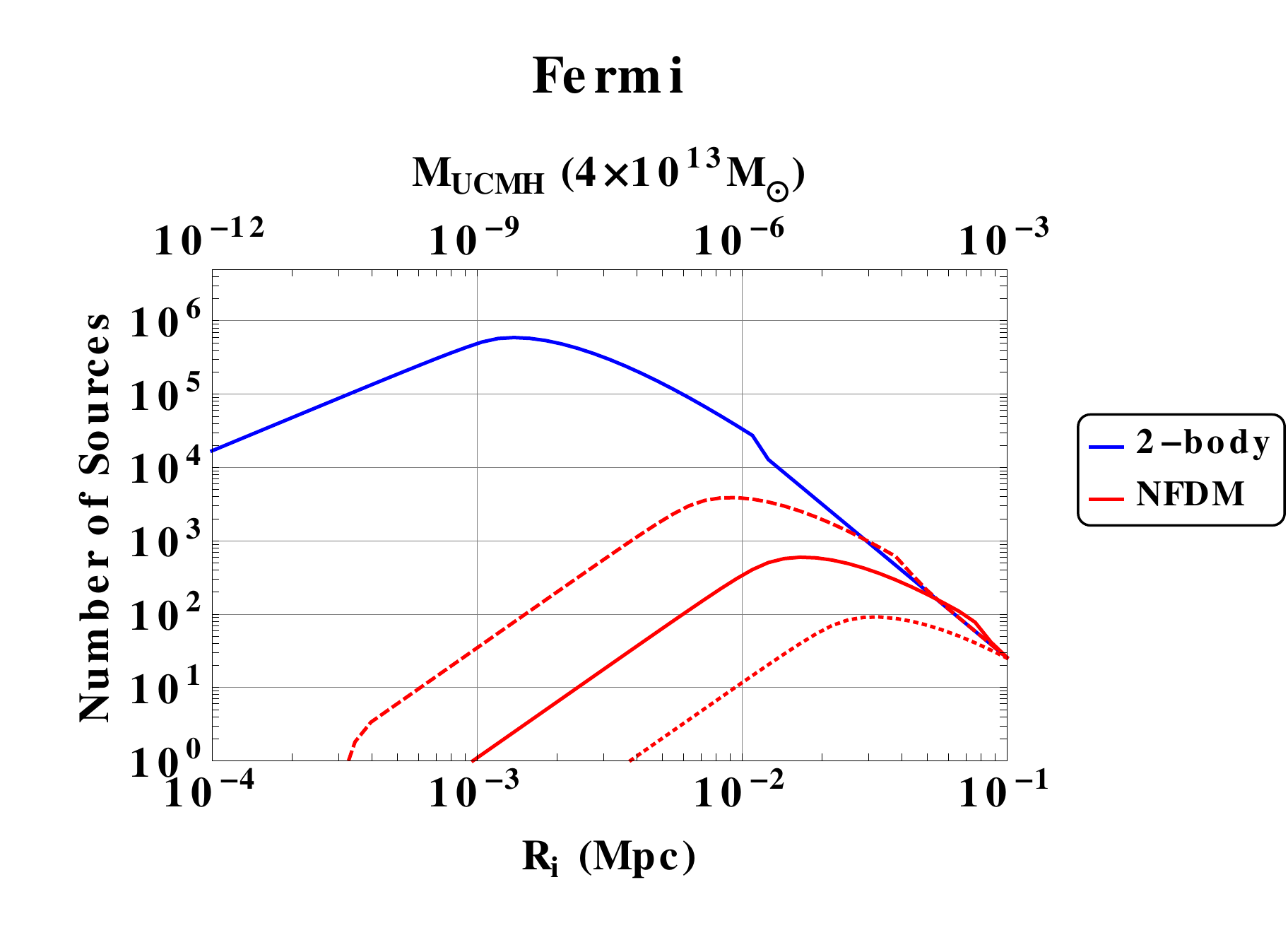}
 \caption{Estimated maximum number of detectable UCMHs in point source searches by Chandra (left) and Fermi (right) for a generic model of 33 MeV thermal relic DM with 3-body annihilation, and a 500 MeV thermal NFDM benchmark point (see text for details). Notice that the expected signal for 3-body thermal case is below the Fermi's sensitivity. For comparison, we also show the corresponding curve for 1 TeV thermal relic DM with 2-body annihilation for Fermi. To illustrate the effect of varying brightness of sources, we set numbers of photon per annihilation for NFDM to be 10 (dashed red line), 1 (solid red line), 0.1 (dotted red line).}
\label{UCMHs}
\end{figure}

Our estimates for the possible number of detectable sources in these two surveys are presented as a function of UCMH mass and $R_i$ in Fig.~\ref{UCMHs} for $\rho \propto r^{-9/4}$. We plot the expected number of observable sources for the maximally visible case where $\eta_\text{UCMH} = 1$; the result for other (more realistic) choices of $\eta_\text{UCMH}$ can be obtained by a trivial rescaling. We have also tested the halo density profile described in Eq.~\eqref{eq:density1-5} for both 3-body benchmarks, and have found that in this case the expected number of observable sources is always below 1 (although such sources could potentially contribute to the Galactic diffuse background). From Fig.~\ref{UCMHs}, we can see that for the 3-body case, if fractions of the DM as small as $10^{-3}-10^{-2}$ were comprised of UCMHs with masses corresponding to $R_i \gtrsim 10^{-4}$ Mpc scales $ \left ( \text{explicitly,}\, M_{\text{UCMH}} \gtrsim 10^2 \, M_{\odot} \right)$ and steep density profiles, then simple thermal relic models could give rise to potentially observable point sources in our Galaxy.

In this calculation, we consider only galactic sources because including extragalactic sources would only be relevant if massive UCMHs (about $10^{10} M_\odot$) constitute a large fraction of the DM for the NFDM model we consider. This possibility is excluded by the observed distribution of DM in galaxies and dwarf galaxies. In contrast, for the 2-body thermal model more UCMHs with mass greater than $10^6 M_\odot$ can be observed when including extragalactic sources. 

As a cross-check, we also compute the detectable number of sources from 2-body annihilation with $m_\chi$ = 1 TeV and $\langle \sigma v \rangle = 3\times10^{-26} \text{cm}^3\text{s}^{-1}$, assuming a sensitivity of $4\times10^{-9}$ photons $\text{cm}^{-2}\text{s}^{-1}$ everywhere in the sky, which matches the assumptions of Ref.~\cite{Bringmann:2011ut}. We find that in this case the sensitivity peaks for $R_i \sim 10^{-3}$ Mpc, and $\mathcal{O}(1)$ visible source would be predicted for $\eta_\text{UCMH} = 10^{-6}$, similar to the estimates of  Ref.~\cite{Bringmann:2011ut}. The limit presented in that work is  stronger by a factor of several than one would expect from our results; the discrepancy is likely due to the $b\bar{b}$ annihilation final state assumed in that work, which produces copious photons (which is appropriate for heavy DM), whereas by default our pipeline assumes 1 photon per annihilation.

We also test a benchmark thermal relic model for 4-body annihilation; from the estimates in Sec.~\ref{sec:parametric}, we choose $m_\chi = 38 \, \text{keV}$ and $\xi = 4.2\times10^{-18} \, \text{keV}^{-11}$. In this case we find that the number of detectable sources is below 1, even in the maximally optimistic $\eta_\text{UCMH}=1$ scenario. Examining a lower-mass (10 keV) 4-body benchmark, we find that it is likewise undetectable; the increase in $\xi$ associated with a lower mass scale is counteracted by an increase in the core size (in this case set by $r_L$) and corresponding reduction in the signal.

\subsection{Early ionization and heating from UCMHs}

The possible impact of annihilation within UCMHs on the early universe has been studied by Ref.~\cite{2011MNRAS.418.1850Z}. We extend this analysis using the methodology of Refs.~\cite{Slatyer2015a,Slatyer2015} to determine the modifications to the ionization and thermal history, and the resulting impact on the CMB. Using the experimental limits on the optical depth and IGM temperature discussed in Sec.~\ref{sec:CMB}, we find constraints that are always weaker than the CMB bounds, although again this could change with future measurements (or confirmations of current claims \cite{Bowman:2018yin}) of the redshifted 21cm line of neutral hydrogen. 

We use Eq.~\eqref{S_n} to determine the redshift dependence of energy injection from UCMHs, and then perform a principal component analysis over a set of injection models corresponding to photons and $e^+ e^-$ pairs injected at different energies. For each of the two benchmark scenarios discussed above, we consider the resulting redshift dependence of the annihilation signal. In contrast to the analysis of Sec.~\ref{sec:CMB}, here the energy injection does not scale with the DM density raised to a high power, as the signal comes from small dense objects that are formed prior to the recombination epoch.

\begin{figure}[h]
\centering
  \includegraphics[width=7.cm]{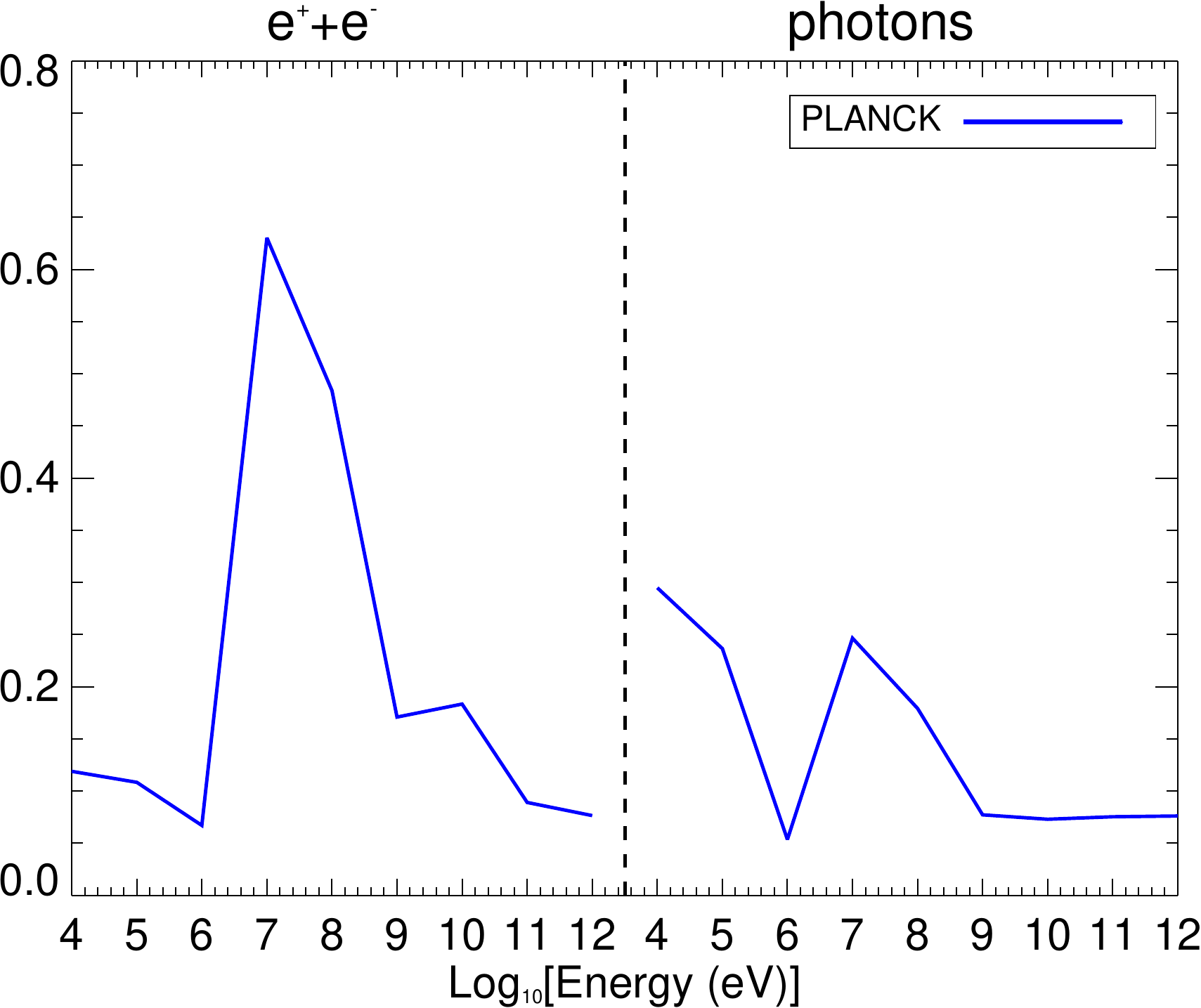} \\
 \caption{First principal component for different injection energies and species, in the case of annihilation in UCMHs. Changing $R_i$ or the benchmark model gives indistinguishable results.}   
\label{UCMHcmbpca}
\end{figure}

We have checked that the first principal component contributes over 90\% of the variance shown in Fig.~\ref{UCMHcmbpca}, so it is a reasonable approximation to show results for one injection energy and extrapolate to the others using the first principal component.  For large $R_i$ the core radius is determined by $r_v$, while for small $R_i$ it is determined by $r_L$,  in the two cases considered here. Both core radii are independent of redshift (in the case of $r_L$, there is an apparent redshift dependence due to $A(z)$, but as mentioned previously in fact $A(z)$ is approximately constant with redshift). Since the signal mainly comes from the UCMH core, this implies that different models (2-body, 3-body and 4-body annihilation) and different choices of $R_i$ give rise to similar redshift dependences for the energy injection, and consequently the first principal component is nearly indistinguishable between these cases. Since the redshift scaling of the injected power is dominated by the evolution of the UCMH number density, proportional to $(1+z)^3$, the redshift dependence is also the same as one obtains from conventional decaying DM; the first principal component shown here is consequently very similar to that derived previously for decaying DM \cite{Slatyer:2016qyl}. Similarly, the relative strength of constraints from probes of different redshifts should trace the results for decaying dark matter; in particular, confirmation of the claimed 21cm absorption signal from the EDGES experiment \cite{Bowman:2018yin} could improve on the CMB bounds by up to two orders of magnitude \cite{Liu:2018uzy}.

\begin{figure}[h]
\hspace{2.6cm}
 \includegraphics[width=11.cm]{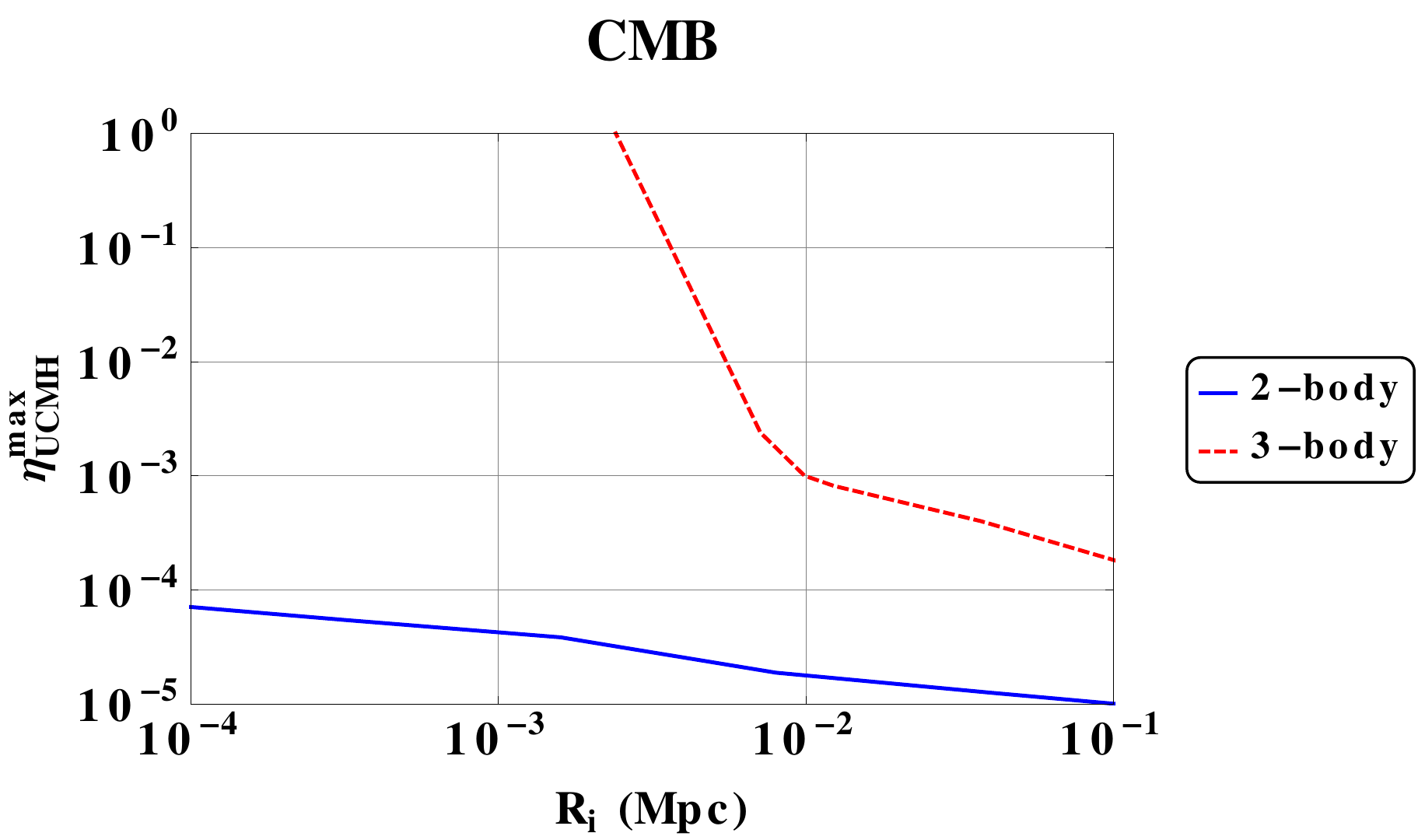}
 \caption{CMB limit on the UCMH fraction $\eta_\text{UCMH}$ for 2-body and 3-body annihilation for thermal models; we use the deposition efficiency for an $e^+e^-$ pair with energy 100 MeV (the results for other energies/species can be obtained from Fig.~\ref{UCMHcmbpca}).}
\label{UCMHcmb}
\end{figure}

In Fig.~\ref{UCMHcmb}, we show the CMB limits on the fraction of DM that could be comprised of UCMHs for the thermal 3-body benchmark models, as well as for the 1 TeV thermal benchmark for 2-body annihilation, using the efficiency appropriate to injection of $\sim$ 100 MeV electrons and positrons (other annihilation channels will relax the constraints by $\mathcal{O}(1)$ factors). The limit for different energies and photon injections can be extrapolated by the first principal component. For NFDM and  4-body models, the CMB gives essentially no constraints for the cases we consider.

We see that for these benchmark models, where there is no low-velocity enhancement, UCMHs constituting a percent-level fraction of the DM could potentially leave detectable signals in the ionization and thermal history. We note that this statement is dependent on the steep profile we have assumed for the UCMHs; we have checked that with the halo profile of Eq.~\eqref{eq:density1-5}, the CMB would not be able to test the scenario with $\eta_\text{UCMH}=1$ for either benchmark.

\section{A case study: Not-forbidden dark matter}
\label{sec:models} 

\subsection{Review of the model}
Ref.~\cite{Cline:2017tka} demonstrated that a simple dark photon model can have 3-body annihilations dominant during freezeout, and labeled this scenario ``not forbidden dark matter'' (NFDM). In this model the DM is a Dirac fermion of mass $m_\chi$, coupled to a massive dark photon $A^\prime$ with mass $m_{A^\prime}$ such that $1.5 m_\chi < m_{A^\prime} < 2 m_\chi$. The dark Higgs responsible for breaking the dark $U(1)$ gauge symmetry is assumed to be heavier. The dark photon kinetically mixes with the SM photon with a mixing $\varepsilon \ll 1$. Thus the part of the model Lagrangian relevant to DM annihilations is \cite{Cline:2017tka}:
\begin{equation}   \mathcal{L} \supset -\frac{1}{4} F_{\mu\nu} F^{\mu\nu} - \frac{1}{4}F_{\mu\nu}' {F'}^{\mu\nu} + 
\frac{1}{2}  m_{A'}^2 {A'}_\mu {A'}^{\mu} \\
   + \bar{\chi} ( i \slashed{D} - m_\chi ) \chi   
         + e J_{\text{EM}}^\mu (A_\mu + \varepsilon {A'}_{\mu}),  \end{equation}
where $\slashed{D} = \slashed{\partial} -ig' \slashed{A'}$.
The annihilation of $\bar{\chi} \chi$ into two dark photons is Boltzmann suppressed, and the annihilation of $\bar{\chi} \chi$ into SM particles is $\varepsilon$-suppressed. Consequently, the three-body annihilation processes $\bar{\chi} \chi \chi \rightarrow A^\prime \chi$ and $\bar{\chi} \chi A^\prime \rightarrow A^\prime A^\prime$ can generically dominate DM depletion during freezeout. Ignoring the exponentially-suppressed kinematically forbidden two-body annihilation at late times (e.g. $\bar{\chi} \chi \rightarrow A' A'$), the competing two-body channel is the $\varepsilon$-suppressed $\bar{\chi} \chi \rightarrow (A')^* \rightarrow$ SM$^+$ SM$^-$, where ``SM'' denotes a charged SM species.

The process $\bar{\chi} \chi \chi \rightarrow A^\prime \chi$ can still occur at late times, as it does not require the presence of an $A^\prime$ bath. At zero velocity, and assuming the width of the $A^\prime$ can be neglected, the cross section for $\bar{\chi} \chi \chi \rightarrow A^\prime \chi$ is given by:
\begin{align} \langle \sigma v^2 \rangle_{\bar{\chi} \chi \chi \rightarrow A^\prime \chi} & =\frac{g'^6 (r_\text{NFDM}-4)(r_\text{NFDM}+4)(-32r_\text{NFDM}^8 + 167r_\text{NFDM}^6 - 534r_\text{NFDM}^4 + 668r_\text{NFDM}^2 - 512)}{36 m_\chi^2 (r_\text{NFDM}^2 - 4)^4 (r_\text{NFDM}^2 + 2)^2} \nonumber \\
& \times \frac{\sqrt{r_\text{NFDM}^4 - 20r_\text{NFDM}^2 + 64}}{96 \pi m_\chi^3},
\label{eq:nfdmxsec}\end{align}
where $r_\text{NFDM} = m_{A^\prime}/m_\chi$. 

Let us define:
\begin{equation} y(r_\text{NFDM}) \equiv  \frac{(4 - r_\text{NFDM})^{3/2}(r_\text{NFDM}+4)^{3/2}(32r_\text{NFDM}^8 - 167r_\text{NFDM}^6 + 534r_\text{NFDM}^4 - 668r_\text{NFDM}^2 + 512)}{3456 (r_\text{NFDM}^2 + 2)^{2}}.\end{equation} 
Then $y$ is a slowly-varying $\mathcal{O}(1)$ function of $r_\text{NFDM}$, with $y(2) = 3 \sqrt{3}/4 \approx 1.3$ and $y(r_\text{NFDM})$ varying between 0.5 and 1.3 for $r_\text{NFDM}$ between 1.5 and 2. We can write the 3-body cross section as:
\begin{align} \langle \sigma v^2 \rangle_{\bar{\chi} \chi \chi \rightarrow A^\prime \chi} & =  y(r_\text{NFDM}) \frac{g'^6}{\pi m_\chi^5} \frac{1}{(r_\text{NFDM}^2 - 4)^{7/2}} \nonumber \\
& = y(r_\text{NFDM}) \frac{64 \pi^2 (\alpha^\prime)^3}{m_\chi^5} \frac{1}{(4 - r_\text{NFDM}^2)^{7/2}} \nonumber \\
& \approx 630 y(r_\text{NFDM}) (4 - r_\text{NFDM}^2)^{-7/2} \frac{\alpha^{\prime 3}}{m_\chi^5}, \end{align}
where $\alpha' =g'^2/4\pi$.
We see that the 3-body cross section in this model is naturally a few orders of magnitude larger than our parametric estimate of $\alpha^3/m_\chi^5$, and can be tuned to be even larger by taking $r_\text{NFDM}$ close to 2.
However, taking this limit also increases the DM self-interaction rate ($\propto \alpha^{\prime 2} (r_\text{NFDM}^2 -4)^{-2}$), which provides stringent constraints on this scenario if the $\chi$ and $\bar{\chi}$ fermions constitute 100\% of the DM, and the 2-body annihilation to the SM ($\propto (r_\text{NFDM}^2 -4)^{-2}$), both of which proceed through a near-resonant $s$-channel $A^\prime$ exchange in this case. The 3-body process increases more rapidly as $r_\text{NFDM} \rightarrow 2$ than either of these 2-body processes; this reflects an enhanced low-velocity scaling in the 3-body case, which we will discuss in the next section. Nonetheless, constraints from the 2-body processes can be significant, particularly if we demand a thermal history for the DM, although they may be relaxed by taking the NFDM component to constitute only a small fraction of the DM (in the case of self-interactions) or by suppressing the coupling $\varepsilon$. 

Generally the DM mass scale preferred for this model is between 100 MeV and 1-2 GeV, due to stringent constraints on the dark photon parameter space at lighter masses, provided $\varepsilon$ is not too small and $r$ is not too close to 2. As a result, we generally need not concern ourselves with constraints on warm DM or extra relativistic degrees of freedom during and after Big Bang Nucleosynthesis.

Since the dominant 3-body annihilation processes in this model are $\chi \chi \bar{\chi} \rightarrow \chi A^\prime$ and the conjugate process $\bar{\chi} \bar{\chi} \chi \rightarrow \bar{\chi} A^\prime$, the power going into SM final states per annihilation (assuming non-relativistic initial states) is $(m_{A^\prime}^2 + 8 m_\chi^2)/(6 m_\chi) = m_\chi (8 + r_\text{NFDM}^2)/6$. The annihilation rate per unit volume per unit time, summing the two conjugate processes, is $\langle \sigma v^2\rangle (\rho_\text{tot}/m_\chi)^3/8$, where $\rho_\text{tot}= \rho_\chi + \rho_{\bar{\chi}}$ (we assume $\rho_\chi = \rho_{\bar{\chi}}$), and  $\langle \sigma v^2\rangle$ describes the rate for one of the two conjugate processes, given in Eq.~\eqref{eq:nfdmxsec}  (equivalently, the rate is $\langle \sigma v^{n-1}\rangle \times \left[ (\rho_\chi^2/2)\rho_{\bar{\chi}} + (\rho_{\bar{\chi}}^2/2)\rho_\chi \right]/m_\chi^3$). Thus we obtain, for this model:
\begin{align} \xi = \langle \sigma v^2\rangle m_\chi \frac{8 + r_\text{NFDM}^2}{6} \dfrac{(\rho_\text{tot}/m_\chi)^3}{8} \frac{1}{\rho_\text{tot}^3} = \frac{\langle\sigma v^2\rangle}{m_\chi^2} \times \frac{1 + r_\text{NFDM}^2/8}{6}.\end{align} 

\begin{figure}
\hspace{4cm}
\includegraphics[scale=0.45]{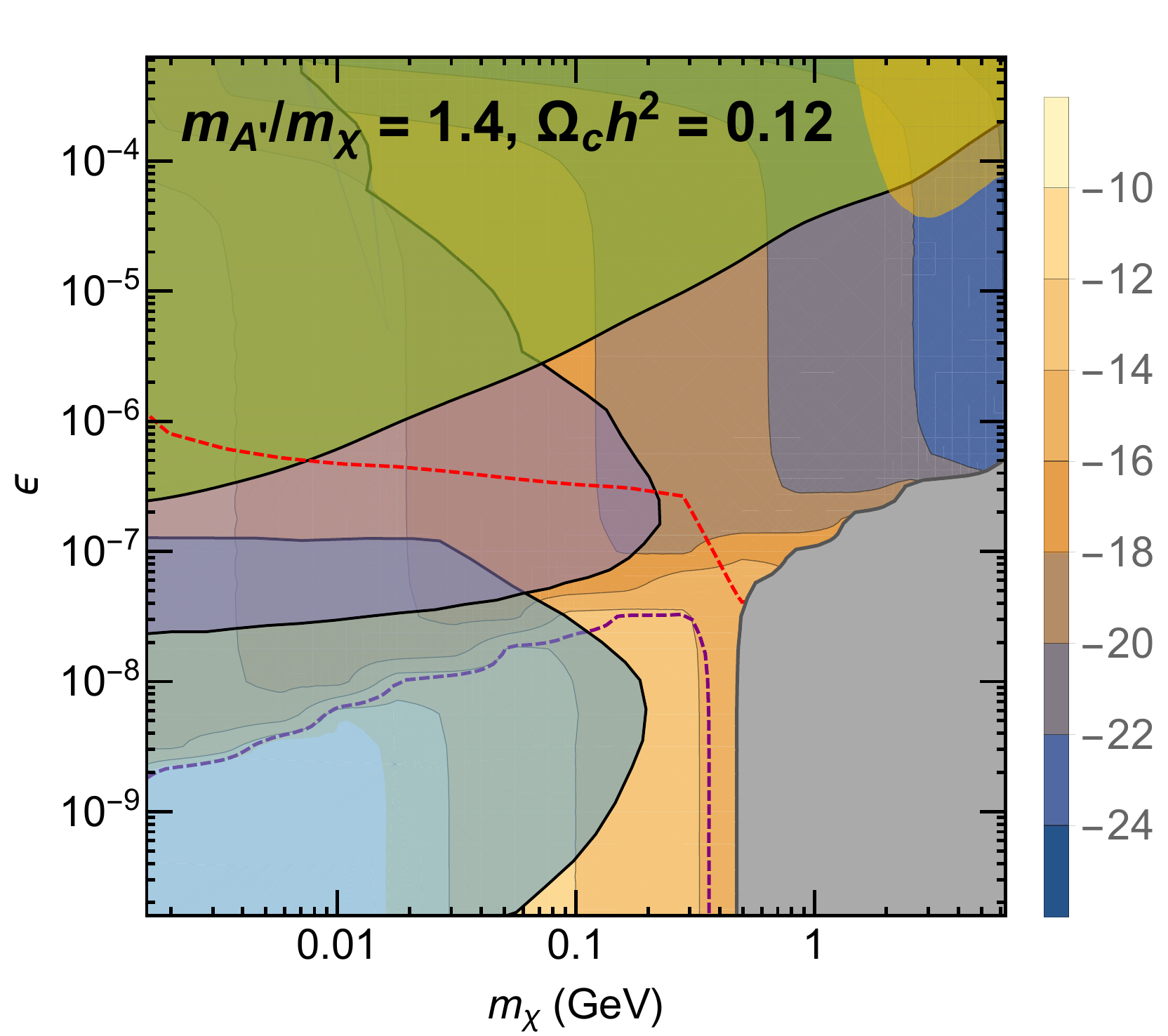}
\\ \vspace{.7cm}
\includegraphics[scale=0.45]{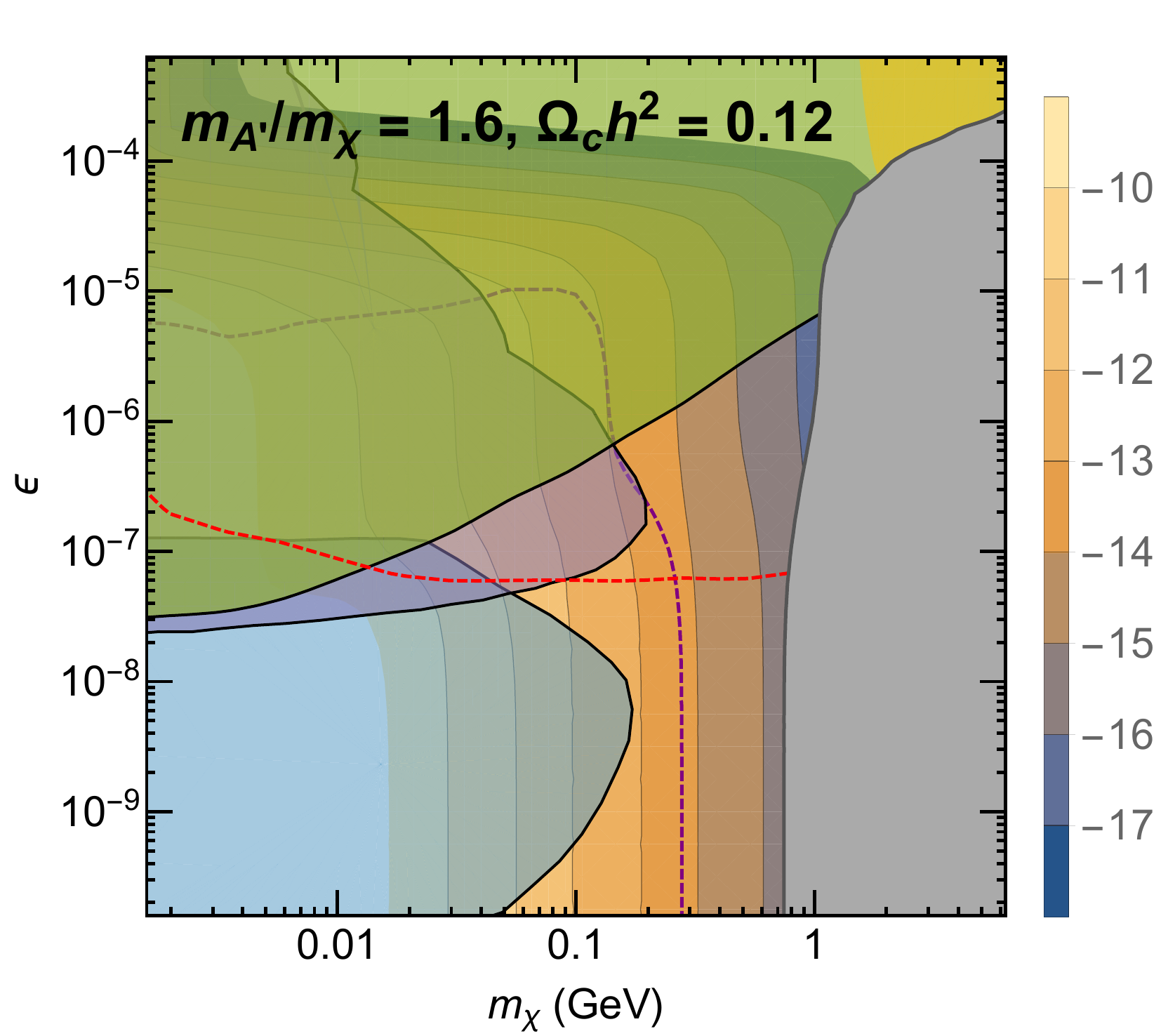} 
\hspace{0.5cm}
\includegraphics[scale=0.45]{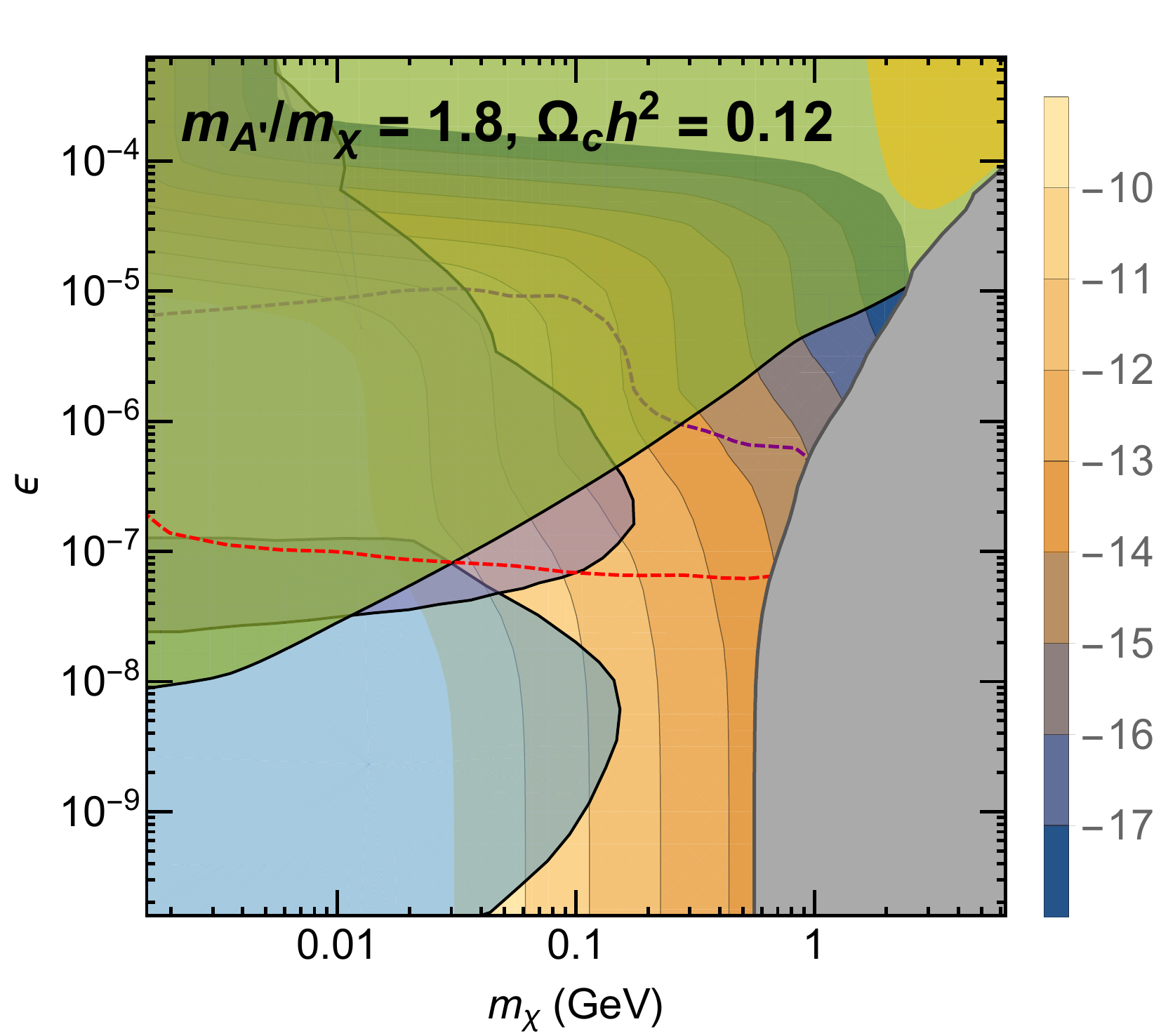}
\caption{Allowed parameter space and range of annihilation parameter $\xi$ for the NFDM model. Colored contours indicate log$_{10}(\xi \times \text{MeV}^7)$. Regions above the red-dashed line correspond to thermal equilibrium between the dark sector and SM during freezeout; regions below the purple-dashed line are excluded by self-interaction constraints if the species in question constitutes all the DM. Other shaded regions correspond to exclusions from perturbativity (gray), beam dump experiments (purple), supernova cooling (blue), CMB bounds on 2-body annihilation (green), and CDMS direct-detection limits (yellow). The panels correspond to $m_{A^\prime}/m_\chi$ of 1.4 (upper), 1.6 (lower-left) and 1.8 (lower-right)}
\label{fig:nfdm-p}
\end{figure}

In Fig.~\ref{fig:nfdm-p} we show the allowed values of $\xi$, and previous experimental constraints on the parameter space, for several non-fine-tuned choices of $r$ and $\varepsilon$ large enough to keep the dark and visible sectors kinetically coupled through freezeout; the dark-sector coupling $\alpha^\prime$ is chosen to yield the correct relic density. We see that in general the allowed values of $\xi$ are $\lesssim 1/(10 \text{MeV})^7$ or smaller, in large part because for these values of $\varepsilon$, a range of stringent limits forces the mass scale for the $\chi$ and $A^\prime$ to be 10s of MeV or greater. Comparing this range for $\xi$ to the limits of Fig.~\ref{limitsmass}, we see that the NFDM parameter space for these choices of $r_\text{NFDM}$ is not constrained by indirect-detection searches, even for optimistic choices of the small-scale structure model. However, as discussed in Sec.~\ref{sec:ucmh}, there is the potential for a visible signal in the presence of a substantial UCMH population.

It is possible that there may be more low-mass parameter space open in variations of the NFDM scenario where the mediator is not a dark photon; for example, light Dirac-fermion DM coupled to the SM via the Higgs portal may allow for open parameter space down to $\sim 1$ MeV mediator masses \cite{Krnjaic:2015mbs}, although this scenario has not been studied in the NFDM context where 3-body annihilation is included. We leave an exploration of alternate models for subsequent work.

\subsection{Low-velocity enhancement in the $m_{A^\prime} \rightarrow 2 m_\chi$ region}

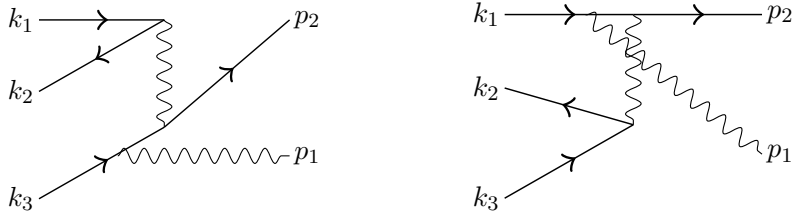
\begin{figure}
\begin{center}
\resizebox{0.3\textwidth}{!}{
\begin{tikzpicture}
\begin{scope}[shift={(0,-4.5)}]
\draw[photon] (1,1.5)--(1,0);
\draw[fermion] (-0.75,1.5)--(1,1.5);
\node at (-1,1.5) {$k_1$};
\draw[fermion] (1,1.5)--(-0.75,0.5);
\node at (-1,0.5) {$k_2$};
\draw[fermion] (-0.75,-1)--(1,0);
\node at (-1,-1) {$k_3$};
\draw[fermion] (1,0)--(2.75,1.5);
\node at (3,1.5) {$p_2$};
\draw[photon] (0.35,-0.40)--(2.75,-0.4);
\node at (3,-0.4) {$p_1$};
\end{scope}
\end{tikzpicture} 
}
\qquad \qquad
\resizebox{0.3\textwidth}{!}{\begin{tikzpicture}
\begin{scope}[shift={(0,-4.5)}]
\draw[photon] (1,1.5)--(1,0);
\draw[fermion] (-0.75,1.5)--(1,1.5);
\node at (-1,1.5) {$k_1$};
\draw[fermion] (1,0)--(-0.75,0.5);
\node at (-1,0.5) {$k_2$};
\draw[fermion] (-0.75,-1)--(1,0);
\node at (-1,-1) {$k_3$};
\draw[fermion] (1,1.5)--(2.75,1.5);
\node at (3,1.5) {$p_2$};
\draw[photon] (0.35,1.5)--(2.75,-0.4);
\node at (3,-0.4) {$p_1$};
\end{scope}
\end{tikzpicture} 
}
\end{center}
\caption{
Feynman diagrams providing the leading contribution to the amplitude for $\chi \chi \bar{\chi} \rightarrow \chi A^\prime$ in the low-velocity limit. 
}
\label{fig:nfdmdivergence}
\end{figure}

In the region where $r_\text{NFDM} \rightarrow 2$, the cross sections for 2-body annihilation to SM particles, 3-body annihilation and DM-DM elastic scattering can all be greatly enhanced at low velocities. Let us first neglect the width of the $A^\prime$ and consider only the diagrams with the strongest low-velocity divergence in the limit $r\rightarrow 2$. The leading contribution in the $r\rightarrow 2$ limit is given by the diagrams in Fig.~\ref{fig:nfdmdivergence}. The divergence originates from the propagator structure $\frac{1}{(k_3 - p_1)^2 - m_\chi^2} \times \frac{1}{(k_1 + k_2)^2 - m_{A^\prime}^2}$ in the left diagram, with a similar expression for the right one obtained by the replacement $k_1 \leftrightarrow k_3$. Note that in the non-relativistic limit with $m_{A^\prime} \rightarrow 2 m_\chi$, the final-state particles are also non-relativistic, as the sum of masses in the initial and final states are nearly equal. We see that both the $A^\prime$ and $\chi$ propagators go on-shell in the non-relativistic limit, as $k_3 - p_1 \rightarrow (m_\chi - m_{A^\prime}, 0, 0, 0) = -(m_\chi, 0, 0, 0)$, and $k_1 + k_2 \rightarrow (2 m_\chi, 0, 0, 0) = (m_{A^\prime}, 0, 0, 0)$.

The $A^\prime$ propagator will be regulated by the width of the $A^\prime$ at $s=(k_1 + k_2)^2$. Note that even if the interaction with the Standard Model is turned off completely, rendering the $A^\prime$ stable at its pole mass, at $s=(k_1 + k_2)^2 \ge (2 m_\chi)^2$ there is a contribution to the decay width from the $\bar{\chi} \chi$ and $e^+e^-$ final state. Near the pole, the corresponding imaginary part of the 1PI diagrams is given by: 
\begin{eqnarray}
\Gamma_{A^{\prime} \rightarrow \chi\bar{\chi}} (s) &=& 2 m_{\chi} \dfrac{\alpha^\prime}{3}\left( 1+\dfrac{2m_\chi^2}{s}\right)\sqrt{1-\dfrac{4 m_\chi^2}{s}} \nonumber \\ 
\Gamma_{A^{\prime} \rightarrow e^+e^-}(s)  &=& 2 m_{\chi} \dfrac{\varepsilon^2 \alpha_{em}}{3} \left(1+\dfrac{2m_e^2}{s} \right) \sqrt{1-\dfrac{4 m_e^2}{s}}.
\label{eq:Adecay}
\end{eqnarray}
Note that this width goes to zero in the non-relativistic limit where the initial particle velocities go to zero; it regulates the naive $1/v^2$ dependence of the propagator to $1/v$.

However, the same regulation does not occur for the $\chi$ propagator; the $\chi$ is absolutely stable, and while a dark-sector $\chi \rightarrow \chi A^\prime$ vertex could regulate the propagator at sufficiently high $s$, in this case $s = (k_1 - p_1)^2 \approx m_\chi^2$, and this channel is closed. The creation of a nearly-on-shell stable particle induces an effective long-range interaction, which is responsible for a large enhancement at low velocities. This behavior is similar to that studied for muon colliders \cite{Melnikov:1996iu}, albeit in this case the mediator is truly stable rather than just long-lived, and unlike in the muon-collider case, the singularity does not actually enter the physical region.

This latter point can be proved by noting that the fermion propagator denominator $(k_3 - p_1)^2 - m_\chi^2 = k_3^2 + p_1^2 - 2 k_3 \cdot p_1 - m_\chi^2 = p_1 \cdot (p_1 - 2 k_3)$ is Lorentz-invariant, and if evaluated in the frame with $p_1 = (m_A^\prime, 0, 0, 0)$, becomes $m_A^\prime (m_A^\prime - 2 E_{k_3})$; since the energy of the outgoing fermion with momentum $k_3$ is greater than or equal to $m_\chi > m_{A^\prime}/2$, the denominator is always negative, and never crosses through zero for any physical choice of momenta. Likewise, for the vector propagator with denominator $(k_1 + k_2)^2 - m_{A^\prime}^2$, we can work in the frame where $\vec{k}_1 + \vec{k}_2 = 0$, and then $(k_1 + k_2)^2 - m_{A^\prime}^2 = (E_{k_1} + E_{k_2})^2 -  m_{A^\prime}^2 > (2 m_\chi)^2 - m_{A^\prime}^2 > 0$. Since neither denominator can ever pass through zero, there is no singularity in the physical region.

Because of the absence of a physical singularity, there is no obvious necessity that the divergence be regulated, but one might still ask if the effects that regulate the singularity in the muon-collider case could soften the low-velocity behavior in this scenario. However, while beam-width effects are important in the muon-collider context \cite{Melnikov:1996iu,Dams:2002uy}, in our case the effective beam width is always very large (corresponding to astrophysical/cosmological scales) and we do not expect it to meaningfully regulate the cross section.

Thus we expect the matrix element to scale as $1/v^4$ (from the two nearly-on-shell propagators) down to the point where the $A^\prime$ width becomes significant, and then to scale as $1/v^3$ until saturating due to $r_\text{NFDM} \ne 2$ (there may also be a further softening of the scaling prior to saturation, if the decay width of the $A^\prime$ becomes dominated by decays to SM particles that are not phase-space suppressed). The rate coefficient $\langle \sigma v^2\rangle$ carries an additional phase-space factor, as the final-state particles are non-relativistic, and thus we expect scaling of $\langle \sigma v^2\rangle \propto v^{-7} \propto T^{-3.5}$ at intermediate velocities, regulated to $v^{-5} \propto T^{-2.5}$ once the $A^\prime$ width becomes relevant, and finally flattening out to a constant saturated value set by $r_\text{NFDM}$ at sufficiently low velocities.

The regulated matrix element for the $\chi\chi\bar{\chi}\rightarrow \chi A^\prime$ process, dividing by the number of degrees of freedom in both the initial and final states, and working in the non-relativistic limit with $r_\text{NFDM} \rightarrow 2$, then takes the form:
\begin{eqnarray}
\overline{|M|}^2 &= & 8 g'^6 m_\chi^{6}\left( 6 P\left(m_{12}, k_3 \right)^2+6 P\left(m_{23}, k_1\right)^2 - 5\left( P\left(m_{12},k_3\right) P\left(m_{23},k_1\right)^* + h.c. \right)\right), \,\,\,\,\,
\end{eqnarray}
where $m_{12}^2 = \left(k_1+k_2\right)^2$ and $P\left(m_{12}, k_3 \right)$ is the combination of propagators with the $A^\prime$ propagator regularized by the decay width:

\begin{align}
P\left(m_{12}, k_3 \right) &= \dfrac{1}{ m_{12}-4 m_\chi^2- 2 i m_\chi\left(\Gamma_{A^\prime \rightarrow e^+e^-} (m_{12})+\Gamma_{A^\prime \rightarrow \chi\overline{\chi}} (m_{12}) \right)+i \epsilon } \nonumber \\
& \times \dfrac{1}{ (k_3-p_1)^2-m_\chi^2+i\epsilon},
\end{align}
and similar for $P\left(m_{23}, k_1 \right) $. The thermally-averaged cross section can then be written as:
\begin{eqnarray}
\langle \sigma v^2 \rangle_{\chi \chi \overline{\chi} \rightarrow \chi A'} & = & \dfrac{g_\chi^4 g_{A^\prime}}{3!}m_\chi^{-9/2} e^{3m_\chi/T}\int d \Pi_2 d\Pi_3 \dfrac{d^4 p_0}{\left(2 \pi \right)^4} e^{-E_0/T} \, \overline{|M|}^2, \nonumber \\
& = & \dfrac{g_\chi^4 g_{A^\prime}}{3!} m_\chi^{-9/2} e^{3m_\chi/T} \int d \Pi_2 d\Pi_3 \int_{9m_\chi^2}^{\infty} ds \int_{\sqrt{s}}^{\infty} dE_0 \sqrt{E_0^2-s}\, e^{-E_0/T}\, \overline{|M|}^2  
\end{eqnarray}
where the phase space integrals defined in Eq.~\eqref{eq:pils} can be simplified as:
\begin{eqnarray}
\int d\Pi_2 &=& \dfrac{1}{8\pi}\sqrt{\left(1-\dfrac{9m_\chi^2}{s}\right)\left(1-\dfrac{m_\chi^2}{s}\right)}, \nonumber \\
\int d\Pi_3 &=& \dfrac{1}{8\left(2 \pi\right)^3}\int_0^{4\pi}\int_0^{4\pi}\int_{2m_\chi}^{\sqrt{s}-m_\chi} \dfrac{\bold{|k_1^*||k_3|}}{\sqrt{s}} dm_{12}d\Omega_1^*d\Omega_3.
\end{eqnarray}
The quantities $\left(\bold{|k_1^*|}, d\Omega_1^*\right)$ describe the momentum of particle 1 in the rest frame of 1 and 2, written explicitly as
\begin{eqnarray}
|\bold{k_1}^*| &=& \dfrac{ \left(m_{12}^2 -4m_\chi^2 \right)^{1/2}}{2},\nonumber \\
|\bold{k_3}| &=& \dfrac{\left[ \left(s -(m_{12}+m_\chi)^2 \right) \left(m_{12}^2 -(m_{12}-m_\chi)^2 \right) \right]^{1/2}}{2\sqrt{s}}
\end{eqnarray}
and the other kinematics in the COM frame can be determined by $\bold{k_1} = \bold{k_1}^*-\bold{k_3}/2$, $\bold{k_2} = -\bold{k_1}^*-\bold{k_3}/2$, with the outgoing momentum $\bold{p}$ determined by $\sqrt{s} = \sqrt{\bold{p}^2+m_\chi^2}+\sqrt{\bold{p}^2+m_{A^\prime}^2} $.

Incidentally, while in this case the tuning between the $A^\prime$ mass and twice the $m_\chi$ mass is purely artificial, similar resonance effects could arise naturally in the context of bound states, where the bound state $\mathcal{B}$ has a mass slightly lighter than the sum of its constituents' masses. Replacing $A^\prime$ with a bound state $\mathcal{B}$ of $\bar{\chi} \chi$, the $\chi \chi \bar{\chi} \rightarrow \chi A^\prime$ process would correspond to 3-body recombination into the bound state. The strong scaling of 3-body recombination with scattering length, and the necessity for regulation to preserve unitarity, has previously been discussed in the condensed-matter literature \cite{GREENE2004119}. The formation of dark matter bound states via 3-body recombination has been studied recently \cite{Braaten:2018xuw}.

\subsection{Example benchmarks}

\begin{figure}
\hspace{2.7cm}
\includegraphics[scale=.6]{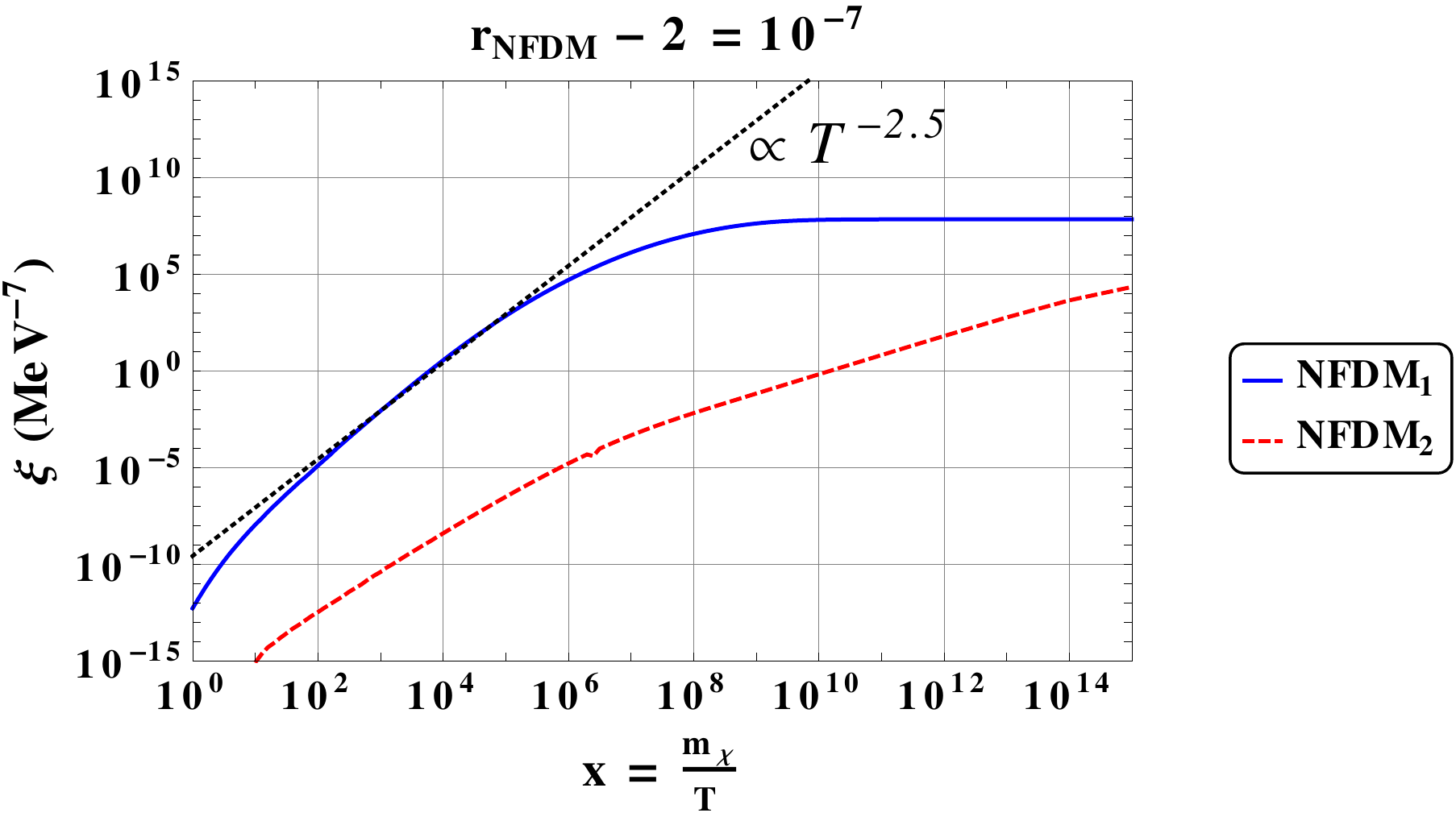}
\caption{$\xi$ as a function of x assuming NFDM$_1$: $m_\chi = 30 \, \text{MeV}, \varepsilon = 10^{-5}, \alpha^\prime =7 \times 10^{-3}$, NFDM$_2$: $m_\chi = 2 \, \text{GeV}, \varepsilon = 10^{-11}, \alpha^\prime = 10$.}
\label{p1}
\end{figure}

Let us first consider an example scenario labeled $\text{NFDM}_1$, with the following parameters: $m_\chi = 30 \text{MeV}, \varepsilon = 10^{-5}, \alpha^\prime = 7 \times 10^{-3} , r_\text{NFDM}=2-10^{-7}$. In this case, the correct relic density is obtained and $\xi \approx 5.0 \times 10^{7} \, \text{MeV}^{-7}$ at late times. The temperature evolution of the parameter $\xi$ is shown in Fig.~\ref{p1}. We observe the expected $T^{-2.5}$ scaling followed by saturation at low temperatures, corresponding to $v \lesssim 10^{-4}$. 

However, this parameter point is strongly excluded by limits on the resonantly enhanced 2-body annihilation channel $\chi \bar{\chi} \rightarrow e^+ e^-$ from the cosmic microwave background; the predicted cross section is $\langle \sigma v \rangle/m_\chi \approx 6 \times 10^{-13} \text{cm}^3/\text{s}/\text{GeV}$.

More generally, the cross section for this annihilation channel is \cite{Cline:2017tka}:
\begin{align} \langle \sigma v \rangle_{\bar{\chi} \chi \rightarrow e^+ e^-} & =  \frac{4 e^2 \varepsilon^2 g'^2 \left(2 + m_e^2/m_\chi^2\right)}{(r_\text{NFDM}^2 - 4)^2+\dfrac{m_{A^\prime}^2}{m_\chi^4}\left(\Gamma_{A \rightarrow e^+e^-} (s)+\Gamma_{A \rightarrow \chi\overline{\chi}} (s)\right)^2} \frac{\sqrt{1 - m_e^2/m_\chi^2}}{8 \pi m_\chi^2}. \label{eq:epmxsec} \end{align}
We can approximate the CMB annihilation limits by $\langle \sigma v \rangle_{\bar{\chi} \chi \rightarrow e^+ e^-} /m_\chi \lesssim 4\times 10^{-27}$ cm$^3$/s/GeV, where we take the limit from the Planck 2015 result \cite{Ade:2015xua} and assume the efficiency parameter to be $f_\text{eff}\gtrsim 0.1$, based on the results for annihilation to electrons in \cite{Slatyer2015a}. Assuming $m_\chi \gg m_e$,  and that the decay of the $A^\prime$ to $e^+ e^-$ is strongly suppressed by $\varepsilon$ and hence gives a subdominant contribution to the denominator of Eq.~\eqref{eq:epmxsec},\footnote{If the $\Gamma_{A\rightarrow e^+ e^-}$ term instead dominates the denominator of Eq.~\eqref{eq:epmxsec}, satisfying the CMB bounds requires a very large value of $\varepsilon$ that is ruled out by null searches for the $A^\prime$.} this limit can be estimated as:
\begin{align} 1.3\times10^{-26}  \text{cm}^3\text{/s/GeV}  & \gtrsim  \frac{e^2 \varepsilon^2 g'^2}{(r_\text{NFDM}^2 - 4)^2+r_\text{NFDM}^2 \left(\dfrac{\Gamma_{A \rightarrow \chi\overline{\chi}}}{m_\chi}\right)^2} \frac{1}{ m_\chi^3}. \label{eq:nfdm_cmb_lim} \end{align}
On one hand suppose the second term in the denominator of Eq.~\eqref{eq:nfdm_cmb_lim} is negligible, then we have in the limit of $r_\text{NFDM} \rightarrow 2$:
\begin{eqnarray}
\varepsilon \lesssim 6.35 \times10^{-7} {\alpha^\prime}^{-1/2} \left(2-r_\text{NFDM}\right) \left( \dfrac{m_\chi}{30 \text{MeV}}\right)^{3/2}. \label{eq:epsilonlimit}
\end{eqnarray}
On the other hand, suppose that $\alpha^\prime$ is large enough that the first term in the denominator is subdominant and can be ignored. The decay width for $A^\prime$ into $ \chi\overline{\chi}$ is given by Eq.~\eqref{eq:Adecay}. Taking $r_\text{NFDM} \rightarrow 2$ and $s \simeq 4 m_\chi^2 (1+p^2/m_\chi^2)$ -- where $p$ is the momentum of either of the incoming particles in the COM frame -- and substituting this result into the CMB limit above, we obtain:
\begin{eqnarray}
\varepsilon \lesssim 5.50 \times10^{-13} {\alpha^\prime}^{1/2} \left( \dfrac{m_\chi}{30 \text{MeV}}\right)^{3/2}  \left(\dfrac{10^{12}}{x}\right)^{1/2} 
\end{eqnarray}
where $x = m_\chi/T$ and we have set $p^2/2m_\chi \simeq 3T/2$. Thus tuning $r_\text{NFDM}$ close to 2 requires $\varepsilon$ to become smaller and smaller in order to evade CMB limits from 2-body annihilation.

To avoid the strong constraints from the cosmic microwave background in the presence of resonantly-enhanced annihilation to SM particles, we are led to posit a very small coupling $\varepsilon$ between the dark and visible sectors. This also increases the lifetime of the $A^\prime$, but requiring that the $A^\prime$ decays before BBN only requires a lifetime shorter than $\sim 200$ seconds, which sets:
\begin{eqnarray}
\varepsilon & \gtrsim 4.74\times 10^{-12} \left(\dfrac{30 \,\text{MeV}}{m_{\chi}}\right)
\end{eqnarray}
Even if this condition is not satisfied, the abundance of $A^\prime$ is generally rather small after freezeout due to its heavy mass, and thus limits from energy injection at BBN can be evaded.

 A much stronger condition is that $\varepsilon$ is large enough that the dark and visible sectors remain in kinetic equilibrium throughout freezeout; this condition in general will not be satisfied if Eq.~\eqref{eq:epsilonlimit} holds, as shown in Fig.~\ref{fig:nfdm-p}. 
If the relevant content of the dark sector is solely the $A^\prime$ and $\chi$, $\bar{\chi}$ fields, and the dark sector is decoupled from the SM, then it will undergo a cannibalization process \cite{Pappadopulo:2016pkp} after kinetic decoupling from the SM, as in ELDER models \cite{Kuflik:2015isi, Kuflik:2017iqs}. In this case the late-time relic density depends sensitively on the DM-SM scattering, and hence on $\varepsilon$. Alternatively, the presence of additional light degrees of freedom (which decay to SM particles prior to nucleosynthesis) in the dark sector can provide an effective dark radiation bath, which allows a standard thermal freezeout (albeit with a different number of relativistic degrees of freedom) to proceed in the dark sector. This possibility was discussed and labeled as the ``secluded'' scenario in Ref.~\cite{Cline:2017tka}; it can be implemented by adding a light scalar field that couples to the DM through a higher-dimension operator. Annihilation of the DM into the light scalars is not resonantly enhanced, and consequently can easily be made unobservable at late times as well as subdominant during freezeout. The cross-section estimate in \cite{Cline:2017tka}, requiring that this channel be unimportant during freezeout, corresponds to $\langle \sigma v\rangle \sim 10^{-18}-10^{-12} \text{GeV}^{-2} \sim 10^{-35}-10^{-29}$ cm$^3$/s for GeV-scale DM, and $\langle \sigma v\rangle \sim 10^{-15}-10^{-8} \text{GeV}^{-2} \sim 10^{-32}-10^{-25}$ cm$^3$/s for MeV-scale DM; even for MeV-scale DM, the low end of this cross-section range is unobservable in indirect detection. 

In such a secluded scenario, we can obtain a second thermal relic benchmark, which we label $\text{NFDM}_2$: $m_\chi = 2 \,\text{GeV}, \varepsilon = 10^{-11}, \alpha^\prime = 10, r_\text{NFDM} = 2-10^{-7}$ gives the correct relic abundance and has $\xi \approx 6\times10^{4} \, \text{MeV}^{-7}$ in the saturation region, while the two-body annihilation rate evades the CMB limit. We show the evolution of $\xi$ with DM temperature in Fig.~\ref{p1}. At this parameter point, the $A^\prime$ decays with a lifetime of $\tau \sim 1$ seconds. This lifetime is shorter than the age of the universe at BBN, and so is unconstrained by cosmological probes, and it is also very short compared to all relevant timescales for indirect detection.

As well as the resonantly enhanced number-changing processes, $\chi \bar{\chi}$ elastic scattering through $s$-channel exchange of a $A^\prime$ also experiences resonant enhancement at low velocities, explicitly as
\begin{equation}
\dfrac{\sigma_{SI}}{m_\chi} = \dfrac{3 g'^4}{16 \pi m_\chi^3}\dfrac{16-16r_\text{NFDM}^2+5r_\text{NFDM}^4}{r_\text{NFDM}^4\left((r_\text{NFDM}^2 - 4)^2+\dfrac{m_{A^\prime}^2}{m_\chi^4}\left(\Gamma_{A \rightarrow e^+e^-} (s)+\Gamma_{A \rightarrow \chi\overline{\chi}} (s)\right)^2\right)}.
\end{equation}
For the $\text{NFDM}_2$ parameters above, $\dfrac{\sigma_{SI}}{m_\chi} \approx 43 \left(\dfrac{v}{100 \, \text{km/s}}\right)^{-2} \text{cm}^2/\text{g}$, to a good approximation, for $10^{-7} < v/c < 0.1$. This self-interaction cross section is small enough at typical cluster velocites ($\sim 1000$ km/s) to evade limits originating from clusters \cite{Wittman:2017gxn,Tulin:2017ara}; however, it would be very large at dwarf-galaxy scales compared to recently estimated bounds \cite{Bondarenko:2017rfu}, and so for consistency with these limits, the NFDM component may need to be a subdominant component of the DM in this scenario (although we caution that these limits do not include baryonic effects, which could substantially modify the impact of DM self-interactions, albeit to a lesser degree in systems like dwarfs which are not baryon-dominated \cite{Kaplinghat:2013xca}). In a thermal-origin scenario, this should only suppress the indirect-detection signal by one power of the small fraction, as discussed in Sec.~\ref{sec:parametric}. 

Note that in this model, it is the interplay of 2-body and 3-body annihilation that determines the relic density; in fact, at this benchmark, the 3-body annihilation rate remains fast relative to Hubble well after the asymptotic relic density is attained. The effect of the rapid 3-body annihilations is to force $Y_{A^\prime} Y_\chi Y_{\chi,eq}^2 / Y_{A^\prime,eq} \approx Y_\chi^3$, where  $Y$ indicates the number density normalized to the entropy density, and the ``eq'' subscript indicates an equilibrium value. Since $m_{A^\prime} \approx 2 m_\chi$, the equilibrium number density of $A^\prime$ scales approximately as $e^{-2 m_\chi/T}$, and this requirement becomes $Y_\chi^2 \propto Y_{A^\prime}$, where the proportionality factor is almost constant with $T$. While this condition holds, the rates of $\chi \chi \bar{\chi} \rightarrow \chi A^\prime$ and its inverse process are comparable, and the net effect on both the $\chi$ and $A^\prime$ comoving densities is small, even though both are far from their equilibrium solutions. The Boltzmann equation thus leads to nearly-constant values for the comoving number densities of both $\chi$ and $A^\prime$. Provided that the 3-body annihilation decouples before the $A^\prime$'s decay away, the late-time $\chi$ abundance is thus set by freezeout of the 2-body annihilation. This is a special case of the general argument in \cite{Cline:2017tka} that the freezeout of the second-slowest process controls the late-time DM density.

Finally, more extreme versions of this scenario, with a closer tuning to the resonance, could lead to extremely large self-interaction cross sections at low velocities. In such a scenario, the distribution of this strongly-interacting component within halos would likely be significantly modified from the discussion in Sec.~\ref{sec:boost}; strong attractive interactions can give rise to a ``gravothermal catastrophe'' where the strongly-interacting component of the halo collapses to a dense cusp, potentially even forming a black hole \cite{Pollack:2014rja}. We leave a detailed study of this scenario to later work, but note that such a mechanism could potentially enhance the expected population of steeply cusped DM minihalos discussed in Sec.~\ref{sec:ucmh}, even if it is confirmed (as argued in \cite{Delos:2017thv,Gosenca:2017ybi}) that such UCMHs do not form and persist in standard CDM.

\subsection{Assisted dark matter}

A related class of models was suggested in Ref.~\cite{Dey:2016qgf}, which also invokes multi-body annihilation of light DM to obtain the correct relic density, with a focus on keV-MeV scalar DM. In particular, this work claims to identify viable parameter space for keV-scale DM where 4-body annihilation dominates freezeout. (They also present a model where 3-body annihilation dominates freezeout, but this scenario would be expected to have a very suppressed indirect detection signal, as the relevant 3-body annihilation channel involves a non-DM particle that is absent at late times.)

This model posits a real scalar DM candidate $\phi$ and real scalar ``assister'' field $S$, with a dark-sector $Z_2\times Z_2^\prime$ symmetry. The SM fields are even under both   $Z_2$ and $Z_2^\prime$, whereas the $\phi$ and $S$ fields are odd under $Z_2$ and $Z_2^\prime$ respectively. The dark-sector Lagrangian contains the operators:
\begin{equation} \mathcal{L}_\text{dark} \supset \frac{1}{2} m_\phi^2 \phi^2  + \frac{1}{2} m_S^2 S^2 + \frac{\lambda_{\phi S}}{4} \phi^2 S^2 + \frac{\lambda_\phi}{4!} \phi^4 + \frac{\lambda_S}{4!} S^4.\end{equation}
The $Z_2^\prime$ symmetry is then broken by adding an explicit interaction term between the $S$ field and the SM, which destabilizes $S$; Ref.~\cite{Dey:2016qgf} uses the leptophilic operator:
\begin{equation} \mathcal{L}_\text{int} \supset \lambda_i S \bar{l}_i l_i,\end{equation}
where $l_i$ denotes the SM leptons (and $i$ labels the flavor). Ref.~\cite{Dey:2016qgf} focuses on the case where the coupling is flavor-dependent, with $\lambda_\tau \gg \lambda_e, \lambda_\mu$, in order to evade constraints from measurements of $g-2$. This choice allows $\lambda_\tau$ as large as $0.3$ \cite{Dey:2016qgf}. The $S$ scalar in this case decays to two photons through a loop of $\tau$'s, with lifetime roughly $t_S = 1$ second for $\lambda_\tau = 0.3$.\footnote{Note that the related one-loop decay to tau neutrinos suffers a chirality suppression.
} In order to evade constraints on DM self-interactions, the authors of Ref.~\cite{Dey:2016qgf} also choose $\lambda_\phi \lesssim 10^{-6}$.
A number of 4-body annihilation channels involving $\phi$ and/or $S$ particles in the initial state contribute to freezeout. However, for $t \gg t_S$, i.e. after the $S$ scalars have decayed away, only the $\phi \phi \phi \phi \rightarrow S S$ channel will contribute to indirect detection signals. 

However, in our attempts to reproduce the results of this model, we identified an apparently significant omission in Ref.~\cite{Dey:2016qgf}, to wit the neglect of the ``forbidden'' 2-body channel $\phi \phi \rightarrow S S$. For $S$ heavier than $\phi$ this channel is kinematically forbidden at zero velocity, and consequently the rate of annihilation per volume per time experiences a Boltzmann suppression (including exponential factors only) of $\sim e^{-2 m_S/T}$, rather than the $e^{-2 m_\phi/T}$ factor one would expect from $\rho_\phi^2$. However, any 4-body process involving any number of $\phi$ or $S$ particles experiences a Boltzmann suppression (in the rate per unit volume per unit time) of at least $\rho_\phi^4 \propto e^{-4 m_\phi/T}$. (In both cases we are assuming the densities can be approximated by their equilibrium values up until freezeout.) In order for the $\phi \phi \phi \phi \rightarrow S S$ process to be open, we require $2 m_\phi > m_S$, and so the factor $e^{-4 m_\phi/T}$ constitutes a stronger exponential suppression than $e^{-2 m_S/T}$. Thus we generically expect the forbidden 2-body process to dominate over the open 4-body process. We have also checked numerically that including this forbidden process in the evolution equations dramatically changes the parameters required to obtain the correct relic density, indicating that it plays a dominant role in freezeout. The effect is to greatly reduce the coupling needed to obtain the correct relic density, which in turn strongly suppresses late-time signals from the 4-body annihilation.
        
\section{Conclusion}
\label{sec:conclusion}

We have explored the possible indirect-detection signatures of $n$-body DM annihilation processes that lead to the production of SM particles, either directly or through a decaying mediator. We have argued that such processes can be more strongly enhanced at low velocities than standard 2-body annihilation without violating partial-wave unitarity constraints; if the DM follows an approximately Maxwell-Boltzmann distribution, the unitarity-saturating temperature scaling is $\langle \sigma v^{n-1}\rangle \propto T^{(5 - 3n)/2}$.

We have calculated the average enhancement of the annihilation rate (i.e. the ``boost factor'') arising from inhomogeneities in the DM density distribution for 3- and 4-body processes, as relevant to both the early universe and to large DM halos in the present day. These results may have applications beyond the case of $n$-body DM annihilation. For example, 2-body annihilation or scattering followed by prompt interaction of SM products with the gas could yield signals scaling as the square of the DM density multiplied by the gas density; to the degree that the gas density approximately traces the DM density, such signals could experience a similar 3-body scaling.

We have estimated general constraints on 3- and 4-body annihilation processes via their modifications to the cosmic microwave background and photon line signals from the Galactic Center, the Galactic halo, and galaxy clusters. In all cases, there are potentially large uncertainties associated with the DM density profile and the degree of small-scale structure; while these uncertainties are also present for standard 2-body annihilation, their importance is increased for $n > 2$ due to the enhanced scaling of the signal with density. We have provided both optimistic and conservative sensitivity estimates, based on a broad range of models for the small-scale structure; in particular, we have shown that in the most optimistic case, such constraints cannot probe unitarity-saturating cross sections for DM masses above $\sim 1$ GeV in the 3-body case, or above  $\sim 1$ MeV in the 4-body case, unless the typical DM velocity in the system of interest is below $\sim 10^{-5} c$. Conversely, there are robust limits on unitarity-saturating cross sections for mass scales below $\sim 10-100$ MeV for 3-body annihilation and below  $\sim 10-30$ keV in the case of 4-body annihilation.

The relative sensitivity of different constraints varies depending on the degree of small-scale structure assumed; unless there are very few small halos or their concentration is low, limits from the CMB and galaxy clusters tend to dominate those from the central Milky Way (where substructure is assumed to be depleted), and from the Galactic halo (where substructures are neglected). This conclusion may not hold true if the Milky Way density profile is steeper than Einasto or one takes into account substructures within the halo.

We have mapped out the redshifts that dominate the CMB signal, and demonstrated that while for 2-body annihilation the signal is very generically dominated by high redshifts (and thus insensitive to the modeling of small halos), for 3-body and 4-body annihilation the signal can readily be dominated by either low or high redshifts depending on the small-scale structure model.

An alternate (but potentially overlapping) class of scenarios occurs where the $n$-body annihilation determines the late-time DM density through thermal freezeout. The needed cross section scales approximately as $\langle \sigma v^{n-1}\rangle \propto m_\chi^{2(2-n)}$; thus for $n > 2$, there is no mass-independent ``thermal relic cross section'' as in the 2-body case, and the sensitivity to thermal DM signals is greatly improved at low DM masses. In general, detecting thermal relic signals in the CMB or in local targets without low-velocity enhancement requires both a relatively low DM mass, below $\sim 200$ keV in the 3-body case and $\sim 30$ keV in the 4-body case, and an optimistic model for the degree of small-scale structure.

We have studied the ``not forbidden dark matter'' class of models as an example of a scenario where 3-body annihilation can dominate freezeout, and the annihilation produces particles that decay visibly in the late universe. We find that if the annihilation is not substantially enhanced at low velocities, the searches we have considered do not set any robust limits on the model parameter space that is not already excluded by other experiments; however, if there is an appreciable population ($\sim 1\%$ of the DM) of ultra-compact DM minihalos, we could potentially observe indirect signals from 3-body annihilation in the form of a new point source population and/or effects on the CMB.

This class of models possesses a resonant regime where the annihilation signal is strongly enhanced at low velocities -- even to a degree that exceeds the partial-wave unitarity bound, as many partial waves can contribute due to an effective long-range interaction. In this regime, we have demonstrated that it is possible to obtain visible signals in the cosmic microwave background, the Milky Way and galaxy clusters. However, in this class of scenarios such large enhancements also imply a resonant 2-body annihilation to SM particles, and a large DM-DM self-interaction cross section; the regions of parameter space that have not already been ruled out correspond to a secluded dark sector with a long-lived massive mediator to the Standard Model, where the 3-body-annihilating component may need to constitute only a small fraction of the DM in order to evade self-interaction limits.

\acknowledgments{We thank Eric Braaten for pointing out the potential for enhanced low-velocity scaling of 3-body processes, Ranjan Laha for early discussions on ultracompact minihalos, Jesse Thaler for introducing us to the muon-collider literature on $t$-channel resonances, Yu-Dai Tsai and Maxim Perelstein for discussions on the calculation of the unitarity bound, and JiJi Fan, Rebecca Leane, Hongwan Liu and Jes\'us Zavala for helpful conversations. This work was supported by the Office of High Energy Physics of the U.S. Department of Energy under grant Contract Numbers DE-SC0012567 and DE-SC0013999. Wu is partially supported by the Taiwan Top University Strategic Alliance (TUSA) Fellowship. TRS is partially supported by a John N. Bahcall Fellowship. TRS thanks the Galileo Galilei Institute for Theoretical Physics and the Kavli Institute for Theoretical Physics for hospitality during the completion of this work, and acknowledges partial support from the INFN, the Simons Foundation (341344, LA), and the National Science Foundation under Grant No. NSF PHY-1748958.}

\appendix 
\section{Calculation of the unitarity limit for $n \to 2$ annihilation}
\label{app:unitarity}
In this appendix, we calculate the unitarity bound on $\langle \sigma v^{n-1} \rangle$ starting from the optical theorem, Eq.~\eqref{eq:optical1}, which we reiterate here:
\begin{equation} \sum_i \frac{1}{S_i} \int d\Pi_n |\mathcal{M}_{f \rightarrow i}|^2 \le 2 \text{Im} \mathcal{M}(f \rightarrow f).\end{equation}
Note we have included the symmetry factor explicitly on the left-hand side, so the phase space needs no implicit restriction on the region of integration to avoid double-counting. Summing over the degrees of freedom in the final state as well, and then dividing by the number of degrees of freedom in both the initial and final states, we obtain:
\begin{equation} \frac{1}{S_i} \int d\Pi_n \overline{|\mathcal{M}_{f \rightarrow i}|^2} \le \frac{1}{g_1 \cdots g_{n+m}} \sum_f 2 \text{Im} \mathcal{M}(f \rightarrow f),\label{eq:optical} \end{equation}

In the non-relativistic limit, and assuming a Boltzmann distribution, the phase space distribution function $f_i$ can be related to the number density $(\rho_i/m_i)$ by: 
\begin{equation} g_i f_i = (\rho_i/m_i) \left(\frac{2\pi}{m_i T} \right)^{3/2} e^{-(E_i - m_i)/T}.\end{equation}

Following \cite{Kuflik:2017iqs}, we can substitute this result into Eq.~\eqref{eq:temp_average} to obtain, for $n\rightarrow m$ annihilation:
\begin{align}\langle \sigma v^{n-1}\rangle & = \frac{\prod_{i=n+1}^{m+n} g_i}{S_f} \left(\prod_{k=1}^n  \left(\frac{2\pi}{m_k T} \right)^{3/2} \right) e^{\sum_{k=1}^n m_k/T}  \nonumber \\
& \times \int \prod_{i=1}^{n+m} \frac{d^3 p_i}{(2\pi)^3 (2 E_i)} (2\pi)^4 \delta^{4}\left(\sum_{j=1}^n p_j - \sum_{j=n+1}^{n+m} p_j \right)  e^{-\sum_{k=1}^n \frac{E_k}{T}} \overline{|\mathcal{M}_{i\rightarrow f}|^2}, \nonumber \\
& = \frac{\prod_{i=n+1}^{m+n} g_i}{S_f} \left(\prod_{k=1}^n  \left(\frac{2\pi}{m_k T} \right)^{3/2} \right) e^{\sum_{k=1}^n m_k/T} \int \prod_{i=n+1}^{n+m} \frac{d^3 p_i}{(2\pi)^3 (2 E_i)} e^{-\sum_{k=n+1}^{m+n} \frac{E_k}{T}}  \nonumber \\
& \times \int d\Pi_n  \overline{|\mathcal{M}_{f \rightarrow i}|^2}, \end{align}
where we have used the delta function to rewrite the exponential terms from the distribution functions in terms of the $m$ final-state energies, rather than the $n$ initial-state energies. Substituting Eq.~\eqref{eq:optical} into this expression, specializing to the case of $m=2$, and inserting the identity $1 = \int \frac{d^4 p_0}{(2\pi)^4} (2\pi)^4 \delta^{4}(p_0 - p_{n+1} - p_{n+2})$, we obtain:
\begin{align}\langle \sigma v^{n-1}\rangle & \le \frac{2 S_i}{S_f \prod_{i=1}^{n} g_i} \left(\prod_{k=1}^n  \left(\frac{2\pi}{m_k T} \right)^{3/2} \right) e^{\sum_{k=1}^n m_k/T} \sum_f \int \frac{d^4 p_0}{(2\pi)^4} e^{-p_0^0/T} \nonumber \\
& \times \int  \frac{d^3 p_{n+1} d^3 p_{n+2}}{(2\pi)^6 (2 E_{n+1}) (2 E_{n+2})} (2\pi)^4 \delta^{4}(p_0 - p_{n+1} - p_{n+2})   \text{Im} \mathcal{M}(f \rightarrow f). \end{align}
The term on the second line is just the usual two-body phase space integral, which as previously can be rewritten as $d\Pi_2 = \frac{1}{16\pi^2} |\vec{p}_{n+1}|_\text{COM}/\sqrt{s} \int d\Omega_\text{COM}$, where ``COM'' subscripts indicate the center-of-mass frame. If we assume that the $s$-wave scattering dominates the elastic forward-scattering cross section (which is generic at low velocities, but need not hold if a long-range interaction is present), and the matrix element has no angular dependence, then our expression becomes:
\begin{align}\langle \sigma v^{n-1}\rangle & \le \frac{2 S_i}{S_f \prod_{i=1}^{n} g_i} \left(\prod_{k=1}^n  \left(\frac{2\pi}{m_k T} \right)^{3/2} \right) e^{\sum_{k=1}^n m_k/T} \sum_f \frac{1}{(2\pi)^3} \int_{s_\text{min}}^\infty ds \int_{\sqrt{s}}^\infty dE_0 e^{-E_0/T} \sqrt{E_0^2 - s}  \nonumber \\
& \times \frac{1}{8\pi} \sqrt{1 - \frac{2 (m_{n+1}^2 + m_{n+2}^2)}{s} + \frac{(m_{n+1}^2 - m_{n+2}^2)^2}{s^2}}  \text{Im} \mathcal{M}(f \rightarrow f)(s). \end{align}
Here we have used the fact that the center-of-mass momentum of the particles in the 2-body final state can be written $|\vec{p}_{n+1}|_\text{COM} = \frac{1}{2\sqrt{s}} \sqrt{s^2 - 2 (m_{n+1}^2 + m_{n+2}^2) s + (m_{n+1}^2 - m_{n+2}^2)^2}$, and the rewriting of $\int \frac{d^4 p_0}{(2\pi)^4} = \frac{1}{(2\pi)^3} \int_{s_\text{min}}^\infty ds \int_{\sqrt{s}}^\infty dE_0 \sqrt{E_0^2 - s}$. (Note we do not need to include a factor of $1/S_f$ when performing the angular integral over $d\Omega_\text{COM}$, as we have explicitly included this factor in the definition of the cross section.)

Now performing the integral over $E_0$, we obtain $\int_{\sqrt{s}}^\infty dE_0 e^{-E_0/T} \sqrt{E_0^2 - s} = \sqrt{s} T K_1(\sqrt{s}/T) \approx  \sqrt{s} T \sqrt{\pi T/(2 \sqrt{s})} e^{-\sqrt{s}/T}$, using the asymptotic expansion of the Bessel function $K_1$ for large argument, since we are working in the non-relativistic limit and hence can assume $T \ll \sqrt{s}$. Thus finally we obtain:
\begin{align}\langle \sigma v^{n-1}\rangle & \le \frac{1}{16 \pi^2} \left( \frac{T}{2\pi}\right)^{3(1 - n)/2} \frac{S_i}{S_f \prod_{i=1}^{n} g_i}   \prod_{k=1}^{n} m_k^{-3/2} e^{m_k/T} \nonumber \\
& \times \sum_f   \int_{(\sum_{i=1}^n m_i)^2}^\infty ds e^{-\sqrt{s}/T} s^{1/4}  \sqrt{1 - \frac{2 (m_{n+1}^2 + m_{n+2}^2)}{s} + \frac{(m_{n+1}^2 - m_{n+2}^2)^2}{s^2}}  \text{Im} \mathcal{M}(f \rightarrow f)(s). \end{align}

In the specific case studied in \cite{Kuflik:2017iqs}, $n=3$, $m_1 = m_2 = m_3 = m_4 = m_5 = m_\chi$, and all $g_i$ are set equal to 1 (so also there is only one relevant state $f$), and consequently we obtain:
\begin{align} \langle \sigma v^{n-1}\rangle & \le \frac{\pi\sqrt{15} }{6}  m_\chi^{-4} T^{-3} \frac{S_i}{S_f}   e^{3 m_\chi/T} \int_{9 m_\chi^2}^\infty ds e^{-\sqrt{s}/T} \text{Im} \mathcal{M}(f \rightarrow f)(s). \end{align}
Here we have approximated the non-exponential terms inside the integral by their values at $s_\text{min}$, since in this case they converge to a non-zero value in that limit. This result is in agreement with the result derived in \cite{Kuflik:2017iqs} up to the symmetry factors; we have checked with the authors of that work that they now agree with the $1/S_f$ factor. Our cross-section is defined differently than in that reference, by a factor of $S_i$, which accounts for the presence of the $S_i$ factor in the bound.  

To set a constraint on the forward-scattering matrix element, we can consider Eq.~\eqref{eq:optical1} for the $s$-wave forward scattering, where the initial and final states are both set equal to $f$ and each particle has COM 3-momentum of magnitude $|\vec{k}|$:
\begin{equation} \frac{1}{4\pi S_f} (|\vec{k}|/\sqrt{s}) |\mathcal{M}(f\rightarrow f)|^2 \le 2 \text{Im}\mathcal{M}(f\rightarrow f) \le 2 |\mathcal{M}(f\rightarrow f)|\end{equation}
Consequently, we obtain:
\begin{equation} \text{Im}(\mathcal{M}(f\rightarrow f))\le |\mathcal{M}(f\rightarrow f)| \le 8 \pi S_f \sqrt{s}/|\vec{k}|.\end{equation}

Writing the $s$-wave contribution to the elastic forward-scattering cross-section as $\sigma_{f \rightarrow f} = \frac{1}{S_f} \frac{1}{16\pi s} |\mathcal{M}_{f\rightarrow f}|^2 \le \frac{4\pi S_f}{|\vec{k}|^2}$, we recover the usual two-body partial-wave unitarity bound.

Substituting into our equation for $\langle \sigma v^{n-1}\rangle$, we obtain:
\begin{align}\langle \sigma v^{n-1}\rangle & \le \frac{S_i}{\pi} \left( \frac{T}{2\pi}\right)^{3(1 - n)/2} \frac{g_4 g_5}{\prod_{i=1}^{n} g_i}   \prod_{k=1}^{n} m_k^{-3/2} e^{m_k/T} \int_{(\sum_{i=1}^n m_i)^2}^\infty ds e^{-\sqrt{s}/T} s^{1/4}. \end{align}
Again working in the non-relativistic approximation, the integral can be approximated as $2 T (\sum_{i=1}^n m_i)^{3/2} e^{-\sum_{i=1}^n m_i/T}$, yielding:
\begin{align}
\langle \sigma v^{n-1}\rangle & \le 2^{-\frac{1}{2} + \frac{3}{2}n} \left(\frac{T}{\pi} \right)^{-(3n - 5)/2} S_i \frac{g_4 g_5}{g_1\cdots g_n}   \left( \frac{m_1 + \cdots + m_n}{m_1 \cdots m_n} \right)^{3/2}. 
\end{align}

\newpage
\bibliographystyle{JHEP0}
\bibliography{3body}

\end{document}